\documentclass[12pt]{article}

\usepackage[utf8x]{inputenc}      
\usepackage{hyperref}
\usepackage{amsmath,amssymb,mathtools}
\usepackage{cleveref}
\usepackage[T1]{fontenc}          
\usepackage{booktabs,tabularx}
\usepackage{graphicx,subfig}
\usepackage{xspace}
\usepackage[usenames]{xcolor}\definecolor{fscolor}{RGB}{44,118,255}
\usepackage{geometry}
\usepackage{todonotes}
\usepackage{listings}
\usepackage[absolute]{textpos}
\usepackage[many]{tcolorbox}
\usepackage{xparse}
\usepackage[font=small,labelfont=bf,format=plain,margin=0.05\textwidth]{caption}
\usepackage{bbm}
\usepackage{tabularx}
\usepackage{cite}
\usepackage{soul}

\crefname{figure}{Fig.}{Figs.}
\crefname{equation}{Eq.}{Eqs.}
\crefname{section}{Sec.}{Secs.}
\crefname{appendix}{Appendix}{Appendices}
\crefname{subappendix}{Appendix}{Appendices}

\allowdisplaybreaks

\evensidemargin \oddsidemargin
\newcommand{\mf}{\hat \mu}
\newcommand{\mfC}{\hat \mu^*}
\newcommand{\amf}{\vert \hat \mu \vert}

\newcommand{\sbe}{s_\beta}
\newcommand{\stb}{s_{2\beta}}
\newcommand{\cbe}{c_\beta}
\newcommand{\ctb}{c_{2\beta}}
\newcommand{\tbe}{t_\beta}
\newcommand{\sbb}{s_\beta^2}
\newcommand{\cbb}{c_\beta^2}

\newcommand{\DR}{{\ensuremath{\overline{\text{DR}}}}\xspace}

\newcommand{\MS}{{\ensuremath{\overline{\text{MS}}}}\xspace}

\newcommand{\SM}{{\text{SM}}}

\newcommand{\THDM}{{\text{THDM}}}
\newcommand{\MSSM}{{\text{MSSM}}}

\newcommand{\SUSY}{{\text{SUSY}}}

\newcommand{\EFT}{{\text{EFT}}}
\newcommand{\HET}{{\text{HET}}}

\newcommand{\Pdd}{\Phi_1^\dagger\Phi_1}
\newcommand{\Puu}{\Phi_2^\dagger\Phi_2}
\newcommand{\Pdu}{\Phi_1^\dagger\Phi_2}
\newcommand{\Pud}{\Phi_2^\dagger\Phi_1}

\newcommand{\FH}{\mbox{{\tt FeynHiggs}}\xspace}

\newcommand{\Eq}[1]{Eq.~(\ref{#1})}

\renewcommand{\Re}{\text{Re}}
\renewcommand{\Im}{\text{Im}}

\newcommand{\mtL}{m_{\tilde{t}_L}}
\newcommand{\mtR}{m_{\tilde{t}_R}}

\newcommand{\cp}{\ensuremath{{\cal CP}}}

\newcommand{\msusy}{\ensuremath{M_\SUSY}\xspace}

\newcommand{\hmg}{\ensuremath{\widehat M_3}\xspace}
\newcommand{\hmgC}{\ensuremath{\widehat M_3^*}\xspace}

\newcommand{\xt}{\ensuremath{\widehat X_t}\xspace}

\newcommand{\at}{\ensuremath{\widehat A_t}\xspace}
\newcommand{\atC}{\ensuremath{\widehat A_t^{*}}\xspace}
\newcommand{\aat}{\ensuremath{\vert \widehat A_t \vert}\xspace}

\newcommand{\tev}{\,\, \mathrm{TeV}}
\newcommand{\gev}{\,\, \mathrm{GeV}}

\newcommand{\order}[1]{\ensuremath{{\cal O}(#1)}}
\newcommand{\DiLog}[1]{\ensuremath{\text{Li}_2(#1)}}
\newcommand{\al}{\alpha}
\newcommand{\als}{\al_s}
\newcommand{\alt}{\al_t}

\newcommand{\Hm}{{H^-}}
\newcommand{\Hp}{{H^+}}

\newcommand{\Gm}{{G^-}}
\newcommand{\Gp}{{G^+}}

\newcommand{\li}{{\lambda_1}}
\newcommand{\lii}{{\lambda_2}}
\newcommand{\liii}{{\lambda_3}}
\newcommand{\liv}{{\lambda_4}}
\newcommand{\lv}{{\lambda_5}}
\newcommand{\lvi}{{\lambda_6}}
\newcommand{\lvii}{{\lambda_7}}
\newcommand{\htp}{{h_t'}}

\newcommand{\phiI}{{\phi_1}}
\newcommand{\phiII}{{\phi_2}}
\newcommand{\chiI}{{\chi_1}}
\newcommand{\chiII}{{\chi_2}}
\newcommand{\phiIp}{{\phi_1^+}}
\newcommand{\phiIIp}{{\phi_2^+}}
\newcommand{\phiIm}{{\phi_1^-}}
\newcommand{\phiIIm}{{\phi_2^-}}

\begin{document}

\thispagestyle{empty}
\def\thefootnote{\fnsymbol{footnote}}

\begin{flushright}
DESY 20-153
\end{flushright}
\vspace{3em}
\begin{center}
{\Large\bf
Two-loop matching of renormalizable operators: \\
general considerations and applications
}
\\
\vspace{3em}
{
Henning Bahl$^{a\,\hspace{-0.25mm}}$\footnote{email: henning.bahl@desy.de},
Ivan Sobolev$^{a,b\,\hspace{-0.25mm}}$\footnote{email: ivan.sobolev@desy.de}
}\\[2em]
{\sl $^{a}$ Deutsches Elektronen-Synchrotron DESY, Notkestra{\ss}e 85, D-22607 Hamburg, Germany} \\[1em]
{\sl $^{b}$ {Institute for Nuclear Research of the Russian Academy of Sciences,\\ 60th October Anniversary prospect 7a, Moscow 117312, Russia}}
\def\thefootnote{\arabic{footnote}}
\setcounter{page}{0}
\setcounter{footnote}{0}
\end{center}
\vspace{2ex}
\begin{abstract}
{}

Low-energy effective field theories (EFT) encode information about the physics at high energies---i.e., the high-energy theory (HET). To extract this information the EFT and the HET have to be matched to each other. At the one-loop level, general results for the matching of renormalizable operators have already been obtained in the literature. In the present paper, we take a step towards a better understanding of renormalizable operator matching at the two-loop level: Focusing on the diagrammatic method, we discuss in detail the various contributions to two-loop matching conditions and compare different approaches to derive them. Moreover, we discuss which observables are best suited for the derivation of matching conditions. As a concrete application, we calculate the \order{\alt\als} and \order{\alt^2} matching conditions of the scalar four-point couplings between the Standard Model (SM) and the Two-Higgs-Doublet Model (THDM) as well as the THDM and the Minimal Supersymmetric Standard Model (MSSM). We use the derived formulas to improve the prediction of the SM-like Higgs mass in the MSSM using the THDM as EFT.

\end{abstract}

\newpage
\tableofcontents
\newpage
\def\thefootnote{\arabic{footnote}}


\section{Introduction}
\label{sec:01_intro}

No direct evidence for physics beyond the Standard Model (SM) has been found at the Large Hadron Collider (LHC) yet. In the absence of a clear beyond SM (BSM) signal, effective field theories (EFTs) are an increasingly popular approach to interpret the constraints set by LHC measurements. To obtain constraints on the parameters of the HET theory in an EFT framework, the heavy HET particles are integrated out at their mass scale. At this matching scale, the EFT parameters are related to the HET parameters via matching conditions. Using renormalization-group equations (RGEs), the EFT parameters are then run down to a low-energy scale (typically the electroweak scale). At this scale, physical observables are calculated and compared to experimental data. Due to the plethora of imaginable HET theories and also of EFTs (i.e., if not all BSM particles are heavy), the automation of this process is highly desirable.

An automated calculation of the needed RGEs at the one- and two-loop level (and the three-loop level for gauge couplings) is already available in the form of the computer codes \texttt{SARAH}~\cite{Staub:2009bi,Staub:2010jh,Staub:2012pb,Staub:2013tta} and \texttt{PyR@te}~\cite{Lyonnet:2013dna,Lyonnet:2016xiz,1808891}, which implement the results of~\cite{Machacek:1983tz,Machacek:1983fi,Machacek:1984zw,Luo:2002ti,Sperling:2013eva,Sperling:2013xqa,Bednyakov:2018cmx,Schienbein:2018fsw,Poole:2019kcm}. Also generally applicable formulas for the derivation of one-loop matching conditions have been derived. While in recent years, most efforts concentrated on the one-loop matching of higher-dimensional operators has been considered in~\cite{Henning:2014wua,Drozd:2015rsp,delAguila:2016zcb,Boggia:2016asg,Henning:2016lyp,Ellis:2016enq,Fuentes-Martin:2016uol,Zhang:2016pja,Ellis:2017jns,Summ:2018oko,Bakshi:2018ics,Kramer:2019fwz,Cohen:2019btp,Ellis:2020ivx}, also the one-loop matching of renormalizable operators~\cite{Braathen:2018htl,Gabelmann:2018axh} can provide information about the HET. In contrast, no general two-loop results are available yet.\footnote{For EFTs with a SM-like Higgs sector, the calculation of the two-loop matching condition for the Higgs self-coupling has been automatized by matching the Higgs-boson mass~\cite{Kwasnitza:2020wli}. This procedure partially includes terms which would normally be associated with higher-dimensional operators.} Especially for renormalizable operators, their impact is, however, often relevant.

One prominent example is the SM as an EFT of the Minimal Supersymmetric Standard Model (MSSM). Due to supersymmetry (SUSY), the SM Higgs self-coupling is determined by the other MSSM parameters at the SUSY scale. This property can be used to predict the Higgs mass in terms of the MSSM parameters, making it possible to derive bounds on the SUSY scale even if all SUSY particles are far beyond the reach of direct LHC searches. To fully exploit the experimental precision, the calculation of higher-order matching conditions is mandatory. Correspondingly, many efforts have been dedicated to derive the full one-loop~\cite{Giudice:2011cg,Draper:2013oza,Bagnaschi:2014rsa,Bagnaschi:2017xid} as well as partial two-loop~\cite{Draper:2013oza,Bagnaschi:2014rsa,Vega:2015fna,Bagnaschi:2017xid,Bagnaschi:2019esc,Bahl:2020tuq} and three-loop~\cite{Harlander:2018yhj} corrections in the simplest case of the SM as an EFT. Still, the remaining theoretical uncertainty is considered to be significantly higher than the experimental uncertainty~\cite{Bagnaschi:2014rsa,Vega:2015fna,Bagnaschi:2017xid,Allanach:2018fif,Bahl:2019hmm}. Less precise results are available if the EFT is not the SM but, e.g., a Two-Higgs-Doublet Model (THDM) allowing to consider the effect of relatively light non-SM Higgs bosons~\cite{Haber:1993an,Lee:2015uza,Bagnaschi:2015pwa,Athron:2017fvs,Benakli:2018vqz,Bahl:2018jom,Murphy:2019qpm}.

The aim of this paper is to take the first step into the direction of systematizing the calculation of higher-order matching conditions of renormalizable operators and to discuss various applications. Focusing on diagrammatic matching, we explain in \cref{sec:02_method} the various contributions to a matching condition (providing explicit one- and two-loop formulas). Moreover, we discuss how Ward identities ensure that matching conditions derived using different observables agree with each other and which observables are best suited for the derivation of matching conditions. Also, different approaches to treat the light masses are compared. In \cref{sec:03_application}, we discuss a couple of simple examples highlighting the main points of \cref{sec:02_method}. Moreover, we discuss non-trivial two-loop applications: the \order{\alt\als} and \order{\alt^2} matching conditions for the Higgs four-point couplings between the SM and the THDM as well as the THDM and the MSSM are calculated. The obtained expressions are used in \cref{sec:04_num_results} to improve the calculation of the SM-like Higgs mass in the MSSM using the THDM as EFT, which is implemented into the public code \FH~\cite{Heinemeyer:1998yj,Heinemeyer:1998np,Hahn:2009zz, Degrassi:2002fi,Frank:2006yh,Hahn:2013ria,Bahl:2016brp,Bahl:2017aev,Bahl:2018qog}.


\section{Matching effective field and high-energy theories}
\label{sec:02_method}

In this Section, we review the theoretical foundations of matching a high energy theory (HET) to an effective field theory (EFT). We explicitly discuss the calculation of matching conditions for renormalizable operators at the two-loop level and how it can be simplified.

\medskip

Without loose of generality, we assume the HET to contain  ``light'' scalar fields, $\phi_i$, and ``heavy'' scalar fields, $\Phi_i$. We assume that all ``light'' fields have masses close to the scale $m$; the ``heavy'' fields, to the scale $M$ with $m \ll M$. The ``heavy'' fields are decoupled at a scale $Q\sim M$. The resulting EFT involves only the ``light'' fields. To ensure that the low-energy EFT and the HET yield the same physical predictions at the matching scale $Q$ in the limit of the heavy masses going to infinity, both theories have to be matched. This means that the parameters of the low-energy EFT are not free parameters but are fixed in terms of the HET parameters.

We restrict us here to the case that the gauge symmetries of the HET remain unbroken when integrating out the heavy fields. Moreover, we focus on the calculation of matching conditions for parameters with a mass dimension of zero or higher. This means that we do not discuss the calculation of matching conditions for dimension-five (or higher) operators, but for operators of dimension-four or lower.

The matching conditions for the EFT parameters can be derived in different ways. One possibility is to use the effective action: the integral over the ``heavy'' degrees of freedom in the path integral can be performed explicitly; alternatively, the effective action can be calculated up to the quartic order, and after that, equations of motions can be used to integrate out the ``heavy'' fields (see~\cite{Braathen:2018htl} for a detailed discussion of these approaches at the one-loop level). Instead of employing the effective action, we focus in this paper on the diagrammatic method restricting us to the matching of renormalizable operators. Diagrammatic matching means that the results for low-energy physical observables, calculated in terms of amplitudes involving only light external fields, are compared between the HET and the EFT. The matching conditions are then derived by requiring the results to be equal order by order in the expansion around $m/M \sim 0$.


\subsection{Diagrammatic matching}
\label{sec:matching}

The calculation of physical observables consists out of two parts: the truncated connected Green's function and external leg corrections, $Z$. For the purpose of calculating matching conditions, it is sufficient to consider only Green's functions which are one-particle irreducible in the ``light'' particle lines. If not specified otherwise, we will call these 1LPI graphs. We assume all these components to be renormalized in the \MS scheme (or the \DR scheme for supersymmetric theories).

We denote a 1LPI Green's function with $n$ external light fields by $\Gamma_{i_1..i_n}$, where the indices $i_{1..n}$ are used to specify the external fields. Since we are interested in low-energy processes when using an EFT, we can assume the external momenta, on which $\Gamma_{i_1..i_n}$ depends, to be of order $\sim m$. In our notation, the dependence of the 1LPI Green's function on them is suppressed.

We denote the 1LPI Green's functions of the HET as
\begin{align}
\Gamma_{i_1..i_n}^\text{\HET} = \Gamma_{i_1..i_n}^\text{\HET}(g^\HET_1,..,g^\HET_q, p_{i_1},..,p_{i_n},\mu_R),
\end{align}
where $q$ is the number of parameters in the HET (including couplings, mass parameters and trilinear couplings). $p_i$ is the external momentum associated with the $i$-th external field (with $p_i\sim m$ as mentioned above). $\mu_R$ is the renormalization scale which is set equal to the matching scale $Q$. The 1LPI Green's functions of the EFT are denoted as
\begin{align}
\Gamma_{i_1..i_n}^\text{\EFT} = \Gamma_{i_1..i_n}^\text{\EFT}(g^\EFT_1,..,g^\EFT_p, p_{i_1},..,p_{i_n},\mu_R),
\end{align}
where $p$ is the number of parameters in the EFT, which can be smaller---e.g.\ in case a much smaller number of fields in the EFT---or larger than the number of HET parameters (even if no higher-dimensional operators are considered)---e.g.,\ in case of the HET being more symmetric than the EFT.

As a consequence of the decoupling theorem~\cite{Appelquist:1974tg}, the difference of all Green's function with a negative mass dimension between the HET and the EFT will be suppressed by powers of $m/M$.\footnote{The HET Green's function can also depend on powers of $M/m$. These terms are, however, canceled by the EFT Green's function after reparametrizing it in terms of HET parameters. One example of such a case is the matching of the MSSM to the MSSM without gluino (see e.g. \cite{Muhlleitner:2008yw,Bahl:2019wzx}).} Since we focus only on the matching of dimension-four (or lower) operators, this corresponds to
\begin{align}\label{eq:NegDimZero}
\Gamma_{i_1..i_n}^\text{\HET} - \Gamma_{i_1..i_n}^\text{\EFT} = 0, \text{ if } \left[\Gamma_{i_1..i_n}\right] < 0,
\end{align}
where the square brackets are used to denote the mass dimension of the Green's functions.


\subsubsection{Field normalization}
\label{sec:FieldNormalization}

The external leg corrections ensure that the external fields fulfill proper on-shell conditions. I.e., they relate the fields used for the calculation of the Feynman diagrams to the external asymptotically free fields,
\begin{align}\label{eq:phi_phys}
\phi_i^\text{physical} = \sum_{j=1}^l\sqrt{Z_i}Z_{ij}\phi_i,
\end{align}
where $l$ is the number of fields mixing with $\phi_i$, $Z_i$ is the LSZ-factor and $Z_{ij}$ accounts for the mixing of the fields. For abbreviation we will use the $\mathbf{Z}$-matrix,
\begin{align}
\mathbf{Z}_{ij} = \sqrt{Z_i}Z_{ij} = \delta_{ij} + \Delta^{(1)}Z_{ij} + \Delta^{(2)}Z_{ij} + \ldots,
\end{align}
where the ellipsis denotes higher order terms in the loop expansion. The elements of the $\mathbf{Z}$ are calculated in terms of renormalized two-point vertex functions (see e.g.~\cite{Frank:2006yh,Fuchs:2016swt,Fuchs:2017wkq}), which are equivalent to the two-point 1LPI Green's functions. Note that the $\mathbf{Z}$ is in general not Hermitian, since the two-point functions are evaluated at the on-shell momentum of the respective external particle.

The two-point 1LPI Green's functions can be organized in a matrix, $\mathbf{\hat\Gamma}$, with the elements\footnote{For simplicity, we will work here in the mass eigenstate basis. The presented arguments are, however, also valid if the mass matrix is not diagonal.}
\begin{align}
\widehat\Gamma_{ij}(p^2) = \delta_{ij} (p^2 - m_{i}^2) + \widehat\Sigma_{ij}(p^2),
\end{align}
where the symbol ``$\;\widehat{\;}\;$'' is used to denote the renormalized self-energies. The two-point functions are evaluated at the physical masses of the external ``light'' fields $\phi_i$ and $\phi_j$ which are $\sim m$ (implying that $p^2\sim m^2$). In the HET, mixing of ``light'' fields with ``heavy'' fields can occur. This mixing is, however, always suppressed by powers of $m/M$ and can, therefore, be neglected.\footnote{The mixing appears in the form $\widehat\Sigma_{ij}(p^2)/(p^2 - M_j^2)$. Since  $\widehat\Sigma_{ij}(p^2)$ contains no terms proportional to positive powers of the heavy mass scale $M$ after taking into account tadpole contributions (and $p^2\sim m^2$), the whole term is suppressed by the heavy mass appearing in the denominator.}

The dependence of the renormalized two-point vertex functions on the field renormalization can be made explicit by
\begin{align}
\mathbf{\widehat\Gamma}(p^2) = \left(\mathbf{1} + \frac{1}{2}\mathbf{\delta Z}\right)^\dagger\cdot\mathbf{\widetilde\Gamma}(p^2)\cdot\left(\mathbf{1} + \frac{1}{2}\mathbf{\delta Z}\right)
\end{align}
where the symbol ``$\;\widetilde{\;}\;$'' is used to denote the renormalized two-point function including no field renormalization but all other renormalization constants (e.g.\ mass renormalization constants).

In the limit $p,m \ll M$, the matrix $\mathbf{\Gamma}$ is Hermitian. Since we are only interested in the relative difference in the field normalizations of the two theories, we can, moreover, set the external momenta to zero for our purpose of matching two theories.\footnote{In the limit $p,m \ll M$, all $p^2$-dependent terms in the difference of the field normalizations between the two theories cancel.} Therefore, we can also assume the field renormalization matrix, $\mathbf{\delta Z}$, to be Hermitian.

Correspondingly, the renormalized self-energy is given by
\begin{align}\label{eq:ren_se}
\widehat\Sigma_{ij}(p^2) =&{} \widetilde\Sigma_{ij}(p^2) + (p^2 - m_i^2) \delta Z_{ij}  + \frac{1}{4} \delta Z_{ia} (p^2 - m_a^2) \delta Z_{aj} \nonumber\\
& + \frac{1}{2} \delta Z_{ia} \widetilde\Sigma_{aj}(p^2) + \frac{1}{2} \widetilde\Sigma_{ia}(p^2) \delta Z_{aj} + \frac{1}{4} \delta Z_{ia} \widetilde\Sigma_{ab}(p^2) \delta Z_{bj},
\end{align}
where the $\delta Z_{ij}$'s are the elements of $\mathbf{\delta Z}$ and summation over the indices $a$ and $b$ is implied.

For the calculation of matching conditions, we need to make sure that the fields of the EFT and the HET share a common normalization. Typically, this is done by matching the derivatives of the two-point vertex functions ,
\begin{align}\label{eq:dTwopoint}
\frac{\partial}{\partial p^2}\widehat{\Gamma}_{ij} =&{} \delta_{ij} + \widetilde\Sigma^{\prime}_{ij}(p^2) + \delta Z_{ij} + \frac{1}{4}\delta Z_{ia}\delta Z_{aj} + \frac{1}{2} \delta Z_{ia} \widetilde\Sigma^{\prime}_{aj} + \frac{1}{2} \widetilde\Sigma^{\prime}_{ia} \delta Z_{aj} + \frac{1}{4} \delta Z_{ia} \widetilde\Sigma^{\prime}_{ab} \delta Z_{bj}
\end{align}
where the prime is used to denote the derivative with respect to the external momentum. The difference of higher-order derivatives with respect to the external momentum between the HET and the EFT are suppressed by powers of $m/M$ since the mass dimension is negative. Therefore, the momentum at which the derivative of the two-point function is matched is irrelevant. Consequently, after performing this matching (and also the matching of the mass parameters) the two-point vertex function of the EFT and the HET are equal up to powers of $m/M$.

In the case of no mixing, \cref{eq:dTwopoint} simplifies to
\begin{align}
\frac{\partial}{\partial p^2}\widehat{\Gamma}(p^2) =&{} \left(1 + \widetilde\Sigma^{\prime}(p^2)\right)\left(1 + \frac{1}{2}\delta Z\right)^2 ,
\end{align}
In this case, the finite part of field renormalization in the HET theory can be chosen according to\footnote{We assume the UV-divergent parts of the EFT and HET self-energies to be already absorbed by appropriate UV-infinite renormalization constants.}
\begin{align}
\delta Z^\HET &= 2\left(\sqrt{\frac{1 + \widetilde\Sigma^{\EFT\prime}}{1 + \widetilde\Sigma^{\HET\prime}}} - 1\right),
\end{align}
where we have set the finite part of $\delta Z^\EFT$ to zero. This choice guarantees that the HET and the EFT fields share the same normalization and corresponds to the ``heavy-OS'' scheme defined in~\cite{Bahl:2018ykj}.\footnote{In~\cite{Bahl:2018ykj}, the ``heavy-OS'' renormalization condition looks slightly different, since the subloop-renormalization contribution is not written out explicitly but assumed to be contained implicitly in $\widetilde\Sigma^\prime$.}

At the one- and two-loop level, the field renormalization constants are given by
\begin{align}
\delta^{(1)} Z^\HET &= - \Delta\widetilde\Sigma^{(1)\prime}, \\
\delta^{(2)} Z^\HET &= - \Delta\widetilde\Sigma^{(2)\prime} + \frac{3}{4}\left(\Delta\widetilde\Sigma^{(1)\prime}\right)^2 + \Delta\widetilde\Sigma^{(1)\prime}\cdot\widetilde\Sigma^{\EFT,(1)\prime},
\end{align}
where $\Delta\widetilde\Sigma = \widetilde\Sigma^\HET - \widetilde\Sigma^\EFT$.

In the case of mixing, we obtain at the one- and two-loop level
\begin{align}
\delta^{(1)} Z_{ij}^\HET &= - \Delta\widetilde\Sigma^{(1)\prime}_{ij}, \\
\delta^{(2)} Z_{ij}^\HET &= - \Delta\widetilde\Sigma^{(2)\prime}_{ij} - \frac{1}{4}\Delta\widetilde\Sigma^{(1)\prime}_{ia}\Delta\widetilde\Sigma^{(1)\prime}_{aj} + \frac{1}{2}\Delta\widetilde\Sigma^{(1)\prime}_{ia}\widetilde\Sigma^{\HET,(1)\prime}_{aj} + \frac{1}{2}\widetilde\Sigma^{\HET,(1)\prime}_{ia}\Delta\widetilde\Sigma^{(1)\prime}_{aj}.
\end{align}
As argued above, using the ``heavy-OS'' scheme for the field renormalization in the HET (and the matching of the mass parameters) implies that the two-point vertex functions are identical in the EFT and the HET. As a direct consequence, also the elements of the $\mathbf{Z}$-matrix, calculated out of two-point vertex functions, are equal to each other in both theories. This justifies matching the derivatives of the two-point vertex functions. Even though the field renormalization drops out the calculation of physical observables, using the ``heavy-OS'' scheme provides a convenient method to implement this matching in practice and to avoid the calculation of the $\mathbf{Z}$-matrix when matching two theories. In the literature, this contribution to matching conditions is often called ``wave-function renormalization'' (WFR) contribution.


\subsubsection{Calculation of matching condition}
\label{sec:calcMatching}

The matrix element for a given process is then obtained by multiplying the Green's functions with the external $\mathbf{Z}$-factors,
\begin{align}
\mathcal{M}_{i_1..i_n} =&{} \sum_{j_{1}=1}^{l} \sum_{j_{2}=1}^{l}\ldots \sum_{j_{n} = 1}^{l}\left(\prod_{a=1}^{n}\mathbf{Z}_{i_a j_a}\right) \Gamma_{j_1..j_n}.
\end{align}
Expanding this expression up to the tree, the one-loop and the two-loop level reads
\begin{align}
\mathcal{M}_{i_1..i_n}^{(0)} =&{} \Gamma_{i_1..i_n}^{(0)}, \\
\mathcal{M}_{i_1..i_n}^{(1)} =&{} \Gamma_{i_1..i_n}^{(1)} + \sum_{a=1}^{n}\sum_{j_a=1}^{l}\Delta^{(1)}Z_{i_a j_a}\cdot\Gamma_{i_1..i_n,i_a\rightarrow j_a}^{(0)},\\
\mathcal{M}_{i_1..i_n}^{(2)} =&{} \Gamma_{i_1..i_n}^{(2)} + \sum_{a=1}^{n}\sum_{j_a=1}^{l}\Delta^{(1)}Z_{i_a j_a}\cdot\Gamma_{i_1..i_n,i_a\rightarrow j_a}^{(1)}  + \sum_{a=1}^{n}\sum_{j_a=1}^{l}\Delta^{(2)}Z_{i_a j_a}\cdot\Gamma_{i_1..i_n,i_a\rightarrow j_a}^{(0)} \nonumber\\
& + \sum_{\substack{a,b=1, \\ a\neq b}}^{n}\sum_{j_a,j_b=1}^{l}\Delta^{(1)}Z_{i_a j_a}\cdot\Delta^{(1)}Z_{i_b j_b}\cdot\Gamma_{i_1..i_n,i_a\rightarrow j_a,i_b\rightarrow j_b}^{(0)},
\end{align}
where we use the notation $i\rightarrow j$ to denote that the index~$i$ should be replaced by the index~$j$. As discussed in \cref{sec:FieldNormalization}, the $Z$-factors can be omitted in the matching calculation if the light fields of the HET theory are renormalized in the ``heavy-OS'' scheme.

The matching conditions for an EFT parameter $g^\EFT_r$ can then be derived by demanding that physical amplitudes, depending on $g^\HET_r$ ($1 \le r \le p$), are equal in the HET and the EFT at the scale $Q$ in the limit of $M\rightarrow\infty$ or equivalently $p^2,m\rightarrow 0$,\footnote{Actually, the squared matrix elements enter the calculation of physical observables. Consequently, overall phases of the matrix element drop out of the calculation. Unphysical phases can be absorbed by redefining the fields and therefore play no role in the calculation of the matching condition. For physical phases, however, it is always possible to find a process sensitive to the phase even after taking the absolute value of the amplitude. Alternatively, one can just match the complex amplitude (as done in \cref{eq:AmplitudeMatching}).}
\begin{align}\label{eq:AmplitudeMatching}
\mathcal{M}_{i_1..i_n}^\EFT(g^\EFT_1,..,g^\EFT_r,..,g^\EFT_p) = \mathcal{M}_{i_1..i_n}^\HET(g^\HET_1,..,g^\HET_r,..,g^\HET_q).
\end{align}
Writing the matching condition for the coupling $g^\EFT_r$ in the form
\begin{align}
g^\EFT_r = g^\text{tree}_r + \Delta^{(1)} g_r + \Delta^{(2)} g_r + \ldots.
\end{align}
we obtain
\begin{align}
\label{eq:matching1L}
\Delta^{(1)} g_r = \bigg(\frac{\partial}{\partial g_r}\mathcal{M}^{\EFT,(0)}_{i_1..i_n}\bigg)^{-1} \Bigg[& \mathcal{M}^{\HET,(1)}_{i_1..i_n} - \mathcal{M}^{\EFT,(1)}_{i_1..i_n} - \sum_{i=1,i \ne r}^p \Delta^{(1)} g_i \cdot \frac{\partial}{\partial g_i}\mathcal{M}^{\EFT,(0)}_{i_1..i_n}\Bigg], \\
\label{eq:matching2L}
\Delta^{(2)} g_r = \bigg(\frac{\partial}{\partial g_r}\mathcal{M}^{\EFT,(0)}_{i_1..i_n}\bigg)^{-1} \Bigg[& \mathcal{M}^{\HET,(2)}_{i_1..i_n} - \mathcal{M}^{\EFT,(2)}_{i_1..i_n} \nonumber\\
& - \sum_{i=1,i \ne r}^p \Delta^{(2)} g_a \cdot \frac{\partial}{\partial g_a}\mathcal{M}^{\EFT,(0)}_{i_1..i_n} - \sum_{i=1}^p \Delta^{(1)} g_a \cdot \frac{\partial}{\partial g_a}\mathcal{M}^{\EFT,(1)}_{i_1..i_n}\nonumber\\
& - \frac{1}{2}\sum_{a,b=1}^p \Delta^{(1)} g_a \Delta^{(1)} g_b \cdot \frac{\partial^2}{\partial g_a\partial g_b}\mathcal{M}^{\EFT,(0)}_{i_1..i_n}\Bigg],
\end{align}
where all quantities on the right side have to be evaluated using the HET parameters (i.e., the EFT parameters are replaced by their tree-level matching condition). The prefactors before the square brackets parametrize the dependence of the tree-level amplitude on $g_r$. The terms containing partial derivatives arise through the reparametrization of the EFT amplitudes in terms of HET couplings. While this reparametrization is always possible, expressing the HET amplitudes in terms of EFT couplings is not always possible (see \cite{Kwasnitza:2020wli} for an extensive discussion on the parameterization of threshold corrections).

Finally, note that the expressions given by \Eq{eq:matching1L} and \Eq{eq:matching2L} are written in terms of amplitudes for physical processes and do not depend on the choice of the renormalization scheme for the field-renormalization constants. However, as we already noticed above in this Section, in the ``heavy-OS'' scheme the external $Z$ factors have the same form both in the EFT and in the HET. So, one can extract the matching coefficients, $\Delta g_r$, by matching the 1LPI Green's functions instead of the full physical amplitudes. We will employ this procedure throughout this paper.


\subsection{Choice of observable}
\label{sec:ObsChoice}

The matching condition for a specific EFT parameter can often be derived using more than one physical observable. In order to ensure the consistency of the EFT framework, calculations based on different observables have to result in the same matching condition. This is ensured by the symmetries of the theory (see e.g.~\cite{Steinhauser:2002rq}).

These symmetries are often obscured by gauge fixing but become manifest when the background field method (BFM) is used (see~\cite{Denner:1994xt,Grassi:1999nb} for detailed discussions on the application of the BFM to the SM and the THDM). In the BFM, the fields are split up into classical background fields and quantum fields. The background fields appear only as external particles allowing to keep the gauge symmetry with respect to the background fields exact. The quantum fields only appear at the loop level as internal fields. As a consequence, the physical Green's functions fulfill simple QED-like Ward identities.

These Ward identities relate different processes which can be used to calculate the matching relation for a specific EFT parameter to each other. In this way, they guarantee that the EFT is indeed able to reproduce all effects of the HET up to terms suppressed by the heavy mass scale. As a direct consequence, the difference between the number of vertex functions and the number of Ward identities is equal to the number of EFT parameters. Note also that the same Ward identities for ``light'' fields are valid in the EFT as well as the HET, since we assume the gauge symmetries of the HET to be unbroken in the EFT. We will discuss examples in \cref{sec:SMtoMSSMexampleOL,sec:SMtoMSSMexampleTL,sec:SMtoTHDMexample,sec:THDMtoMSSMexample}.

\medskip

The freedom of using different processes to derive the matching condition for a specific coupling can be used to simplify the calculation. For instance, choosing a process in which the respective coupling appears at the tree-level is beneficial. Choosing between processes with different numbers of external legs, which all depend on the respective coupling at the tree level, is, however, more subtle.

A lower number of external legs corresponds to a lower number of Feynman diagrams, which have to be computed. Also the involved loop integrals are easier to evaluate. Typically, however, matching processes with a mass dimension below four aggravates the expansion of the calculated amplitude in $m/M$.\footnote{Setting the masses of all light fields to zero from the beginning is in general not a solution. In this limit, e.g. processes with a mass dimension below four do not depend on dimensionless couplings at the tree level.} In particular, in theories with mixing particles, the particle masses and mixing matrices often depend in a non-trivial way on the ratio $m/M$. Consequently, the final expansion around $m/M\sim 0$ can often become the most complicated step of the overall calculation. We will discuss examples in \cref{sec:03_application}.

In those situations, it can be advantageous to use processes with a mass dimension of zero to obtain matching relations for dimensionless couplings (or of processes with a mass dimension of $x$ for a parameter with the mass dimension $x$). While the number of Feynman diagrams is significantly higher, no expansion of the result in the ratio of $m/M$ is needed since $m$ can be set to zero before evaluating the Feynman diagrams (the treatment of infrared divergences is discussed in \cref{sec:infrared}). For theories in which the mass terms are generated by spontaneous symmetry breaking, this is often equivalent to performing the calculation in the unbroken phase of the theory.

For certain EFTs, it is, moreover, not possible to derive all matching conditions without calculating processes with a mass dimension of zero. In \cref{sec:THDMtoMSSMexample}, we will discuss the THDM as an EFT of the MSSM as an example.


\subsection{Treatment of infrared divergences}
\label{sec:infrared}

In the limit $m\rightarrow 0$, infrared divergences appear in the calculation. These cancel out in the difference between the results in the HET and the EFT (for an extensive discussion at the one-loop level see~\cite{Braathen:2018htl}).

The easiest option to handle these divergences is to use dimensional regularization (or dimensional reduction for supersymmetric theories). This implies in particular that scaleless loop integrals can be set to zero. Consequently, the terms originating from the reparametrization of the EFT result in terms of HET couplings (see \cref{eq:matching1L,eq:matching2L}) vanish if $m$ is set to zero.

\begin{figure}\centering
\includegraphics{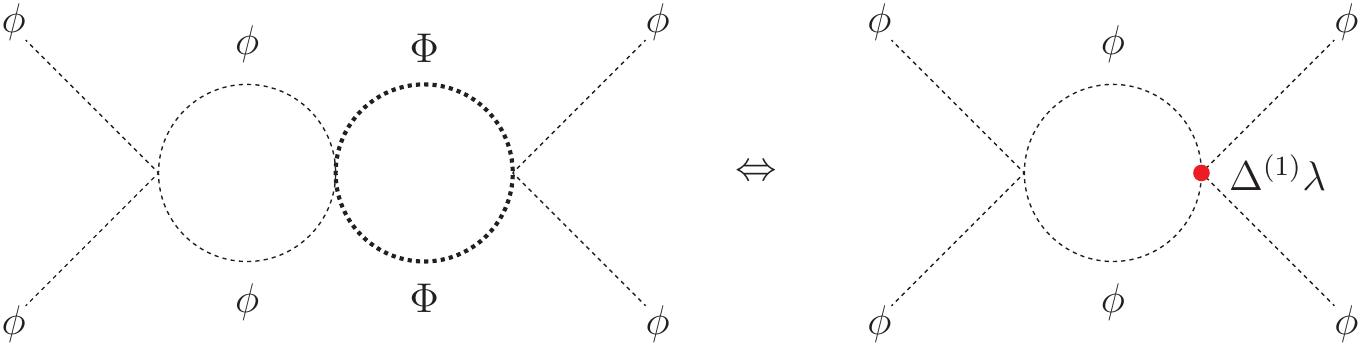}
\caption{\textit{Left:} Exemplary two-loop diagram of the HET with a light field $\phi$ and a heavy field $\Phi$. \textit{Right:} Corresponding diagram in the EFT which cancels the HET diagram if the EFT couplings are reparametrized in terms of HET couplings.}
\label{fig:2LdoubleB0}
\end{figure}

If instead the full dependence on the light masses, $m_i$, is kept explicit, the infrared divergences will manifest as large logarithms of the form $\ln(Q^2/m^2)$. In this scheme, the terms originating from the reparametrization of the EFT result do not vanish. These terms, however, are compensated by additional contributions to the HET vertex functions. A simple example is shown in \cref{fig:2LdoubleB0}.\footnote{This and all other diagrams in this work were produced with \texttt{Axodraw} \cite{Collins:2016aya}.}
The amplitude of the left diagram, showing a two-loop contribution to a scalar four-point function in the HET, is zero in dimensional regularization if $m$, the mass of the light field $\phi$, is set to zero.\footnote{Since we want to match the EFT and the HET, also the external momenta are set to zero.} If $m$ is kept finite, the diagram does not vanish but is canceled in the matching calculation by the EFT diagram shown on the right side of \cref{fig:2LdoubleB0}, in which $\Delta^{(1)}\lambda$ is the one-loop threshold correction for the $\phi^4$ coupling. While it is possible to prove this cancellation up to arbitrary loop order (see e.g.~\cite{Collins:1984xc}), we will restrict us to discussing a more complicated example in \cref{sec:SMtoMSSMexampleTL}.

In this work, we will give all light particles a common mass regulator, $\mu_\text{IR}$, if we set $m$ to zero from the beginning of the calculation. This provides a useful cross-check of the calculation in the form of verifying explicitly the cancellation of infrared divergences in the final matching conditions, but still significantly simplifies the calculation in comparison to taking into account the full dependence on all light masses.

Note that the use of a regulator mass can lead to a violation of the gauge invariance implying the need to include additional counterterms (see e.g.~\cite{Steinhauser:2002rq,Luthe:2017ttg}). For the applications discussed in \cref{sec:03_application}, the gauge invariance is, however, not violated, since we work in the limit of vanishing electroweak gauge couplings.


\section{Application: the SM and the THDM as EFTs}
\label{sec:03_application}

In this Section, we apply the methodology discussed in \cref{sec:02_method} to the matching of the SM, the THDM and the MSSM to each other. We present so far unknown two-loop matching conditions for the scalar quartic couplings.


\subsection{Higgs sectors}

Here, we shortly introduce the Higgs sectors of the various theories whose matching to each other will be discussed.


\subsubsection{The SM Higgs sector}

The Higgs potential of the SM reads
\begin{align}\label{HiggsPotential}
V_{\text{SM}}(\Phi) =& -\mu^2 \Phi^\dagger\Phi + \frac{\lambda}{2}\left(\Phi^\dagger\Phi\right)^2.
\end{align}
The Higgs doublet, $\Phi$, can be expanded around its vacuum expectation value (vev), $v$,
\begin{align}
\Phi =
\begin{pmatrix}
G^+ \\
v + \frac{1}{\sqrt{2}}(h + i G)
\end{pmatrix}
\end{align}
where $h$ is the SM Higgs, $G$ is the neutral Goldstone boson, $G^+$ is the charged Goldstone boson, and
\begin{align}\label{eq:SMvev}
v = \sqrt{\frac{\mu^2}{\lambda}}\simeq 174\gev.
\end{align}
The top-Yukawa part of the SM Lagrangian is given by
\begin{align}
\mathcal{L}_{\rm{Yuk}}^\SM = - y_t \bar t_R \left(-i \Phi^T\sigma_2\right)Q_L + \rm{h.c.},
\end{align}
where $t_R$ is the right-handed top-quark field and $Q_L$ is the third generation left-handed quark doublet. We use $y_t$ to denote the SM top-Yukawa coupling. Often, we will, moreover, use $\alt \equiv y_t^2/(4\pi)$.


\subsubsection{The THDM Higgs sector}

The Higgs sector of the THDM consists out of two Higgs doublets. The Higgs potential is given by
\begin{align}\label{eq:THDM_HiggsPotential}
V_{\text{THDM}}(\Phi_1,\Phi_2) =& m_{11}^2\,\Pdd + m_{22}^2\,\Puu - \left(m_{12}^2 \Pdu + \rm{h.c.}\right) \nonumber\\
& + \frac{1}{2}\lambda_1 (\Pdd)^2 + \frac{1}{2}\lambda_2 (\Puu)^2  + \lambda_3 (\Pdd)(\Puu) + \lambda_4 (\Pdu)(\Pud)\nonumber\\
& + \left(\frac{1}{2}\lambda_5 (\Pdu)^2 + \lambda_6 (\Pdd)(\Pdu) + \lambda_7 (\Puu)(\Pdu) + \rm{h.c.}\right).
\end{align}
$m_{12}^2$, $\lambda_5$, $\lambda_6$ and $\lambda_7$ are potentially complex parameters. Consequently, the Higgs potential in total has 14 free parameters.

The Higgs doublets, $\Phi_1$ and $\Phi_2$, can be expanded around their vevs, $v_1$ and $v_2$,
\begin{align}
\Phi_i =
\begin{pmatrix}
\phi^+_i \\
v_i + \frac{1}{\sqrt{2}}(\phi_i + i \chi_i)
\end{pmatrix}
\end{align}
with $v^2 = v_1^2 + v_2^2$. We use $\tbe \equiv \tan\beta = v_2/v_1$ to denote the ratio of the two vevs.

The diagonal mass parameters, $m_{11}^2$ and $m_{22}^2$, can be eliminated using the minimum conditions of the Higgs potential; $|m_{12}|^2$ can be reexpressed in terms of the charged Higgs tree-level mass,
\begin{align}
& M_{H^\pm}^2 = \frac{\Re(m_{12}^2)}{\sbe\cbe}-v^2(\liv + \Re\lv + \Re\lvi/\tbe + \Re\lvii\tbe),
\end{align}
The mass matrix of the neutral Higgs bosons is then given by
\begin{align}
\mathcal{M}_{\phi\phi}^2 = m_{H^\pm}^2
\begin{pmatrix}
\sbb & -\sbe\cbe & 0 & 0 \\
-\sbe\cbe & \cbb & 0 & 0 \\
0 & 0 & \sbb & -\sbe\cbe \\
0 & 0 & -\sbe\cbe & \cbb
\end{pmatrix}
+
2 v^2 \mathcal{B},
\end{align}
where we introduced the abbreviations
\begin{align}
s_\gamma \equiv \sin\gamma \hspace{.5cm} c_\gamma \equiv \cos\gamma
\end{align}
for a generic angle $\gamma$. The matrix $\mathcal{B}$ is a function of $\beta$ and $\lambda_{1..7}$. Its explicit form can be found e.g.\ in \cite{Murphy:2019qpm}.

In the limit $v\rightarrow 0$ or if the quartic couplings are neglected, the neutral-Higgs mass matrix is diagonalized by the transformations,
\begin{align}\label{eq:THDMneutralBasis}
\begin{pmatrix} h \\ H \end{pmatrix} = \mathbf{R}(\beta - \pi/2) \begin{pmatrix} \phi_1 \\ \phi_2 \end{pmatrix}, \hspace{.2cm}
\begin{pmatrix} A \\ G \end{pmatrix} = \mathbf{R}(\beta) \begin{pmatrix} \chi_1 \\ \chi_2 \end{pmatrix}
\end{align}
with
\begin{align}
\mathbf{R}(\gamma) = \begin{pmatrix} - s_\gamma & c_\gamma \\ c_\gamma & s_\gamma \end{pmatrix},
\end{align}
yielding the mass eigenstates $h$, $H$, $A$ and $G$. Also the charged-Higgs mass matrix (see e.g.~\cite{Murphy:2019qpm}) is diagonalized by the transformation,
\begin{align}\label{eq:THDMchargedBasis}
\begin{pmatrix} H^\pm \\ G^\pm \end{pmatrix} = \mathbf{R}(\beta) \begin{pmatrix} \phi_1^\pm \\ \phi_2^\pm \end{pmatrix},
\end{align}
yielding the mass eigenstates $H^\pm$ and $G^\pm$.

If all parameters are real, often the $A$-boson mass, $M_A$, is used as input instead of $M_{H^\pm}$. Then,
\begin{align}
M_{H^\pm}^2 = M_A^2 + v^2 (\lv - \liv)
\end{align}
can be used to obtain the charged Higgs mass.

Neglecting all Yukawa couplings apart from the top-Yukawa couplings, the Yukawa part of the THDM Lagrangian is given by
\begin{align}
\mathcal{L}_{\rm{Yuk}}^\THDM = - h_t e^{-i\phi_t}\bar t_R \left(-i \Phi_2^T\sigma_2\right)Q_L - \htp e^{-i\phi_t} \bar t_R \left(-i \Phi_1^T\sigma_2\right)Q_L + \rm{h.c.}.
\end{align}
The top-Yukawa couplings $h_t$ and $\htp$ can be complex numbers. In this case, the phase $\phi_t$, which can be absorbed e.g.\ into the right-handed top-quark field, has to be chosen according to $\phi_t = \arg(h_t\sbe + \htp\cbe)$ in order to obtain a real top-quark mass.


\subsubsection{The MSSM Higgs sector}

Similar to the THDM Higgs sector, the MSSM Higgs sector also consist out of two Higgs doublets. The Higgs potential is analogous to the THDM Higgs potential (see \cref{eq:THDM_HiggsPotential}). SUSY, however, fixes the quartic Higgs couplings in terms of gauge couplings. These well-known relations read (see e.g.~\cite{Haber:1993an}),
\begin{align}
& \lambda_1^\MSSM = \lambda_2^\MSSM = \frac{1}{4}(g^2 + g_y^2),\; \lambda_3^\MSSM = \frac{1}{4}(g^2 -g_y^2),\; \lambda_4^\MSSM = -\frac{1}{2}g^2,\nonumber\\
& \lambda_5^\MSSM = \lambda_6^\MSSM = \lambda_7^\MSSM = 0,
\end{align}
where $g$ and $g_y$ are the gauge couplings of the $SU(2)_L$ and $U(1)_Y$ gauge groups, respectively.

The top-Yukawa part of the Lagrangian is given by
\begin{align}
\mathcal{L}_{\rm{Yuk}}^\MSSM = - h_t^\MSSM \bar t_R \left(-i \Phi_2^T\sigma_2\right)Q_L + \rm{h.c.}.
\end{align}
This implies i.e.\ that $h_t = h_t^\THDM = h_t^\MSSM$ and $\htp = 0$ at the tree level if the THDM is matched to the MSSM.


\subsection{The SM as EFT}
\label{sec:SMasEFT}

In this Section, we discuss the SM as EFT of a generic HET. As discussed in \cref{sec:ObsChoice}, a specific EFT coupling can typically be matched using different observables. The fact that all observables will yield the same matching condition is a result of the gauge symmetries. These become apparent in the form of simple Ward identities when using the BFM. Here, we illustrate this using the matching of the SM Higgs self-coupling to a generic HET as an example.

An infinite number of SM Ward identities involving only external Higgs or neutral Goldstone bosons can be derived. If we subtract the Ward identities of the HET, in which the SM-like Ward identities are only a subset of all Ward identities, from the SM ones, the identities greatly simplify. In the limit $m/M\rightarrow 0$, the difference of a specific HET 1LPI Green's function and the corresponding EFT Green's function can be set to zero if the Green's function has negative mass dimension (see \cref{eq:NegDimZero}). As a result, only a finite set of Ward identities remains,
\begin{subequations}
\begin{align}
\Delta\Gamma_{G} &= 0, \\
\Delta\Gamma_{GG} &= \frac{1}{\sqrt{2}}\Delta\left(\frac{\Gamma_h}{v}\right), \\
\Delta\Gamma_{Gh} &= 0, \\
\Delta\Gamma_{hhh} &= 3\Delta\Gamma_{GGh} = \frac{3}{\sqrt{2}}\Delta\left(\frac{\Gamma_{hh}}{v}\right) - \frac{3}{2}\Delta\left(\frac{\Gamma_h}{v^2}\right), \label{eq:WardHHH}\\
\Delta\Gamma_{GGG} &= \Delta\Gamma_{Ghh} = 0, \\
\Delta\Gamma_{GGGG} &= \Delta\Gamma_{hhhh} = 3\Gamma_{GGhh} =\frac{3}{2}\Delta\left(\frac{\Gamma_{hh}}{v^2}\right) - \frac{3}{2\sqrt{2}} \Delta\left(\frac{\Gamma_h}{v^3}\right), \label{eq:WardHHHH}\\
\Delta\Gamma_{GGGh} &= \Delta\Gamma_{Ghhh} = 0,
\end{align}
\end{subequations}
where the notation $\Delta X = X^\HET(\{g_i^{\HET}\}) - X^\SM(\{g_i^{\SM}\})$ is used for a generic quantity $X$ depending on the couplings $g_i$. These relations are valid order-by-order in the perturbative expansion for the renormalized and unrenormalized Green's functions.

In total, these are 12~Ward identities. In the SM, 14~different Green's functions involving only neutral SM scalars with a non-negative mass dimension exist. As explained in \cref{sec:ObsChoice} the difference of the two numbers gives the number of free parameters in the EFT, which can be chosen to be $\lambda$ and $v$ in the case of the SM.

These relations can be applied to the matching of the SM-Higgs self-coupling. When matching the SM to a HET, its matching relation can be obtained e.g.\ by either calculating $\Gamma_h$ and $\Gamma_{hh}$, $\Gamma_{hhh}$, or $\Gamma_{hhhh}$. At the tree-level, these Green's functions are given in the SM by
\begin{align}
\Gamma_{hh}^{\SM,(0)}(p^2=0) = -2 \lambda v_\SM^2, \hspace{.5cm} \Gamma_{hhh}^{\SM,(0)} = -3\sqrt{2}\lambda v_\SM, \hspace{.5cm} \Gamma_{hhhh}^{\SM,(0)} = -3 \lambda.
\end{align}
Using the four-point function, the loop correction to the matching condition of $\lambda$, $\Delta\lambda$, is given by
\begin{align}\label{eq:SMhhhhMatching}
hh \rightarrow hh:\;\Delta\lambda =&{} -\frac{1}{3}\Delta\mathcal{M}_{hhhh} = -\frac{1}{3}\Delta\widehat\Gamma_{hhhh},
\end{align}
where the ``\;$\widehat{\;}\;$'' is used to denote that the Green's function is renormalized using the ``heavy-OS'' scheme for the HET theory. As explained in \cref{sec:FieldNormalization}, this implies that no LSZ factors for the external SM-like Higgs has to be taken into account in the matching calculation and thereby that $\Delta\mathcal{M}_{hhhh} = \Delta \Gamma_{hhhh}$.

Alternatively, we can use the three-point function,
\begin{align}
h \rightarrow hh:\;\Delta\lambda =&{} -\frac{1}{3\sqrt{2}v_\SM}\Delta\mathcal{M}_{hhh} = -\frac{1}{3\sqrt{2}}\Delta\left(\frac{\widehat\Gamma_{hhh}}{v}\right) = \nonumber\\
& = -\frac{1}{3}\Delta\widehat\Gamma_{hhhh},
\end{align}
where we used in the first line that $v_\HET = v_\SM$ in the ``heavy-OS'' scheme and in the second line \cref{eq:WardHHH,eq:WardHHHH} to recover the result of \cref{eq:SMhhhhMatching}.

In a similar way, we can also use the Higgs two-point function to obtain the matching condition for $\lambda$,
\begin{align}\label{eq:SMhhMatching}
h \rightarrow h:\;\Delta\lambda =&{} -\frac{1}{2v_\SM^2}\Delta\left(\mathcal{M}_{hh} - \frac{1}{\sqrt{2}v}\mathcal{M}_h\right) = -\frac{1}{2}\Delta\left(\frac{\widehat\Gamma_{hh}}{v^2}\right) - \frac{1}{2\sqrt{2}}\Delta\left(\frac{\widehat\Gamma_h}{v^3}\right) = \nonumber\\
& = -\frac{1}{3}\Delta\widehat\Gamma_{hhhh}.
\end{align}
In the last line we again used \cref{eq:WardHHH,eq:WardHHHH} to recover the result of \cref{eq:SMhhhhMatching}.


\subsubsection{Example: \texorpdfstring{\order{\alt}}{O(at)}  matching of the SM to the MSSM}
\label{sec:SMtoMSSMexampleOL}

Here, we give an explicit example for the various possibilities to obtain the matching condition for the SM Higgs self-coupling as presented in \cref{sec:SMasEFT}. I.e., we consider the well-known one-loop matching condition between the SM and the MSSM, whose dominant \order{\alt} contribution originates from the stop/top sector and is given by (see e.g.~\cite{Bagnaschi:2014rsa})
\begin{align}\label{eq:lamSMtoMSSM}
\Delta^{(1)}_{\alt}\lambda(Q) = 6 k y_t^4 \bigg(\log \frac{\mtL \mtR}{Q^2} + \widehat X_t^2 \widetilde{F}_1 \left(\frac{\mtL}{\mtR}\right) - \frac{\widehat X_t^4}{12} \widetilde{F}_2\left(\frac{\mtL}{\mtR}\right) \bigg),
\end{align}
where $Q$ is the matching scale between the SM and the MSSM, $k \equiv (4\pi)^{-2}$, $X_t$ is the stop mixing parameter and $\widehat X_t = X_t/\msusy$ with $\msusy = \sqrt{\mtL \mtR}$ and $\mtL$ and $\mtR$ being the soft SUSY-breaking stop mass-parameters. The loop functions $\widetilde{F}_{1,2}$ are given in the Appendix of Ref.~\cite{Bagnaschi:2014rsa}.

Using the two-point function for the matching (see \cref{eq:SMhhMatching}), we obtain
\begin{align}
\Delta&^{(1)}_{\alt}\lambda(Q) = \nonumber\\
=&{} -\frac{1}{2 v^2}\Delta\left[\frac{\widehat\Gamma_{hh}}{v^2} - \frac{1}{\sqrt{2}}\frac{\widehat\Gamma_h}{v^3}\right]_{\order{\alt}} =\nonumber\\
=&{}\frac{1}{2v^2} k \bigg[ - \frac{6 m_t^4}{v^2} \left(B_0(0, m_{\tilde t_1}^2, m_{\tilde t_1}^2) + B_0(0, m_{\tilde t_2}^2, m_{\tilde t_2}^2)\right)\nonumber\\
& \hspace{1.12cm} + \frac{3m_t X_t}{2 v^2}s_{2\theta_{\tilde t}}\left(A_0(m_{\tilde t_1}^2) - A_0(m_{\tilde t_2}^2)\right) \nonumber\\
& \hspace{1.12cm} - \frac{3m_t^2 X_t^2}{2v^2}\left(B_0(0, m_{\tilde t_1}^2, m_{\tilde t_1}^2)s_{2\theta_{\tilde t}}^2 + B_0(0, m_{\tilde t_2}^2, m_{\tilde t_2}^2)s_{2\theta_{\tilde t}}^2  + 2 B_0(0, m_{\tilde t_1}^2, m_{\tilde t_2}^2)c_{2\theta_{\tilde t}}^2\right) \nonumber\\
& \hspace{1.12cm} - \frac{12m_t^3 X_t}{v^2}s_{\theta_{\tilde t}}c_{\theta_{\tilde t}}\left(B_0(0, m_{\tilde t_1}^2, m_{\tilde t_1}^2) - B_0(0, m_{\tilde t_2}^2, m_{\tilde t_2}^2)\right) \bigg],
\end{align}
where $m_{\tilde t_{1,2}}$ are the stop masses and $\theta_{\tilde t}$ is the stop mixing angle. $A_0$ and $B_0$ are the one- and two-point scalar Passarino-Veltman functions, respectively.

This expressions needs to be expanded in the limit $m_t/\mtL,\,m_t/\mtR$ to obtain \cref{eq:lamSMtoMSSM}, where $m_t$ is the top-quark mass. For this expansion, $m_{\tilde t_{1,2}}$ and $\theta_{\tilde t}$ need to be expressed in terms of $\mtL$, $\mtR$ and $X_t$,
\begin{align}
m_{\tilde t_1}^2 &= \mtL^2 + m_t^2 + \frac{m_t^2 X_t^2}{\mtL^2 - \mtR^2} - \frac{m_t^4 X_t^4}{(\mtL^2 - \mtR^2)^3} + \ldots,\\
m_{\tilde t_2}^2 &= \mtR^2 + m_t^2 - \frac{m_t^2 X_t^2}{\mtL^2 - \mtR^2} + \frac{m_t^4 X_t^4}{(\mtL^2 - \mtR^2)^3} + \ldots,\\
c_{\theta_{\tilde t}} &= 1 - \frac{m_t^2 X_t^2}{2(\mtL^2 - \mtR^2)^2} + \frac{11 m_t^4 X_t^4}{8(\mtL^2 - \mtR^2)^4} + \ldots,\\
s_{\theta_{\tilde t}} &= \frac{m_t X_t}{\mtL^2 - \mtR^2} - \frac{3 m_t^3 X_t^3}{2(\mtL^2 - \mtR^2)^3} + \ldots,
\end{align}
where the ellipses denotes terms suppressed by higher powers of $(m_t X_t)/(\mtL^2 - \mtR^2)$. Moreover, we need to insert the explicit formulas for the loop functions.

Whereas the expansion is straightforward in the example discussed here, it can be become cumbersome for more complicated calculations (i.e., if electroweak corrections\footnote{E.g., the calculation of the electroweakino contribution to the one-loop threshold correction to $\lambda$ is really tedious using the two-point function. Already the expansion of the electroweakino mixing matrix is hard to work out if the electroweakino mass parameters are non-degenerate.} or higher-order loop corrections are taken into account). We will discuss corresponding two-loop examples in \cref{sec:03_asat,sec:03_atat}.

Using the four-point function to derive the matching condition (see \cref{eq:SMhhhhMatching}), we can set the SM masses to zero before even evaluating the Feynman diagrams (see discussion in \cref{sec:ObsChoice}) and obtain
\begin{align}
\Delta^{(1)}_{\alt}\lambda(Q) =&{} -\frac{1}{3}\Delta\widehat\Gamma_{hhhh}^{\order{\alt}} = \nonumber\\
={}& -y_t^4 k \left[3 B_0(0, \mtL^2 , \mtL^2) + 3 B_0(0, \mtR^2, \mtR^2) \right.\nonumber\\
&\left.\hspace{1.3cm} + 6 X_t^2 \left( C_0(0, 0, 0, \mtL^2, \mtL^2, \mtR^2) + C_0(0, 0, 0, \mtL^2, \mtR^2, \mtR^2) \right) \right.\nonumber\\
&\left.\hspace{1.3cm} + 3 X_t^4 D_0(0,0,0,0,0,0, \mtL^2, \mtL^2, \mtR^2, \mtR^2)\right],
\end{align}
where $C_0$ and $D_0$ are the scalar three- and four-point Passarino-Veltman functions, respectively. No expansion is needed to arrive at the result given in \cref{eq:lamSMtoMSSM}.\footnote{The loop functions $\widetilde{F}_{1,2}$ in \cref{eq:lamSMtoMSSM} are related to the Passarino-Veltman functions $C_0$ and $D_0$ as follows, $\widetilde{F}_1 \left(\frac{\mtL}{\mtR}\right) = -\mtL \mtR \left( C_0(0, 0, 0, \mtL^2, \mtL^2, \mtR^2) + C_0(0, 0, 0, \mtL^2, \mtR^2, \mtR^2) \right),~\widetilde{F}_2\left(\frac{\mtL}{\mtR}\right) = 6 \mtL^2 \mtR^2 D_0(0,0,0,0,0,0, \mtL^2, \mtL^2, \mtR^2, \mtR^2)$.}


\subsubsection{Example: \texorpdfstring{\order{\alt\als}}{O(atas)} matching of the SM to the MSSM}
\label{sec:SMtoMSSMexampleTL}

The calculation of the \order{\alt\als} threshold corrections between the SM and the MSSM nicely illustrates some of the key differences between the different approaches to treat the ``light'' fields discussed in \cref{sec:infrared} ($\als\equiv g_3^2/(4\pi)$, where $g_3$ is the strong gauge coupling). In particular, we will consider the derivation of the \order{\alt\als} threshold correction to the Higgs quartic coupling between the SM and the MSSM. Originally, it was computed in~\cite{Bagnaschi:2014rsa} and shown to be a polynomial in the ratio $\xt = X_t/\msusy$

\begin{figure}\centering
\includegraphics{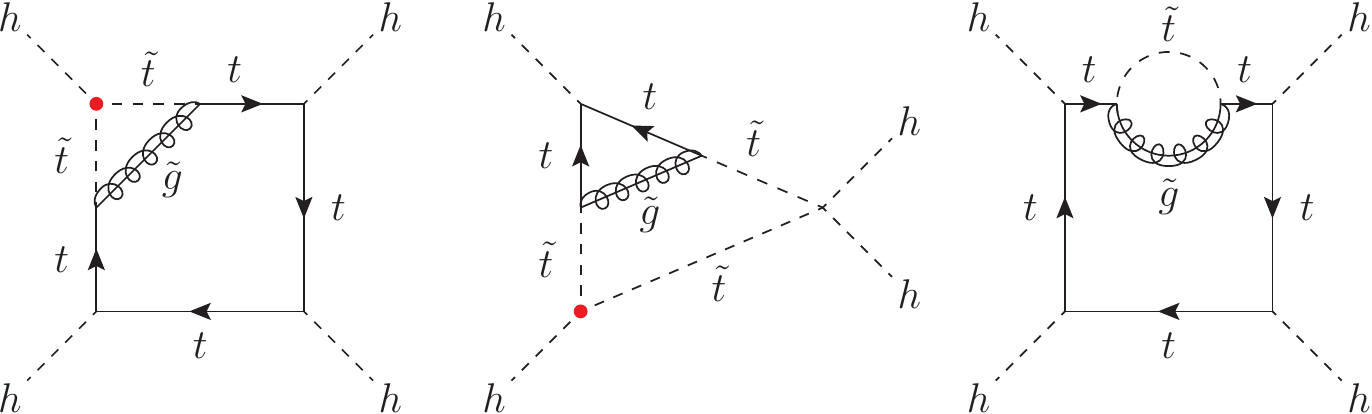}
\includegraphics{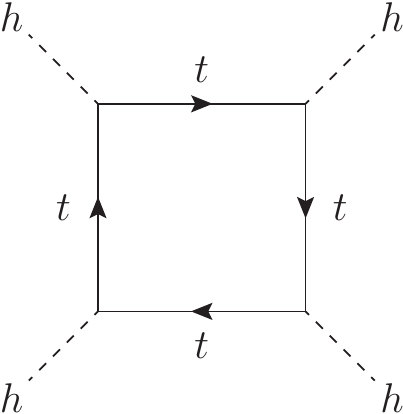}
\caption{
\textit{Top:} Exemplary Feynman diagrams contributing to the \order{\alt\als} matching of the SM Higgs self-coupling between the SM and the MSSM and resulting in terms linear in \xt. The red dot denotes vertices proportional to $X_t$.
\textit{Bottom:} The \order{\alt} SM contribution to the matching of the SM Higgs self-coupling between the SM and the MSSM.
}
\label{fig:2LatasXt1Diagrams}
\end{figure}

The genuine two-loop diagrams contributing to the terms linear in $\xt$ in the broken phase (i.e., if the SM-like vev is not set to zero) are shown in the upper row of \cref{fig:2LatasXt1Diagrams}. The red dot denotes the $h-\tilde{t}_{1,2}-\tilde{t}_{1,2}$ vertices which depend directly on the stop mixing parameter $X_t$. The third diagram in the upper row \cref{fig:2LatasXt1Diagrams} does not contain such a vertex.  The $t-\tilde{t}_{1/2}-\tilde{g}$ vertices are, however, dependent on the stop mixing angle which is proportional to $X_t$. The explicit evaluation of such diagrams---i.e., the one containing a $\tilde t_1$ and the one containing a $\tilde t_2$---in the broken phase and in the limit $m_t/\msusy \to 0$ shows that both of them contain terms proportional to \xt. These terms are in fact canceled by the reparameterization (see \cref{sec:calcMatching}) of the SM \MS top-quark mass in the \order{\alt} SM contribution (see bottom diagram of~\cref{fig:2LatasXt1Diagrams}) in terms of the MSSM \DR top-quark mass (explicit expressions can be found in~\cite{Bagnaschi:2017xid}).

If the SM-like vev is set to zero right from the beginning of the calculation, the stop-mixing angle is equal to  zero and also the stop masses do not depend on $X_t$. Therefore, no \xt dependence can arise from diagrams like the top-right diagram of \cref{fig:2LatasXt1Diagrams}. Since the one-loop SM contribution is a scaleless integral if $v=0$ is set, also no reparameterization contribution exists.

If instead an infrared regulator mass (independent of the Higgs vev) is introduced for the top quark, diagrams like the third one shown in \cref{fig:2LatasXt1Diagrams} still yield no \xt-dependent terms. The one-loop SM contribution is, however, non-zero. In this case, it is important to realize that the infrared regulator mass used in the SM and the MSSM are not equal at the one-loop level and a one-loop matching condition is needed. This matching condition differs from the matching condition of the regular top-quark mass, since the vev is set to zero in the couplings entering the one-loop correction. As a consequence, the matching condition of the regulator mass is not dependent on \xt in contrast to the matching condition of the regular (vev-dependent) top-quark mass.


\subsubsection{Application: \texorpdfstring{\order{\alt^2}}{O(atat)} matching between the SM and the THDM}
\label{sec:SMtoTHDMexample}

As a further concrete application for the matching of the SM Higgs self-coupling to a HET, we consider matching the SM to the THDM. While the one-loop corrections to the matching of $\lambda$ have already been calculated in~\cite{Bahl:2018jom} in the case of the real THDM and extended to the complex THDM in~\cite{Bahl:2020mjy}, we here calculate the previously unknown \order{\alt^2} two-loop threshold correction, $\Delta^{(2)}_{\alt^2}\lambda$. As explained in the previous Sections, it is advantageous to calculate this threshold correction by matching the scalar four-point function.\footnote{Using the two-point function, already the calculation of the one-loop $\mathcal{O}(\lambda_i^2)$ threshold correction is very complicated (especially in the presence of \cp-violating phases).}

Since the tree-level and one-loop level matching condition for $\lambda$ between the SM and the THDM is a combination of $\lambda_{1..7}$ and trigonometric functions of $\beta$, the difference in the field normalizations between the SM and the THDM does not contribute to the \order{\alt^2} threshold correction. Consequently, only the genuine \order{\alt^2} corrections to $\widehat\Gamma_{hhhh}$ have to be calculated in both theories and subtracted from each other.

We generate the required Feynman diagrams with the help of \texttt{FeynArts}~\cite{Kublbeck:1990xc,Eck:1992ms,Hahn:2000kx}. The needed model file has been generated with the help of \texttt{SARAH}~\cite{Staub:2009bi,Staub:2010jh,Staub:2012pb,Staub:2013tta}. The two-loop diagrams are processed with \texttt{TwoCalc}~\cite{Weiglein:1992hd,Weiglein:1993hd}. Subloop renormalization diagrams are evaluated using \texttt{FormCalc}~\cite{Hahn:1998yk}. As interface between the different tools, we use an adapted version of the scripts presented in~\cite{Hahn:2015gaa}.\footnote{Since the vevs of the THDM-Higgs doublets do not enter any of the couplings at order \order{\alt^2} using an infrared regulator for the top-quark mass (and setting $v = 0$) is equivalent to keeping the vevs non-zero. Both approaches lead to a comparable number of diagrams in this case.}

We checked explicitly that all infrared and ultaviolet divergences cancel. As result, we obtain
\begin{align}
\Delta^{(2)}_{\alt^2}\lambda(Q) = \frac{3}{2}k^2 \vert h_t\sbe + \htp\cbe \vert^4 \vert h_t\cbe - \htp\sbe \vert^2 \left(2 \pi ^2-7 + 6\ln^2\frac{M_{H^\pm}^2}{Q^2} + 14 \ln\frac{M_{H^\pm}^2}{Q^2}\right),
\end{align}
where $Q$ is the matching scale between SM and THDM. Instead of $M_{H^\pm}$, $M_A$ can be used without further modifications if all THDM parameters are real. This expression assumes that the corresponding one-loop threshold correction is parameterized in terms of THDM parameters evaluated at the scale $Q$.\footnote{Since the number of SM-Yukawa couplings is smaller than the number of THDM-Yukawa couplings, a parameterization in terms of SM parameters is actually not possible.}


\subsection{The THDM as EFT}
\label{sec:THDMtoMSSMexample}

As a next application, we discuss the case of the THDM as EFT concentrating again on the Higgs sector. Using the BFM with the additional simplification of setting Green's functions with a negative mass dimension to zero (see \cref{eq:NegDimZero}), we derived a set of 124~Ward identities involving 138~Green's functions with external Higgs fields. They are listed in \cref{app:06_THDMward}. Correspondingly, at least 14~Green's functions have to be calculated to obtain the matching conditions for the parameters of the THDM Higgs sector. Moreover, this means that the THDM Higgs sector has 14~free parameters which need to be matched.\footnote{These are $m_{H^\pm}$, $\arg(m_{12}^2)$, $v$, $\tan\beta$, $\lambda_1$, $\lambda_2$, $\lambda_3$, $\lambda_4$, $\Re\lambda_5$, $\Im\lambda_5$, $\Re\lambda_6$, $\Im\lambda_6$, $\Re\lambda_7$, and $\Im\lambda_7$.}

The set of Ward identities is not solvable using only one-, two-, or three-point functions as input. To obtain all matching conditions, at least one four-point function has to be calculated. This is not only true for the complex THDM but also for the real THDM.


\subsubsection{Deriving matching conditions for the Higgs four-point couplings}

We will calculate two-loop corrections to the THDM scalar four-point couplings, $\lambda_{1..7}$. To calculate these, we will use only four-point Green's functions. Working in the THDM basis before mass diagonalization, one possibility to relate the matching conditions to Green's functions is given by
\begin{subequations}
\begin{align}\label{eq:THDMGaugeBasisMatching1}
\Delta\lambda_1 &= -\frac{1}{3}\Delta\widehat\Gamma_{\phiI\phiI\phiI\phiI}, \\
\Delta\lambda_2 &= -\frac{1}{3}\Delta\widehat\Gamma_{\phiII\phiII\phiII\phiII}, \\
\Delta\lambda_3 &= -\Delta\widehat\Gamma_{\phiI\phiI\phiIIp\phiIIm}, \\
\Delta\lambda_4 &= \Delta\widehat\Gamma_{\phiI\phiI\phiIIp\phiIIm} - \Delta\widehat\Gamma_{\phiIp\phiIIm\phiIp\phiIIm}, \\
\Delta\lambda_5 &= \Delta\widehat\Gamma_{\phiIp\phiIIm\phiIp\phiIIm} - \Delta\widehat\Gamma_{\phiI\phiI\phiIIp\phiIIm} - 2\Delta\widehat\Gamma_{\phiI\phiII\phiIIp\phiIm}, \\
\Delta\lambda_6 &= -\Delta\widehat\Gamma_{\chiI\chiI\phiIIp\phiIm}, \\
\Delta\lambda_7 &= -\Delta\widehat\Gamma_{\chiII\chiII\phiIIp\phiIm}\label{eq:THDMGaugeBasisMatching2}.
\end{align}
\end{subequations}
Here, $\Delta X = X^\HET(\{g_i^{\HET}\}) - X^\THDM(\{g_i^{\THDM}\})$ and the ``$\;\widehat{\;}\;$'' symbol is used to denote that the HET 1LPI Green's functions are renormalized in the ``heavy-OS'' scheme.

Taking Green's functions in the basis defined in \crefrange{eq:THDMneutralBasis}{eq:THDMchargedBasis} as input,\footnote{For the calculation of the \order{\alt^2} threshold corrections between the THDM and the MSSM (see below), the tree-level quartic couplings can be considered to be zero. Therefore, the basis defined in \crefrange{eq:THDMneutralBasis}{eq:THDMchargedBasis} is equivalent to the mass eigenstate basis.}  the THDM self-couplings can alternatively derived by calculating
\begin{subequations}
\begin{align}\label{eq:THDMMassBasisMatching1}
\Delta\lambda_1 =&{} - \frac{1}{3}\cbe^4\Delta\widehat\Gamma_{hhhh} - \frac{1}{3}\sbe^4\Delta\widehat\Gamma_{HHHH} - 2\cbb\sbb\Delta\widehat\Gamma_{hh\Hp\Hm} + 4\cbb\sbb\Re(\Delta\widehat\Gamma_{Hh\Gm\Hp}) \nonumber\\
& + 4\sbe\cbe^3\Re(\Delta\widehat\Gamma_{GG\Gm\Hp}) + 4\cbe\sbe^3\Re(\Delta\widehat\Gamma_{AA\Gm\Hp}), \\
\Delta\lambda_2 =&{} - \frac{1}{3}\sbe^4\Delta\widehat\Gamma_{hhhh} - \frac{1}{3}\cbe^4\Delta\widehat\Gamma_{HHHH} - 2\cbb\sbb\Delta\widehat\Gamma_{hh\Hp\Hm} + 4\cbb\sbb\Re(\Delta\widehat\Gamma_{Hh\Gm\Hp}) \nonumber\\
& - 4\cbe\sbe^3\Re(\Delta\widehat\Gamma_{GG\Gm\Hp}) - 4\cbe^3\sbe\Re(\Delta\widehat\Gamma_{AA\Gm\Hp}), \\
\Delta\lambda_3 =&{} - \frac{1}{3}\cbb\sbb\Delta\widehat\Gamma_{hhhh} - \frac{1}{3}\cbb\sbb\Delta\widehat\Gamma_{HHHH} - \frac{1}{4}(3 + c_{4\beta})\Delta\widehat\Gamma_{hh\Hp\Hm} - 4\cbb\sbb\Re(\Delta\widehat\Gamma_{Hh\Gm\Hp}) \nonumber\\
& - \ctb\stb\Re(\Delta\widehat\Gamma_{GG\Gm\Hp}) + \ctb\stb\Re(\Delta\widehat\Gamma_{AA\Gm\Hp}), \\
\Delta\lambda_4 =&{} - \frac{1}{3}\cbb\sbb\Delta\widehat\Gamma_{hhhh} - \frac{1}{3}\cbb\sbb\Delta\widehat\Gamma_{HHHH} + \frac{1}{4}(5 - c_{4\beta}) \Delta\widehat\Gamma_{hh\Hp\Hm} - 4\cbb\sbb\Re(\Delta\widehat\Gamma_{Hh\Gm\Hp}) \nonumber\\
& - \ctb\stb\Re(\Delta\widehat\Gamma_{GG\Gm\Hp}) + \ctb\stb\Re(\Delta\widehat\Gamma_{AA\Gm\Hp}) - \Delta\widehat\Gamma_{\Gp\Hm\Gp\Hm}, \\
\Delta\lambda_5 =&{} - \frac{1}{3}\cbb\sbb\Delta\widehat\Gamma_{hhhh} - \frac{1}{3}\cbb\sbb\Delta\widehat\Gamma_{HHHH} - \frac{1}{4}(3 + c_{4\beta}) \Delta\widehat\Gamma_{hh\Hp\Hm}  + \Delta\widehat\Gamma_{\Gp\Hm\Gp\Hm}\nonumber\\
& + \frac{1}{2}(3+c_{4\beta})\Re(\Delta\widehat\Gamma_{Hh\Gm\Hp}) - \ctb\stb\Re(\Delta\widehat\Gamma_{GG\Gm\Hp}) + \ctb\stb\Re(\Delta\widehat\Gamma_{AA\Gm\Hp}) \nonumber\\
& + 2 i \Im\bigg[\ctb\Delta\widehat\Gamma_{Hh\Gm\Gp} + \cbe\sbe\Delta\widehat\Gamma_{AA\Gm\Hp} - \cbe\sbe\Delta\widehat\Gamma_{GG\Gm\Hp} \bigg], \\
\Delta\lambda_6 =&{} -\frac{1}{3}\cbe^3\sbe\Delta\widehat\Gamma_{hhhh} + \frac{1}{3}\cbe\sbe^3\Delta\widehat\Gamma_{HHHH} + \frac{1}{2}\ctb\stb\Delta\widehat\Gamma_{hh\Hp\Hm} - \ctb\stb\Re(\Delta\widehat\Gamma_{Hh\Gm\Hp}) \nonumber\\
& - (1 + 2\ctb)\sbb\Re(\Delta\widehat\Gamma_{AA\Gm\Hp}) + (1 - 2\ctb)\cbb \Re(\Delta\widehat\Gamma_{GG\Gm\Hp}) \nonumber\\
& - i \Im\bigg[\stb\Delta\widehat\Gamma_{Hh\Gm\Gp} + \sbb\Delta\widehat\Gamma_{AA\Gm\Hp} + \cbb\Delta\widehat\Gamma_{GG\Gm\Hp} \bigg], \\
\Delta\lambda_7 =&{} -\frac{1}{3}\cbe\sbe^3\Delta\widehat\Gamma_{hhhh} + \frac{1}{3}\cbe^3\sbe\Delta\widehat\Gamma_{HHHH} - \frac{1}{2}\ctb\stb\Delta\widehat\Gamma_{hh\Hp\Hm} + \ctb\stb\Re(\Delta\widehat\Gamma_{Hh\Gm\Hp}) \nonumber\\
& + (1 - 2\ctb)\cbb\Re(\Delta\widehat\Gamma_{AA\Gm\Hp}) - (1 + 2\ctb)\sbb \Re(\Delta\widehat\Gamma_{GG\Gp\Hm}) \nonumber\\
& + i \Im\bigg[\stb\Delta\widehat\Gamma_{Hh\Gm\Gp} - \cbb\Delta\widehat\Gamma_{AA\Gm\Hp} - \sbb\Delta\widehat\Gamma_{GG\Gm\Hp} \bigg].\label{eq:THDMMassBasisMatching2}
\end{align}
\end{subequations}

For reference, also the explicit contributions due to the different normalization of the fields in the THDM and the HET are listed in \cref{sec:THDMWFR} for the case that the tree-level matching conditions are zero. In \crefrange{eq:THDMGaugeBasisMatching1}{eq:THDMGaugeBasisMatching2} and \crefrange{eq:THDMMassBasisMatching1}{eq:THDMMassBasisMatching2} these are taken into account implicitly by renormalizing the HET 1LPI Green's functions in the ``heavy-OS'' scheme.


\subsubsection{Application: \texorpdfstring{\order{\alt\als}}{O(atas)} matching between the THDM and the MSSM}
\label{sec:03_asat}

In this Section, we present analytic expressions for the \order{\alt\als} threshold corrections of the THDM Higgs self-couplings when the THDM is matched to the MSSM. I.e., we calculate the corrections generated by integrating out the squarks and/or the gluino. These corrections have been derived before in the case $|M_3| = \msusy$ using the \MS scheme for the renormalization of the stop soft-SUSY-breaking parameters in~\cite{Lee:2015uza,Carena:2015uoe}. For the \DR scheme, expressions have been derived in~\cite{Bahl:2018jom,Bahl:2020mjy} based on the \order{\alt\als} threshold correction for the SM Higgs self-coupling between the SM and the MSSM. The derivation used in~\cite{Bahl:2018jom,Bahl:2020mjy}, however, does not allow to disentangle the threshold corrections of $\lambda_3$, $\lambda_4$ and $\lambda_5$ (only the expression for $\liii + \liv + \lv$ can be derived) and to obtain expressions for the imaginary parts of $\lv$, $\lvi$ and $\lvii$.

Here, we compute the missing parts for the \DR case using two independent calculations, which differ in the treatment of the light masses (see discussion in \cref{sec:infrared}). We use the same technical setup as described in \cref{sec:SMtoTHDMexample}.

For the first calculation, we keep the full dependence on the light masses. In this approach, the expansion for large \msusy, which is assumed to be the mass scale of the squarks and the gluino, is carried out as the last step of the calculation. Infrared divergences appear in the form of large logarithms involving the ratio of \msusy over one of the light THDM masses. We perform this calculation in the mass eigenstate basis using the expressions given in \crefrange{eq:THDMMassBasisMatching1}{eq:THDMMassBasisMatching2}.

For the second calculation, we set all light masses to zero before even generating the expressions for the amplitudes. More explicitly, this means that we set the SM-like vev to zero. To check the cancellation of infrared divergences explicitly, we, however, reintroduce a common regulator mass for all light fields. Due to setting $v=0$, the number of couplings is drastically reduced resulting in a smaller number of Feynman diagrams. To further simplify the calculation, we work in the original THDM Higgs basis (before mass diagonalization) using \crefrange{eq:THDMGaugeBasisMatching1}{eq:THDMGaugeBasisMatching2} to derive the matching conditions.

Concretely, the first calculation requires an evaluation of 42260 Feynman diagrams, while the second one requires only 3861 diagrams. Both of these numbers include genuine two-loop diagrams as well as one-loop diagrams with one-loop counterterm insertions. On top of that, the first calculation also requires an additional step: the expansion for large \msusy. This expansion typically takes several hours on a single CPU core. In the second calculation the heavy mass expansion is much easier since the it needs to be performed only for the loop integrals involving the infrared regulator, while the couplings do not need to be expanded.

For both approaches, we find identical results and a full cancellation of infrared divergences. In the degenerate case of $|M_3| = m_{\tilde t_L} = m_{\tilde t_R} =  \msusy$, the final expressions read
\begin{subequations}
\begin{alignat}{2}
&\Delta^{(2)}_{\alt\als}\li =&& -\frac{4}{3}k^2 g_3^2 h_t^4 \vert \mf \vert ^4, \\
&\Delta^{(2)}_{\alt\als}\lii =&& -\frac{4}{3} k^2 g_3^2 h_t^4 \bigg[\big(\vert \at \vert^2 - 12\big) \vert \at \vert^2-2 \big(\vert \at \vert^2 - 6\big) \big(\atC \hmg + \at \hmgC\big) \bigg], \\
&\Delta^{(2)}_{\alt\als}\liii =&& \frac{4}{3} k^2 g_3^2 h_t^4 \vert \mf \vert^2 \big(3 - \vert \at \vert^2 +  \atC \hmg + \at \hmgC \big), \\
&\Delta^{(2)}_{\alt\als}\liv =&& \frac{4}{3} k^2 g_3^2 h_t^4 \vert \mf \vert^2 \big(3 - \vert \at \vert^2 +  \atC \hmg + \at \hmgC \big) , \\
&\Delta^{(2)}_{\alt\als}\lv =&& - \frac{4}{3} k^2 g_3^2 h_t^4 \at \big(\at - 2 \hmg\big) \mf^2, \\
&\Delta^{(2)}_{\alt\als}\lvi =&& \frac{4}{3} k^2 g_3^2 h_t^4 \mf \vert \mf \vert^2 \big(\at - \hmg \big), \\
&\Delta^{(2)}_{\alt\als}\lvii =&& \frac{4}{3} k^2 g_3^2 h_t^4 \mf\bigg[6 \hmg - 2 \at \big(3 + \atC \hmg\big) + \at^2 \big(\atC - \hmgC\big) \bigg],
\end{alignat}
\end{subequations}
where
\begin{equation}
\mf = \frac{\mu}{\msusy}, \quad \hmg = e^{i \phi_{M_3}}, \quad \at = \frac{A_t}{\msusy}.
\end{equation}
$M_3$ is the gluino mass parameter (and $\phi_{M_3}$ its phase); $A_t$, the stop trilinear coupling; $\mu$, the Higgsino mass parameter. The matching scale is set to $\msusy$. In the limit $M_A\rightarrow \msusy$, we recover the threshold corrections given in~\cite{Bahl:2020tuq}. Moreover, we check that the expressions agree with the ones presented in~\cite{Carena:2015uoe} after conversion to the \MS scheme used in~\cite{Lee:2015uza,Carena:2015uoe}.

In the case $|M_3| \ll \msusy$ the threshold corrections read,
\begin{subequations}
\begin{alignat}{2}
&\Delta^{(2)}_{\alt\als}\li =&& -\frac{8}{3}k^2 g_3^2 h_t^4 \vert \mf \vert ^4, \\
&\Delta^{(2)}_{\alt\als}\lii =&& -\frac{8}{3} k^2 g_3^2 h_t^4 \big[(\vert \at \vert^2 - 12) \vert \at \vert^2 + 9\big], \\
&\Delta^{(2)}_{\alt\als}\liii =&&  -\frac{8}{3} k^2 g_3^2 h_t^4 (\vert \at \vert^2 - 3) \vert \mf \vert^2, \\
&\Delta^{(2)}_{\alt\als}\liv =&&  -\frac{8}{3} k^2 g_3^2 h_t^4 (\vert \at \vert^2 - 3) \vert \mf \vert^2, \\
&\Delta^{(2)}_{\alt\als}\lv =&&  - \frac{8}{3} k^2 g_3^2 h_t^4 \at^2 \mf^2, \\
&\Delta^{(2)}_{\alt\als}\lvi =&& \frac{8}{3} k^2 g_3^2 h_t^4 \at \mf \vert \mf \vert^2, \\
&\Delta^{(2)}_{\alt\als}\lvii =&& \frac{8}{3} k^2 g_3^2 h_t^4 \at \mf (\vert \at \vert^2 - 6).
\end{alignat}
\end{subequations}
Expressions valid for the fully non-degenerate case are distributed as ancillary files alongside with the paper.

Note that the expressions, presented above, are valid only if the one-loop threshold correction is parametrized in terms of MSSM \DR parameters evaluated at the scale $\msusy$. Note moreover that the expressions are in agreement with the partial results given in~\cite{Bahl:2018jom,Bahl:2020mjy}.


\subsubsection{Application: \texorpdfstring{\order{\alt^2}}{O(atat)} matching between the THDM and the MSSM}
\label{sec:03_atat}

In addition to the \order{\alt\als} threshold correction for the Higgs self-coupling between the THDM and the MSSM, we also compute the \order{\alt^2} contributions. These have been previously unknown in the literature.

For the calculation, we use the same technical setup as described in \cref{sec:SMtoTHDMexample} and follow the same two approaches as for the calculation of the \order{\alt\als} corrections: in the first approach, we take all light masses fully into account and only in the last step expand the result in the limit of large \msusy; in the second approach, we set all light masses to zero, but introduce an infrared regulator mass for the top quark. Since at \order{\alt^2} internal Higgses appear in the calculation, we, however, work in the mass eigenstate basis also for the second approach (using \crefrange{eq:THDMMassBasisMatching1}{eq:THDMMassBasisMatching2}).

As for the \order{\alt\als} threshold corrections, we find that the number of diagrams is drastically reduced in the second approach in comparison to the first approach. In particular, the first approach requires the computation of 163168 Feynman diagrams while the second one requires the evaluation of 24392 diagrams.

We again have explicitly verified the cancellation of infrared divergences in the two approaches. And we find full agreement between the final results of the two approaches. In the degenerate case of $m_A \ll m_{\tilde t_L} = m_{\tilde t_R} = \msusy$, we find
\begin{subequations}
\begin{alignat}{2}\label{eq:atat1}
&\Delta^{(2)}_{\alt^2}\li =&& k^2 h_t^6 \amf^4 \bigg[\frac{1}{4}\frac{9+5\amf^2-8\amf^4}{1-\amf^2} + \frac{3}{2}\amf^4\frac{3-2\amf^2}{(1-\amf^2)^2}\ln(\amf^2) - \aat^2 \bigg], \\
&\Delta^{(2)}_{\alt^2}\lii =&& k^2 h_t^6 \bigg[\frac{3}{2}\frac{1+13\amf^2-8\amf^4}{1-\amf^2}+12 \frac{1-2\amf^2-2\amf^4}{(1-\amf)^2}\DiLog{1-\amf^2} \nonumber\\
& && \hspace{1.cm} - 3\amf^2 \frac{4 - \amf^2 + 6\amf^4}{(1-\amf)^2}\ln(\amf^2) \nonumber\\
& && \hspace{1.cm} + \aat^2\bigg(-\frac{3}{2}\frac{28-11\amf^2-9\amf^4}{1-\amf^2}-6\amf^4\frac{5-3\amf^2}{(1-\amf^2)^2}\ln(\amf^2)\bigg) \nonumber\\
& && \hspace{1.cm} + \aat^4\bigg(\frac{1}{4}\frac{51-37\amf^2-8\amf^4}{1-\amf^2}+\frac{3}{2}\amf^4\frac{3-2\amf^2}{(1-\amf^2)^2}\ln(\amf^2)\bigg) - \aat^6\bigg], \\
&\Delta^{(2)}_{\alt^2}\liii =&& k^2 h_t^6 \amf^2 \bigg[-3\amf^2\frac{3-\amf^2}{1-\amf^2} - \amf^4\frac{11-5\amf^2}{(1-\amf^2)^2}\ln(\amf^2) \nonumber\\
& && \hspace{1.7cm} + \aat^2 \bigg(\frac{1}{4}\frac{29-15\amf^2-8\amf^4}{1-\amf^2} + \frac{3}{2}\amf^4\frac{3-2\amf^2}{(1-\amf^2)^2}\ln(\amf^2)\bigg) \nonumber\\
& && \hspace{1.7cm} - \aat^4 \bigg], \\
&\Delta^{(2)}_{\alt^2}\liv =&& k^2 h_t^6 \amf^2 \bigg[-\frac{3}{4}(8+5\amf^2) - \frac{4\amf^4}{1-\amf^2}\ln(\amf^2) \nonumber\\
& && \hspace{1.7cm} + \aat^2 \bigg(\frac{1}{2}\frac{11-4\amf^2-4\amf^4}{1-\amf^2} + \frac{3}{2}\amf^4\frac{3-2\amf^2}{(1-\amf^2)^2}\ln(\amf^2)\bigg) \nonumber\\
& && \hspace{1.7cm} - \aat^4 \bigg], \\
&\Delta^{(2)}_{\alt^2}\lv =&& k^2 h_t^6 \bigg[\at^2\mf^2 \left(\frac{1}{4}\frac{21-7\amf^2-8\amf^4}{1-\amf^2} + \frac{3}{2}\amf^4\frac{3-2\amf^2}{(1-\amf^2)^2}\ln(\amf^2)- \aat^2 \right) \nonumber\\
& && \hspace{1.cm} + \frac{1}{4}\left( (\at \mf)^2 - (\atC \mfC)^2 \right) \left(6 - \amf^2 - \aat^2 \right)\bigg], \\
&\Delta^{(2)}_{\alt^2}\lvi =&& k^2 h_t^6 \bigg[\at \mf \left(-\frac{\amf^2}{4}\frac{15-\amf^2-8\amf^4}{1-\amf^2} -\frac{3\amf^6}{2}\frac{3-2\amf^2}{(1-\amf^2)^2}\ln(\amf^2) + \aat^2 \amf^2 \right) \nonumber\\
& && \hspace{1.cm} + \frac{1}{8} \atC \mfC \left( (\at \mf)^2 - (\atC \mfC)^2 \right) \bigg], \\
&\Delta^{(2)}_{\alt^2}\lvii =&& k^2 h_t^6 \bigg[\at\mf \left(\frac{3}{4}\frac{16+\amf^2-9\amf^4}{1-\amf^2} + 3\amf^4\frac{5-3\amf^2}{(1-\amf^2)^2}\ln(\amf^2) \right) \nonumber\\
& && \hspace{1.cm} + \aat^2 \at \mf \bigg(-\frac{3}{8}\frac{24-15\amf^2-5\amf^4}{1-\amf^2} - \frac{3\amf^4}{2}\frac{3-2\amf^2}{(1-\amf^2)^2}\ln(\amf^2) + \aat^2 \bigg) \nonumber\\
& && \hspace{1.cm} - \frac{1}{8}(\atC \mfC)^3  \bigg],\label{eq:atat2}
\end{alignat}
\end{subequations}
where, as in the previous Section, the matching scale is set to $\msusy$. These expressions are valid only if the one-loop threshold correction is parametrized in terms of MSSM \DR parameters evaluated at the scale $\msusy$. The rather complicated dependence on $|\mf|$ originates from diagrams with internal Higgsinos. Expressions valid for general $M_A$ are distributed as ancillary files alongside with the paper.

Note that one does not recover the SM to MSSM \order{\alt^2} threshold correction for the SM Higgs self-coupling if these expressions are combined with the matching condition for $\lambda$ between the SM and the THDM. This is due to the expansion around $M_{H^\pm}/\msusy \sim 0$. We checked explicitly that the SM to MSSM threshold correction is indeed recovered if the \order{\alt^2} threshold corrections between the THDM and the MSSM are not expanded in the limit $M_{H^\pm}^2/\msusy^2\rightarrow 0$.


\section{Numerical application: calculation of the lightest MSSM Higgs-boson mass}
\label{sec:04_num_results}

In this Section, we use the analytic results obtained in \cref{sec:SMtoTHDMexample,sec:03_asat,sec:03_atat} to improve the prediction of the lightest MSSM Higgs-boson mass. Our calculation is based upon the one presented in~\cite{Bahl:2018jom} (similar calculations have been performed in~\cite{Haber:1993an,Lee:2015uza,Murphy:2019qpm}).

In~\cite{Bahl:2018jom}, all squarks are integrated out of the MSSM at the scale \msusy obtaining the THDM as EFT. At the scale $M_A$, the heavy Higgs bosons are integrated out recovering the SM as EFT. In addition, the electroweakinos and the gluino are integrated out at two independent scales. Eventually, the SM is recovered as EFT. All couplings are evolved down to the electroweak scale, where the SM-like Higgs mass is calculated.

All EFTs are matched to each other using full one-loop threshold corrections. The evolution of the couplings between the different scales is performed using two-loop renormalization group equations. The pure EFT calculation can be merged with a two-loop fixed-order calculation which allows one to take also contributions suppressed by \msusy and/or $M_A$ into account. The calculation has become part of the publicly available code \FH~\cite{Heinemeyer:1998yj,Heinemeyer:1998np,Hahn:2009zz, Degrassi:2002fi,Frank:2006yh,Hahn:2013ria,Bahl:2016brp,Bahl:2017aev,Bahl:2018qog}.

Here, we extend the calculation presented in~\cite{Bahl:2018jom} by implementing the two-loop threshold corrections derived in \cref{sec:SMtoTHDMexample,sec:03_asat,sec:03_atat}. This means especially that the \order{\alt\als} threshold corrections for $\liii$, $\liv$, and $\lv$, as well as the previously unknown \order{\alt^2} threshold corrections for all $\lambda_i$'s, are taken correctly into account.\footnote{We implement the \order{\alt^2} threshold corrections without assuming $M_A \ll \msusy$. While we in this way include formally suppressed terms into the threshold correction of four-dimensional operators, this procedure allows one to extend the validity range of the THDM-EFT calculation also to the case of $M_A\sim\msusy$ properly recovering the SM to MSSM matching conditions.} Moreover, while it was assumed that the electroweakinos have masses lower or equal to \msusy in \cite{Bahl:2018jom}, we also allow for electroweakino masses larger than \msusy. In this case, we integrate out the sfermions and electroweakinos at the same scale (i.e., \msusy) and match the THDM directly to the MSSM (as done e.g.\ in~\cite{Bahl:2020mjy}).

We note that these improvements still do not allow for a full resummation of \order{\alt,\als} NNLL logarithms, since the full three-loop THDM RGEs are still unknown. Based on findings for the SM as an EFT (see e.g.~\cite{Draper:2013oza}), the numerical impact of the three-loop RGEs is, however, expected to be less sizeable than the impact of the two-loop threshold corrections. As an additional cross-check, we implemented the partial THDM three-loop RGEs derived in~\cite{Herren:2017uxn,Bednyakov:2018cmx} finding $M_h$ shifts of $\lesssim 0.1 \gev$.

\medskip

We investigate the size of the \order{\alt^2} threshold corrections in a simplified scenario with three relevant scales: \msusy, $M_A$ and $\mf$. All sfermion masses and the gluino mass are set equal to \msusy; the electroweakino mass parameters $M_1$ and $M_2$, equal to $\mf$. The trilinear sfermion couplings are set to zero apart from $A_t$ which is fixed in terms of $X_t = A_t - \mu/\tbe$. We set all \cp-violating phases to zero and therefore use the mass of the $A$ boson as input. If not stated otherwise all parameters are assumed to be renormalized in the \DR scheme at the scale \msusy. The only exceptions are $M_A$ and $\tbe$ which we assume to be THDM parameters renormalized in the \MS scheme at the scale $M_A$. Some of the considered scenarios resemble the $M_h^{125}(\text{alignment})$ Higgs-benchmark scenario presented in~\cite{Bahl:2018zmf}. For all shown results, $M_h$ is computed in the pure EFT approach without combining it with the fixed-order calculation implemented in \FH.

\medskip

\begin{figure}\centering
\begin{minipage}{.49\textwidth}\centering
\includegraphics[width=\textwidth]{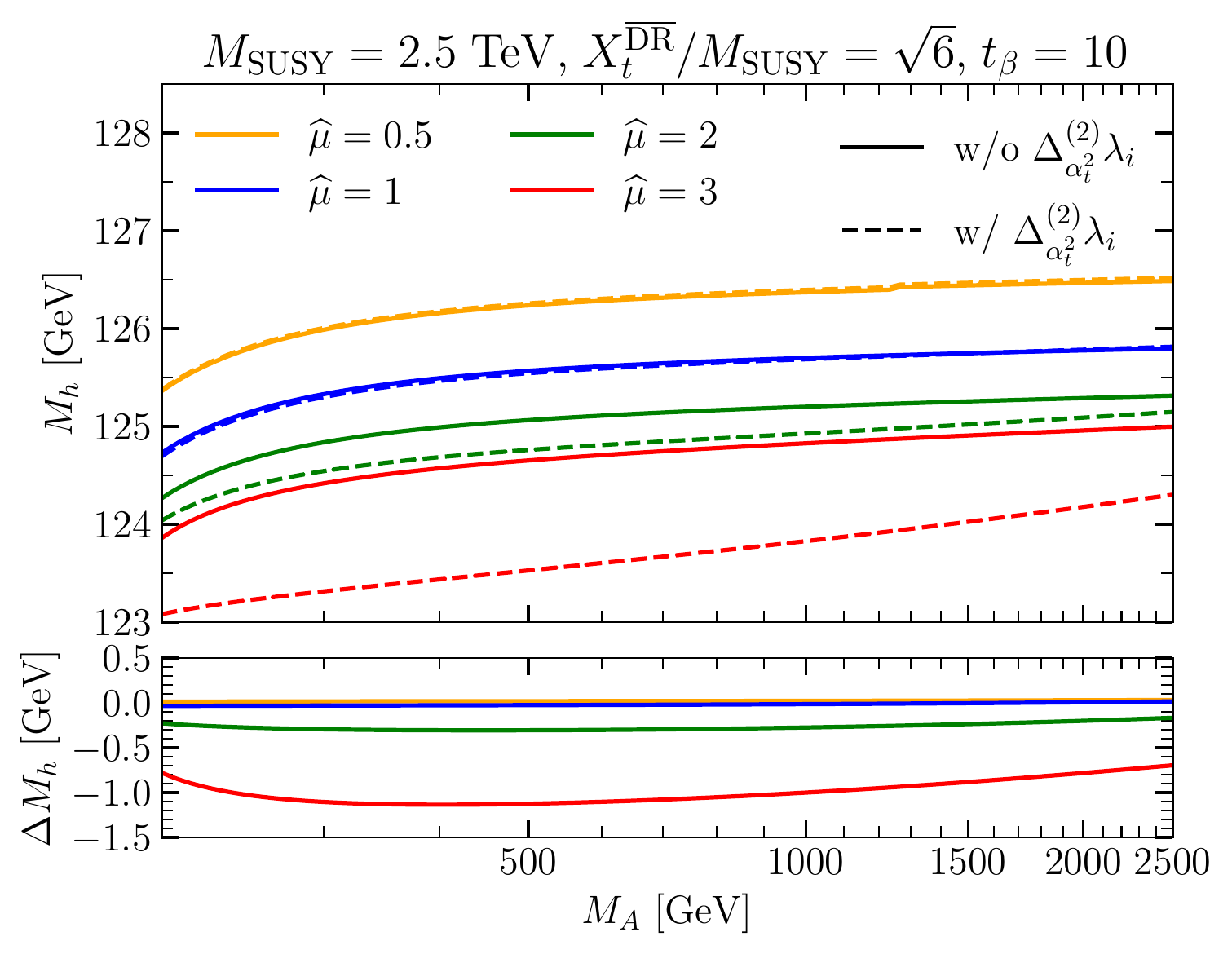}
\end{minipage}\hfill
\begin{minipage}{.49\textwidth}\centering
\includegraphics[width=\textwidth]{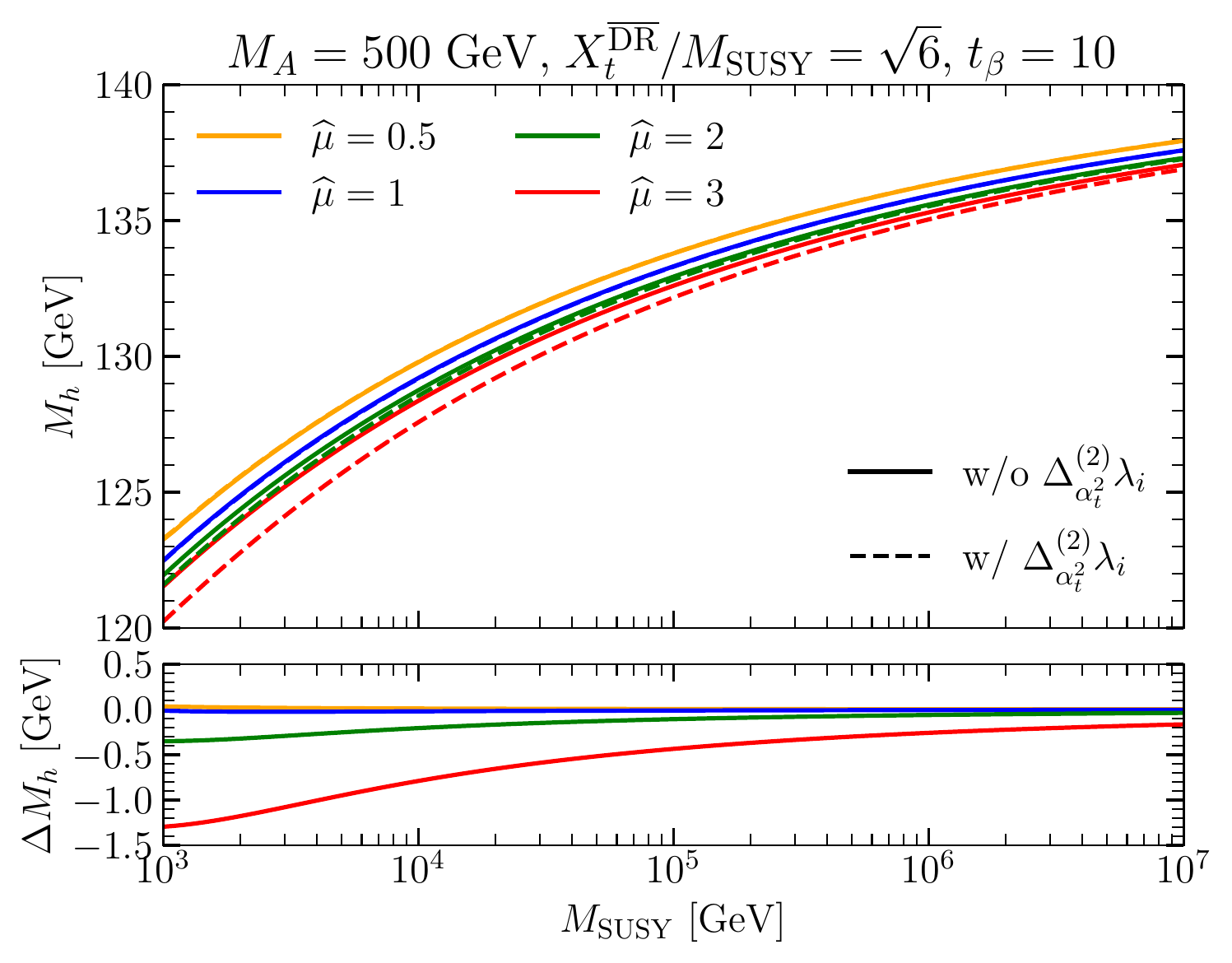}
\end{minipage}
\caption{\textit{Left:}  EFT results for $M_h$ as a function of $M_A$ comparing the THDM-EFT results not including (solid) and including (dashed) the \order{\alt^2} threshold corrections to the Higgs self-couplings. The results are shown for $\mf = 0.5$ (orange), $\mf = 1$ (blue), $\mf = 2$ (green), and $\mf = 3$ (red). $\msusy = 2.5$ TeV, $X_t^\DR/\msusy = \sqrt{6}$, and $\tbe = 10$ are chosen. In the bottom panel the difference between the solid and the dashed lines is displayed. \textit{Right:} Same as left plot, but $M_h$ is shown as a function of \msusy for $M_A = 500$ GeV.}
\label{fig:MA_MSusy_var}
\end{figure}

In \cref{fig:MA_MSusy_var}, we compare the predictions for $M_h$ using the THDM as EFT below \msusy with (dashed) and without (solid) including the \order{\alt^2} threshold corrections for the Higgs self-couplings. The results are shown for four different choices of $\mf = \mu/\msusy$: $\mf = 0.5$ (orange), $\mf = 1$ (blue), $\mf = 2$ (green), and $\mf = 3$ (red). The other parameter are chosen as $\msusy = 2.5\tev$, $X_t^\DR/\msusy = \sqrt{6}$, and $\tbe = 10$.

The left plot of \cref{fig:MA_MSusy_var} shows $M_h$ as a function of $M_A$. The difference between the results with and without the \order{\alt^2} threshold corrections is negligibly small for $\mf = 0.5$ and $\mf = 1$. Different choices for the other parameters, like raising \msusy or lowering $\tbe$, do not lead to larger shifts.\footnote{This holds true especially for the scenarios discussed in~\cite{Bahl:2018jom,Bahl:2019ago} (i.e., for high \msusy, low $\tbe$, and low $\mf$).} The small numerical impact for $\mf\lesssim 1$ can be explained by the parametrization of the one-loop threshold corrections in terms of MSSM couplings. As discussed in detail in~\cite{Kwasnitza:2020wli}, this one-loop parameterization absorbs the most relevant two-loop terms.

The shifts induced by including the two-loop corrections are, however, enlarged for $\mf > 1$: choosing $\mf = 2$ leads to an approximately constant downward shift of $M_h$ by $\sim 0.2\gev$; choosing $\mf = 3$, to a downwards shift of $\sim 0.7 - 1.1 \gev$. As is visible in \crefrange{eq:atat1}{eq:atat2} the two-loop threshold corrections scale with powers of $\mf$ leading to enhancement of them for $\mf > 1$. While this parameter choice also enhances Higgs mixing effects, the dominant contribution to the shifts observed in \cref{fig:MA_MSusy_var} is the two-loop threshold correction to the SM-like Higgs' quartic coupling induced if the two-loop \order{\alt^2} threshold corrections are included. If Higgs mixing would be the dominant effect, the shifts would significantly decrease with rising $M_A$.

In the right plot of \cref{fig:MA_MSusy_var}, the same scenario as in the left plot is shown but \msusy is varied setting $M_A = 500\gev$. As in the left plot, the shifts induced by the \order{\alt^2} threshold correction are negligible for $\mf \lesssim 1$. The shifts for $\mf > 1$ are largest for small \msusy and shrink with raising \msusy (e.g.\ for $\mf = 3$ the shift shrinks from $\sim 1.3\gev$ for $\msusy = 1\tev$ to $\sim 0.2 \gev$ for $\msusy = 10^4 \tev$). This behavior is explained by the RGE running of the top-Yukawa coupling. For rising \msusy, the MSSM top-Yukawa coupling evaluated at \msusy shrinks resulting in a decrease of the \order{\alt^2} threshold corrections.

\medskip

\begin{figure}\centering
\begin{minipage}{.49\textwidth}\centering
\includegraphics[width=\textwidth]{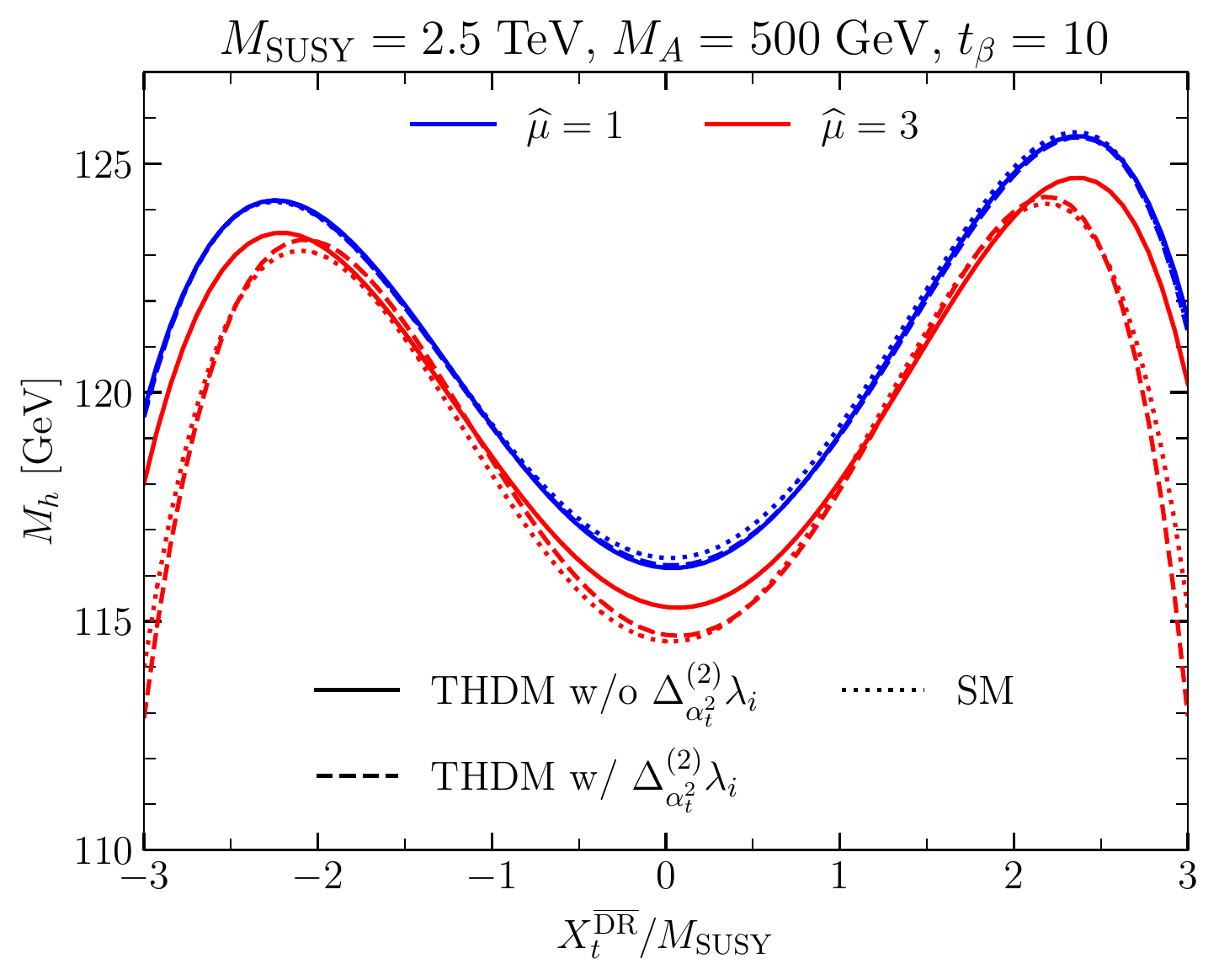}
\end{minipage}\hfill
\begin{minipage}{.49\textwidth}\centering
\includegraphics[width=\textwidth]{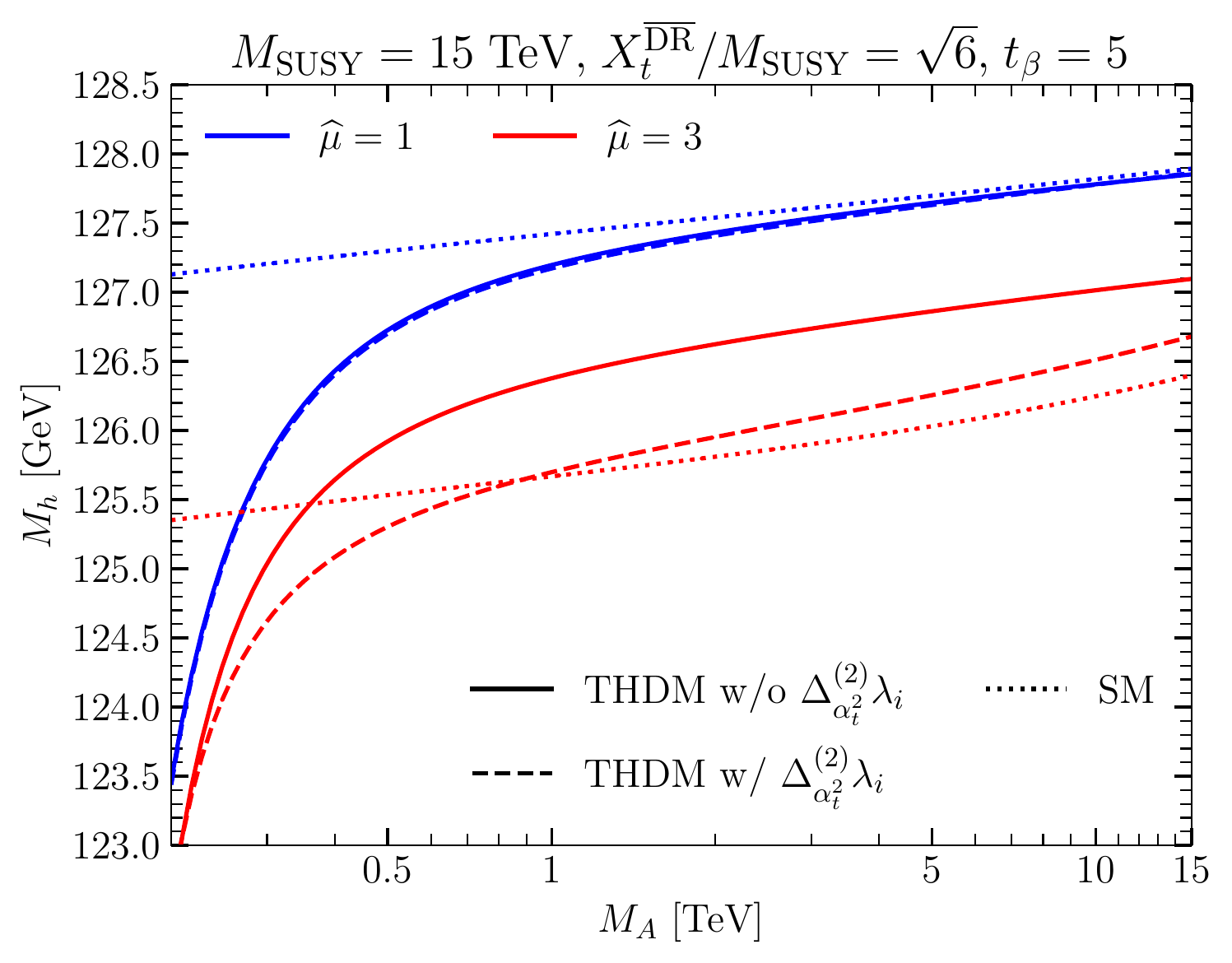}
\end{minipage}
\caption{\textit{Left:} $M_h$ as a function of $X_t^\DR/\msusy$ comparing the THDM-EFT results not including (solid) and including (dashed) the \order{\alt^2} threshold corrections to the Higgs self-couplings as well as the SM-EFT results (dotted). The results are shown for $\mf = 1$ (blue), and $\mf = 3$ (red). $\msusy = 2.5$ TeV, $M_A = 500$ GeV, and $\tbe = 10$ are chosen. \textit{Right:} $M_h$ as a function of $M_A$ comparing the THDM-EFT results not including (solid) and including (dashed) the \order{\alt^2} threshold corrections to the Higgs self-couplings as well as the SM-EFT results (dotted). The results are shown for $\mf = 1$ (blue), and $\mf = 3$ (red). $\msusy = 15$ TeV, $X_t^\DR/\msusy = \sqrt{6}$ and $\tbe = 5$ are chosen.}
\label{fig:xt_var}
\end{figure}

Next, we explore the dependence of the shifts induced by the \order{\alt^2} threshold corrections on $X_t$ and also compare the results using the THDM as EFT below \msusy to the case of using the SM as EFT. In the left plot of \cref{fig:xt_var}, we display $M_h$ as a function of $X_t^\DR/\msusy$ comparing three different results: using the THDM without the \order{\alt^2} threshold corrections as EFT below \msusy (solid), using the THDM with the \order{\alt^2} threshold corrections as EFT below \msusy (dashed), and using the SM as EFT below \msusy (dotted). The results are shown for $\mf = 1$ (blue) and $\mf = 3$ (red). Moreover, $\msusy = 2.5\tev$, $M_A = 500\gev$, and $\tbe = 10$ are chosen.

All three calculations agree very well for $\mf = 1$. For the given scenario, Higgs mixing effects, the resummation of logarithms involving $M_A$ and $\msusy$, as well as the two-loop \order{\alt^2} threshold corrections have a negligible effect. This is slightly different for $\mf = 3$. Here, the THDM result including the \order{\alt^2} threshold corrections and the SM result agree very well, while the THDM result without the \order{\alt^2} threshold correction is shifted to higher $M_h$ values for $X_t^\DR/\msusy\sim 0$ and $|X_t^\DR/\msusy|\gtrsim 2$. This indicates that the impact of the \order{\alt^2} threshold corrections is sizeable for $\mf = 3$, but the Higgs mixing effects and the resummation of logarithms involving $M_A$ and \msusy are of minor importance. Note that before actually calculating the \order{\alt^2} threshold corrections, it is unclear which effect is more dominant and whether the SM-EFT result is a good approximation of the THDM-EFT result including the \order{\alt^2} threshold corrections.

This statement is supported by the right plot of \cref{fig:xt_var}. In this plot, the same three calculations of $M_h$ as in the left plot of \cref{fig:xt_var} are compared as a function of $M_A$ for $\msusy = 15\tev$, $X_t^\DR/\msusy = \sqrt{6}$, and $\tbe = 5$. For $\mf = 1$ (blue), the THDM-EFT results are in good agreement, while the SM-EFT result yields higher $M_h$ values for low $M_A\lesssim 1\tev$. In this region, Higgs mixing effects become relevant, which are not taken into account in the SM-EFT result. For $M_A \gtrsim 5\tev$ all three results are in good agreement indicating that the \order{\alt^2} threshold corrections have only a minor numerical impact. This is different in case of $\mf = 3$ (red): While the THDM-EFT results are only in good agreement for $M_A\sim 200\gev$, the \order{\alt^2} threshold correction shift $M_h$ downwards by $\sim 0.5\gev$ for large $M_A$. While the SM result and the THDM result including the \order{\alt^2} threshold corrections are in relatively good agreement for $M_A\gtrsim 0.8\tev$,\footnote{The disagreement for $M_A\sim\msusy$ can be explained by different parameterization of the threshold corrections. While in the case of the THDM EFT, MSSM couplings are used, SM couplings are employed in case of the SM-EFT calculation. The difference between both options can be regarded as an estimate for the size of higher-order threshold corrections~\cite{Allanach:2018fif,Bahl:2019hmm}.} there is a difference of up to $\sim 2.5\gev$ for lower $M_A$ values originating from Higgs mixing effects. Without the inclusion of the \order{\alt^2} threshold corrections it is again unclear whether the SM- or the THDM-EFT calculation is more trustworthy for low $M_A$.


\section{Conclusions}
\label{sec:05_conclusions}

Without direct evidence for BSM physics, EFT techniques become increasingly popular to constrain high-energy theories (HETs). In this work, we concentrated on the calculation of matching conditions for renormalizable operators using the diagrammatic approach. Giving explicit expressions up to the two-loop level, we highlighted different contributions arising due to 1LPI Green's functions, different field normalizations in the EFT and the HET, and the reparametrization of the EFT result in terms of HET couplings. We also discussed how different observables can be used to derive the same matching condition and that it is often preferable to choose an observable with a higher number of external legs. Moreover, we discussed the treatment of the ``light'' masses pointing out that setting them to zero (if possible) considerably simplifies the calculation. In order to be able to use the cancellation of infrared divergencies as a cross-check but to also simplify the calculation, we proposed to introduce infrared regulator masses, which are independent of the other parameters (i.e.\ the Higgs vev) of the theory.

As illustration and application of the presented general statement, we considered the SM and the THDM as EFTs. For the SM, we for the first time computed the \order{\alt^2} threshold correction for the matching of the SM-Higgs self-coupling to the THDM.

In case of the THDM, we pointed out that the calculation of the matching conditions for the quartic Higgs couplings require the computation of at least one four-point function. Moreover, we presented equations relating a set of four-point functions to the matching conditions of the quartic Higgs couplings. Using them we calculated the previously only partly known \order{\alt\als} threshold corrections for the matching of the THDM quartic Higgs couplings to the MSSM as well as the previously completely unknown \order{\alt^2} corrections. For both calculations, we used two different approaches: in the first approach, we kept the full dependence on the ``light'' masses; in the second approach, we set the Higgs vev to zero and introduced an infrared regulator for the top quark. While we confirmed that both calculations yield identical results, the second calculation was found to be much simpler due to a lower number of Feynman diagrams and a simplified expansion in the large \msusy limit.

As a numerical application, we used the calculated two-loop corrections to improve the calculation of the lightest MSSM Higgs-boson mass using the THDM as EFT finding shifts of up to $\sim 1\gev$ for $\mf > 1$. The precision level of the updated THDM-EFT calculation is now on a similar level as the SM-EFT calculation mitigating the issue of deciding which calculation is more precise for a given parameter setting. The presented improvements will become part of the public code \FH.

While we used the matching of quartic Higgs couplings between the SM and the THDM as well as the THDM and the MSSM as examples, the used techniques are straightforwardly applicable to the matching of other theories or couplings.


\section*{Acknowledgements}
\sloppy{
We thank Pietro Slavich and Georg Weiglein for useful discussions. We acknowledge support by the Deutsche Forschungsgemeinschaft (DFG, German Research Foundation) under Germany's Excellence Strategy -- EXC 2121 ``Quantum Universe'' – 390833306.
}


\appendix


\section{Ward identities of the THDM Higgs sector}
\label{app:06_THDMward}

For deriving the Ward identities of the THDM Higgs sector, we follow the recipe given e.g.\ in~\cite{Denner:1994xt}. To be more specific, we use the invariance of the effective action $\Gamma$ with respect to variations of the $Z$-boson background field $\hat Z$, $\delta\Gamma/\delta\hat Z = 0$. The resulting expression is successively differentiated with respect to the Higgs fields.

After neglecting vertex functions with a negative mass dimension, we arrive at the following set of Ward identities,
\begin{align}
\Gamma_{\chiII}=&{} -\frac{v_1 }{v_2}\Gamma_{\chiI},\\
\Gamma_{\chiII\phiI}=&{} -\frac{v_1 }{v_2}\Gamma_{\chiI\phiI}-\frac{1}{v_2}\Gamma_{\chiI},\\
\Gamma_{\chiII\phiII}=&{} \frac{v_1 }{v_2^2}\Gamma_{\chiI}-\frac{v_1 }{v_2}\Gamma_{\chiI\phiII},\\
\Gamma_{\chiI\chiII}=&{} \frac{1}{v_2}\Gamma_{\phiI}-\frac{v_1 }{v_2}\Gamma_{\chiI\chiI},\\
\Gamma_{\chiII\chiII}=&{} \frac{v_1^2 }{v_2^2}\Gamma_{\chiI\chiI}-\frac{v_1 }{v_2^2}\Gamma_{\phiI}+\frac{1}{v_2}\Gamma_{\phiII},\\
\Gamma_{\phiIp\phiIIm}=&{} -\frac{v_1 }{v_2}\Gamma_{\phiIm\phiIp}-\frac{i }{v_2}\Gamma_{\chiI}+\frac{1}{v_2}\Gamma_{\phiI},\\
\Gamma_{\phiIm\phiIIp}=&{} -\frac{v_1 }{v_2}\Gamma_{\phiIm\phiIp}+\frac{i }{v_2}\Gamma_{\chiI}+\frac{1}{v_2}\Gamma_{\phiI},\\
\Gamma_{\phiIIm\phiIIp}=&{} \frac{v_1^2 }{v_2^2}\Gamma_{\phiIm\phiIp}-\frac{v_1 }{v_2^2}\Gamma_{\phiI}+\frac{1}{v_2}\Gamma_{\phiII},\\
\Gamma_{\chiII\phiI\phiI}=&{} -\frac{v_1 }{v_2}\Gamma_{\chiI\phiI\phiI}-\frac{2 }{v_2}\Gamma_{\chiI\phiI},\\
\Gamma_{\phiI\phiII\phiII}=&{} -\frac{2 v_1 }{v_2}\Gamma_{\phiI\phiI\phiII}+\frac{3 }{v_2}\Gamma_{\phiI\phiII}-\frac{v_1^2 }{v_2^2}\Gamma_{\phiI\phiI\phiI}+\frac{3 v_1 }{v_2^2}\Gamma_{\phiI\phiI}-\frac{3 }{v_2^2}\Gamma_{\phiI},\\
\Gamma_{\chiI\phiI\phiII}=&{} \frac{2 }{v_2}\Gamma_{\chiI\phiI}-\frac{v_1 }{v_2}\Gamma_{\chiI\phiI\phiI},\\
\Gamma_{\chiII\phiI\phiII}=&{} \frac{v_1^2 }{v_2^2}\Gamma_{\chiI\phiI\phiI}-\frac{v_1 }{v_2^2}\Gamma_{\chiI\phiI}-\frac{1}{v_2}\Gamma_{\chiI\phiII}+\frac{1}{v_2^2}\Gamma_{\chiI},\\
\Gamma_{\chiI\chiI\phiI}=&{} \frac{1}{3} \Gamma_{\phiI\phiI\phiI},\\
\Gamma_{\chiI\chiII\phiI}=&{} -\frac{1}{v_2}\Gamma_{\chiI\chiI}-\frac{v_1 }{3 v_2}\Gamma_{\phiI\phiI\phiI}+\frac{1}{v_2}\Gamma_{\phiI\phiI},\\
\Gamma_{\chiII\chiII\phiI}=&{} \frac{2 v_1 }{v_2^2}\Gamma_{\chiI\chiI}+\frac{1}{v_2}\Gamma_{\phiI\phiII}+\frac{v_1^2 }{3 v_2^2}\Gamma_{\phiI\phiI\phiI}-\frac{v_1 }{v_2^2}\Gamma_{\phiI\phiI}-\frac{1}{v_2^2}\Gamma_{\phiI},\\
\Gamma_{\phiI\phiIm\phiIp}=&{} \frac{1}{3} \Gamma_{\phiI\phiI\phiI},\\
\Gamma_{\phiI\phiIp\phiIIm}=&{} -\frac{i }{v_2}\Gamma_{\chiI\phiI}-\frac{v_1 }{3 v_2}\Gamma_{\phiI\phiI\phiI}+\frac{1}{v_2}\Gamma_{\phiI\phiI}-\frac{1}{v_2}\Gamma_{\phiIm\phiIp},\\
\Gamma_{\phiI\phiIm\phiIIp}=&{} \frac{i }{v_2}\Gamma_{\chiI\phiI}-\frac{v_1 }{3 v_2}\Gamma_{\phiI\phiI\phiI}+\frac{1}{v_2}\Gamma_{\phiI\phiI}-\frac{1}{v_2}\Gamma_{\phiIm\phiIp},\\
\Gamma_{\phiI\phiIIm\phiIIp}=&{} \frac{1}{v_2}\Gamma_{\phiI\phiII}+\frac{v_1^2 }{3 v_2^2}\Gamma_{\phiI\phiI\phiI}-\frac{v_1 }{v_2^2}\Gamma_{\phiI\phiI}+\frac{2 v_1 }{v_2^2}\Gamma_{\phiIm\phiIp}-\frac{1}{v_2^2}\Gamma_{\phiI},\\
\Gamma_{\phiII\phiII\phiII}=&{} \frac{3 v_1^2 }{v_2^2}\Gamma_{\phiI\phiI\phiII}-\frac{3 v_1 }{v_2^2}\Gamma_{\phiI\phiII}+\frac{2 v_1^3 }{v_2^3}\Gamma_{\phiI\phiI\phiI}-\frac{6 v_1^2 }{v_2^3}\Gamma_{\phiI\phiI}+\frac{3 }{v_2}\Gamma_{\phiII\phiII} \nonumber\\
& +\frac{6 v_1 }{v_2^3}\Gamma_{\phiI}-\frac{3 }{v_2^2}\Gamma_{\phiII},\\
\Gamma_{\chiI\phiII\phiII}=&{} \frac{v_1^2 }{v_2^2}\Gamma_{\chiI\phiI\phiI}-\frac{v_1 }{v_2^2}\Gamma_{\chiI\phiI}+\frac{3 }{v_2}\Gamma_{\chiI\phiII}-\frac{3 }{v_2^2}\Gamma_{\chiI},\\
\Gamma_{\chiII\phiII\phiII}=&{} -\frac{v_1^3 }{v_2^3}\Gamma_{\chiI\phiI\phiI}+\frac{v_1^2 }{v_2^3}\Gamma_{\chiI\phiI}-\frac{v_1 }{v_2^2}\Gamma_{\chiI\phiII}+\frac{v_1 }{v_2^3}\Gamma_{\chiI},\\
\Gamma_{\chiI\chiI\phiII}=&{} \frac{2 }{v_2}\Gamma_{\chiI\chiI}+\Gamma_{\phiI\phiI\phiII}+\frac{2 v_1 }{3 v_2}\Gamma_{\phiI\phiI\phiI}-\frac{2 }{v_2}\Gamma_{\phiI\phiI},\\
\Gamma_{\chiI\chiII\phiII}=&{} -\frac{v_1 }{v_2^2}\Gamma_{\chiI\chiI}-\frac{v_1 }{v_2}\Gamma_{\phiI\phiI\phiII}+\frac{1}{v_2}\Gamma_{\phiI\phiII}-\frac{2 v_1^2 }{3 v_2^2}\Gamma_{\phiI\phiI\phiI}+\frac{2 v_1 }{v_2^2}\Gamma_{\phiI\phiI}-\frac{1}{v_2^2}\Gamma_{\phiI},\\
\Gamma_{\chiII\chiII\phiII}=&{} \frac{v_1^2 }{v_2^2}\Gamma_{\phiI\phiI\phiII}-\frac{v_1 }{v_2^2}\Gamma_{\phiI\phiII}+\frac{2 v_1^3 }{3 v_2^3}\Gamma_{\phiI\phiI\phiI}-\frac{2 v_1^2 }{v_2^3}\Gamma_{\phiI\phiI}+\frac{1}{v_2}\Gamma_{\phiII\phiII} \nonumber\\
& +\frac{2 v_1 }{v_2^3}\Gamma_{\phiI}-\frac{1}{v_2^2}\Gamma_{\phiII},\\
\Gamma_{\phiIm\phiIp\phiII}=&{} \Gamma_{\phiI\phiI\phiII}+\frac{2 v_1 }{3 v_2}\Gamma_{\phiI\phiI\phiI}-\frac{2 }{v_2}\Gamma_{\phiI\phiI}+\frac{2 }{v_2}\Gamma_{\phiIm\phiIp},\\
\Gamma_{\phiIp\phiII\phiIIm}=&{} -\frac{i }{v_2}\Gamma_{\chiI\phiII}-\frac{v_1 }{v_2}\Gamma_{\phiI\phiI\phiII}+\frac{1}{v_2}\Gamma_{\phiI\phiII}-\frac{2 v_1^2 }{3 v_2^2}\Gamma_{\phiI\phiI\phiI}+\frac{2 v_1 }{v_2^2}\Gamma_{\phiI\phiI}-\frac{v_1 }{v_2^2}\Gamma_{\phiIm\phiIp} \nonumber\\
& +\frac{i }{v_2^2}\Gamma_{\chiI}-\frac{1}{v_2^2}\Gamma_{\phiI},\\
\Gamma_{\phiIm\phiII\phiIIp}=&{} \frac{i }{v_2}\Gamma_{\chiI\phiII}-\frac{v_1 }{v_2}\Gamma_{\phiI\phiI\phiII}+\frac{1}{v_2}\Gamma_{\phiI\phiII}-\frac{2 v_1^2 }{3 v_2^2}\Gamma_{\phiI\phiI\phiI}+\frac{2 v_1 }{v_2^2}\Gamma_{\phiI\phiI}-\frac{v_1 }{v_2^2}\Gamma_{\phiIm\phiIp} \nonumber\\
& -\frac{i }{v_2^2}\Gamma_{\chiI}-\frac{1}{v_2^2}\Gamma_{\phiI},\\
\Gamma_{\phiII\phiIIm\phiIIp}=&{} \frac{v_1^2 }{v_2^2}\Gamma_{\phiI\phiI\phiII}-\frac{v_1 }{v_2^2}\Gamma_{\phiI\phiII}+\frac{2 v_1^3 }{3 v_2^3}\Gamma_{\phiI\phiI\phiI}-\frac{2 v_1^2 }{v_2^3}\Gamma_{\phiI\phiI}+\frac{1}{v_2}\Gamma_{\phiII\phiII} \nonumber\\
& +\frac{2 v_1 }{v_2^3}\Gamma_{\phiI}-\frac{1}{v_2^2}\Gamma_{\phiII},\\
\Gamma_{\chiI\chiI\chiI}=&{} 3 \Gamma_{\chiI\phiI\phiI},\\
\Gamma_{\chiI\chiI\chiII}=&{} \frac{2 }{v_2}\Gamma_{\chiI\phiI}-\frac{3 v_1 }{v_2}\Gamma_{\chiI\phiI\phiI},\\
\Gamma_{\chiI\chiII\chiII}=&{} \frac{3 v_1^2 }{v_2^2}\Gamma_{\chiI\phiI\phiI}-\frac{3 v_1 }{v_2^2}\Gamma_{\chiI\phiI}+\frac{1}{v_2}\Gamma_{\chiI\phiII}-\frac{1}{v_2^2}\Gamma_{\chiI},\\
\Gamma_{\chiI\phiIm\phiIp}=&{} \Gamma_{\chiI\phiI\phiI},\\
\Gamma_{\chiI\phiIp\phiIIm}=&{} -\frac{v_1 }{v_2}\Gamma_{\chiI\phiI\phiI}+\frac{1}{v_2}\Gamma_{\chiI\phiI}-\frac{i }{v_2}\Gamma_{\chiI\chiI}+\frac{i }{v_2}\Gamma_{\phiIm\phiIp},\\
\Gamma_{\chiI\phiIm\phiIIp}=&{} -\frac{v_1 }{v_2}\Gamma_{\chiI\phiI\phiI}+\frac{1}{v_2}\Gamma_{\chiI\phiI}+\frac{i }{v_2}\Gamma_{\chiI\chiI}-\frac{i }{v_2}\Gamma_{\phiIm\phiIp},\\
\Gamma_{\chiI\phiIIm\phiIIp}=&{} \frac{v_1^2 }{v_2^2}\Gamma_{\chiI\phiI\phiI}-\frac{v_1 }{v_2^2}\Gamma_{\chiI\phiI}+\frac{1}{v_2}\Gamma_{\chiI\phiII}-\frac{1}{v_2^2}\Gamma_{\chiI},\\
\Gamma_{\chiII\chiII\chiII}=&{} -\frac{3 v_1^3 }{v_2^3}\Gamma_{\chiI\phiI\phiI}+\frac{3 v_1^2 }{v_2^3}\Gamma_{\chiI\phiI}-\frac{3 v_1 }{v_2^2}\Gamma_{\chiI\phiII}+\frac{3 v_1 }{v_2^3}\Gamma_{\chiI},\\
\Gamma_{\chiII\phiIm\phiIp}=&{} -\frac{v_1 }{v_2}\Gamma_{\chiI\phiI\phiI},\\
\Gamma_{\chiII\phiIp\phiIIm}=&{} \frac{v_1^2 }{v_2^2}\Gamma_{\chiI\phiI\phiI}-\frac{v_1 }{v_2^2}\Gamma_{\chiI\phiI}+\frac{i v_1 }{v_2^2}\Gamma_{\chiI\chiI}-\frac{i v_1 }{v_2^2}\Gamma_{\phiIm\phiIp},\\
\Gamma_{\chiII\phiIm\phiIIp}=&{} \frac{v_1^2 }{v_2^2}\Gamma_{\chiI\phiI\phiI}-\frac{v_1 }{v_2^2}\Gamma_{\chiI\phiI}-\frac{i v_1 }{v_2^2}\Gamma_{\chiI\chiI}+\frac{i v_1 }{v_2^2}\Gamma_{\phiIm\phiIp},\\
\Gamma_{\chiII\phiIIm\phiIIp}=&{} -\frac{v_1^3 }{v_2^3}\Gamma_{\chiI\phiI\phiI}+\frac{v_1^2 }{v_2^3}\Gamma_{\chiI\phiI}-\frac{v_1 }{v_2^2}\Gamma_{\chiI\phiII}+\frac{v_1 }{v_2^3}\Gamma_{\chiI},\\
\Gamma_{\phiI\phiI\phiI\phiII}=&{} \frac{1}{v_2}\Gamma_{\phiI\phiI\phiI}-\frac{v_1 }{v_2}\Gamma_{\phiI\phiI\phiI\phiI},\\
\Gamma_{\chiI\phiI\phiI\phiI}=&{} 0,\\
\Gamma_{\chiII\phiI\phiI\phiI}=&{} -\frac{3 }{v_2}\Gamma_{\chiI\phiI\phiI},\\
\Gamma_{\phiI\phiI\phiII\phiII}=&{} \frac{1}{v_2}\Gamma_{\phiI\phiI\phiII}+\frac{v_1^2 }{v_2^2}\Gamma_{\phiI\phiI\phiI\phiI}-\frac{v_1 }{v_2^2}\Gamma_{\phiI\phiI\phiI},\\
\Gamma_{\chiI\phiI\phiI\phiII}=&{} \frac{1}{v_2}\Gamma_{\chiI\phiI\phiI},\\
\Gamma_{\chiII\phiI\phiI\phiII}=&{} \frac{2 v_1 }{v_2^2}\Gamma_{\chiI\phiI\phiI}-\frac{2 }{v_2^2}\Gamma_{\chiI\phiI},\\
\Gamma_{\chiI\chiI\phiI\phiI}=&{} \frac{1}{3} \Gamma_{\phiI\phiI\phiI\phiI},\\
\Gamma_{\chiI\chiII\phiI\phiI}=&{} \frac{1}{3 v_2}\Gamma_{\phiI\phiI\phiI}-\frac{v_1 }{3 v_2}\Gamma_{\phiI\phiI\phiI\phiI},\\
\Gamma_{\chiII\chiII\phiI\phiI}=&{} \frac{2 }{v_2^2}\Gamma_{\chiI\chiI}+\frac{1}{v_2}\Gamma_{\phiI\phiI\phiII}+\frac{v_1^2 }{3 v_2^2}\Gamma_{\phiI\phiI\phiI\phiI}+\frac{v_1 }{3 v_2^2}\Gamma_{\phiI\phiI\phiI}-\frac{2 }{v_2^2}\Gamma_{\phiI\phiI},\\
\Gamma_{\phiI\phiI\phiIm\phiIp}=&{} \frac{1}{3} \Gamma_{\phiI\phiI\phiI\phiI},\\
\Gamma_{\phiI\phiI\phiIp\phiIIm}=&{} -\frac{i }{v_2}\Gamma_{\chiI\phiI\phiI}-\frac{v_1 }{3 v_2}\Gamma_{\phiI\phiI\phiI\phiI}+\frac{1}{3 v_2}\Gamma_{\phiI\phiI\phiI},\\
\Gamma_{\phiI\phiI\phiIm\phiIIp}=&{} \frac{i }{v_2}\Gamma_{\chiI\phiI\phiI}-\frac{v_1 }{3 v_2}\Gamma_{\phiI\phiI\phiI\phiI}+\frac{1}{3 v_2}\Gamma_{\phiI\phiI\phiI},\\
\Gamma_{\phiI\phiI\phiIIm\phiIIp}=&{} \frac{1}{v_2}\Gamma_{\phiI\phiI\phiII}+\frac{v_1^2 }{3 v_2^2}\Gamma_{\phiI\phiI\phiI\phiI}+\frac{v_1 }{3 v_2^2}\Gamma_{\phiI\phiI\phiI}-\frac{2 }{v_2^2}\Gamma_{\phiI\phiI}+\frac{2 }{v_2^2}\Gamma_{\phiIm\phiIp},\\
\Gamma_{\phiI\phiII\phiII\phiII}=&{} -\frac{3 v_1 }{v_2^2}\Gamma_{\phiI\phiI\phiII}+\frac{3 }{v_2^2}\Gamma_{\phiI\phiII}-\frac{v_1^3 }{v_2^3}\Gamma_{\phiI\phiI\phiI\phiI}+\frac{3 v_1 }{v_2^3}\Gamma_{\phiI\phiI}-\frac{3 }{v_2^3}\Gamma_{\phiI},\\
\Gamma_{\chiI\phiI\phiII\phiII}=&{} \frac{2 }{v_2^2}\Gamma_{\chiI\phiI}-\frac{2 v_1 }{v_2^2}\Gamma_{\chiI\phiI\phiI},\\
\Gamma_{\chiII\phiI\phiII\phiII}=&{} -\frac{v_1^2 }{v_2^3}\Gamma_{\chiI\phiI\phiI}+\frac{v_1 }{v_2^3}\Gamma_{\chiI\phiI}-\frac{1}{v_2^2}\Gamma_{\chiI\phiII}+\frac{1}{v_2^3}\Gamma_{\chiI},\\
\Gamma_{\chiI\chiI\phiI\phiII}=&{} \frac{1}{3 v_2}\Gamma_{\phiI\phiI\phiI}-\frac{v_1 }{3 v_2}\Gamma_{\phiI\phiI\phiI\phiI},\\
\Gamma_{\chiI\chiII\phiI\phiII}=&{} -\frac{1}{v_2^2}\Gamma_{\chiI\chiI}+\frac{v_1^2 }{3 v_2^2}\Gamma_{\phiI\phiI\phiI\phiI}-\frac{2 v_1 }{3 v_2^2}\Gamma_{\phiI\phiI\phiI}+\frac{1}{v_2^2}\Gamma_{\phiI\phiI},\\
\Gamma_{\chiII\chiII\phiI\phiII}=&{} -\frac{v_1 }{v_2^2}\Gamma_{\phiI\phiI\phiII}+\frac{1}{v_2^2}\Gamma_{\phiI\phiII}-\frac{v_1^3 }{3 v_2^3}\Gamma_{\phiI\phiI\phiI\phiI}+\frac{v_1 }{v_2^3}\Gamma_{\phiI\phiI}-\frac{1}{v_2^3}\Gamma_{\phiI},\\
\Gamma_{\phiI\phiIm\phiIp\phiII}=&{} \frac{1}{3 v_2}\Gamma_{\phiI\phiI\phiI}-\frac{v_1 }{3 v_2}\Gamma_{\phiI\phiI\phiI\phiI},\\
\Gamma_{\phiI\phiIp\phiII\phiIIm}=&{} \frac{i v_1 }{v_2^2}\Gamma_{\chiI\phiI\phiI}-\frac{i }{v_2^2}\Gamma_{\chiI\phiI}+\frac{v_1^2 }{3 v_2^2}\Gamma_{\phiI\phiI\phiI\phiI}-\frac{2 v_1 }{3 v_2^2}\Gamma_{\phiI\phiI\phiI}+\frac{1}{v_2^2}\Gamma_{\phiI\phiI} \nonumber\\
& -\frac{1}{v_2^2}\Gamma_{\phiIm\phiIp},\\
\Gamma_{\phiI\phiIm\phiII\phiIIp}=&{} -\frac{i v_1 }{v_2^2}\Gamma_{\chiI\phiI\phiI}+\frac{i }{v_2^2}\Gamma_{\chiI\phiI}+\frac{v_1^2 }{3 v_2^2}\Gamma_{\phiI\phiI\phiI\phiI}-\frac{2 v_1 }{3 v_2^2}\Gamma_{\phiI\phiI\phiI}+\frac{1}{v_2^2}\Gamma_{\phiI\phiI} \nonumber\\
& -\frac{1}{v_2^2}\Gamma_{\phiIm\phiIp},\\
\Gamma_{\phiI\phiII\phiIIm\phiIIp}=&{} -\frac{v_1 }{v_2^2}\Gamma_{\phiI\phiI\phiII}+\frac{1}{v_2^2}\Gamma_{\phiI\phiII}-\frac{v_1^3 }{3 v_2^3}\Gamma_{\phiI\phiI\phiI\phiI}+\frac{v_1 }{v_2^3}\Gamma_{\phiI\phiI}-\frac{1}{v_2^3}\Gamma_{\phiI},\\
\Gamma_{\chiI\chiI\chiI\phiI}=&{} 0,\\
\Gamma_{\chiI\chiI\chiII\phiI}=&{} -\frac{1}{v_2}\Gamma_{\chiI\phiI\phiI},\\
\Gamma_{\chiI\chiII\chiII\phiI}=&{} \frac{2 v_1 }{v_2^2}\Gamma_{\chiI\phiI\phiI}-\frac{2 }{v_2^2}\Gamma_{\chiI\phiI},\\
\Gamma_{\chiI\phiI\phiIm\phiIp}=&{} 0,\\
\Gamma_{\chiI\phiI\phiIp\phiIIm}=&{} 0,\\
\Gamma_{\chiI\phiI\phiIm\phiIIp}=&{} 0,\\
\Gamma_{\chiI\phiI\phiIIm\phiIIp}=&{} 0,\\
\Gamma_{\chiII\chiII\chiII\phiI}=&{} -\frac{3 v_1^2 }{v_2^3}\Gamma_{\chiI\phiI\phiI}+\frac{3 v_1 }{v_2^3}\Gamma_{\chiI\phiI}-\frac{3 }{v_2^2}\Gamma_{\chiI\phiII}+\frac{3 }{v_2^3}\Gamma_{\chiI},\\
\Gamma_{\chiII\phiI\phiIm\phiIp}=&{} -\frac{1}{v_2}\Gamma_{\chiI\phiI\phiI},\\
\Gamma_{\chiII\phiI\phiIp\phiIIm}=&{} \frac{v_1 }{v_2^2}\Gamma_{\chiI\phiI\phiI}-\frac{1}{v_2^2}\Gamma_{\chiI\phiI}+\frac{i }{v_2^2}\Gamma_{\chiI\chiI}-\frac{i }{v_2^2}\Gamma_{\phiIm\phiIp},\\
\Gamma_{\chiII\phiI\phiIm\phiIIp}=&{} \frac{v_1 }{v_2^2}\Gamma_{\chiI\phiI\phiI}-\frac{1}{v_2^2}\Gamma_{\chiI\phiI}-\frac{i }{v_2^2}\Gamma_{\chiI\chiI}+\frac{i }{v_2^2}\Gamma_{\phiIm\phiIp},\\
\Gamma_{\chiII\phiI\phiIIm\phiIIp}=&{} -\frac{v_1^2 }{v_2^3}\Gamma_{\chiI\phiI\phiI}+\frac{v_1 }{v_2^3}\Gamma_{\chiI\phiI}-\frac{1}{v_2^2}\Gamma_{\chiI\phiII}+\frac{1}{v_2^3}\Gamma_{\chiI},\\
\Gamma_{\phiII\phiII\phiII\phiII}=&{} \frac{6 v_1^2 }{v_2^3}\Gamma_{\phiI\phiI\phiII}-\frac{6 v_1 }{v_2^3}\Gamma_{\phiI\phiII}+\frac{v_1^4 }{v_2^4}\Gamma_{\phiI\phiI\phiI\phiI}+\frac{2 v_1^3 }{v_2^4}\Gamma_{\phiI\phiI\phiI}-\frac{9 v_1^2 }{v_2^4}\Gamma_{\phiI\phiI}+\frac{3 }{v_2^2}\Gamma_{\phiII\phiII} \nonumber\\
& +\frac{9 v_1 }{v_2^4}\Gamma_{\phiI}-\frac{3 }{v_2^3}\Gamma_{\phiII},\\
\Gamma_{\chiI\phiII\phiII\phiII}=&{} \frac{3 v_1^2 }{v_2^3}\Gamma_{\chiI\phiI\phiI}-\frac{3 v_1 }{v_2^3}\Gamma_{\chiI\phiI}+\frac{3 }{v_2^2}\Gamma_{\chiI\phiII}-\frac{3 }{v_2^3}\Gamma_{\chiI},\\
\Gamma_{\chiII\phiII\phiII\phiII}=&{} 0,\\
\Gamma_{\chiI\chiI\phiII\phiII}=&{} \frac{2 }{v_2^2}\Gamma_{\chiI\chiI}+\frac{1}{v_2}\Gamma_{\phiI\phiI\phiII}+\frac{v_1^2 }{3 v_2^2}\Gamma_{\phiI\phiI\phiI\phiI}+\frac{v_1 }{3 v_2^2}\Gamma_{\phiI\phiI\phiI}-\frac{2 }{v_2^2}\Gamma_{\phiI\phiI},\\
\Gamma_{\chiI\chiII\phiII\phiII}=&{} -\frac{v_1 }{v_2^2}\Gamma_{\phiI\phiI\phiII}+\frac{1}{v_2^2}\Gamma_{\phiI\phiII}-\frac{v_1^3 }{3 v_2^3}\Gamma_{\phiI\phiI\phiI\phiI}+\frac{v_1 }{v_2^3}\Gamma_{\phiI\phiI}-\frac{1}{v_2^3}\Gamma_{\phiI},\\
\Gamma_{\chiII\chiII\phiII\phiII}=&{} \frac{2 v_1^2 }{v_2^3}\Gamma_{\phiI\phiI\phiII}-\frac{2 v_1 }{v_2^3}\Gamma_{\phiI\phiII}+\frac{v_1^4 }{3 v_2^4}\Gamma_{\phiI\phiI\phiI\phiI} \nonumber\\
& +\frac{2 v_1^3 }{3 v_2^4}\Gamma_{\phiI\phiI\phiI}-\frac{3 v_1^2 }{v_2^4}\Gamma_{\phiI\phiI}+\frac{1}{v_2^2}\Gamma_{\phiII\phiII}+\frac{3 v_1 }{v_2^4}\Gamma_{\phiI}-\frac{1}{v_2^3}\Gamma_{\phiII},\\
\Gamma_{\phiIm\phiIp\phiII\phiII}=&{} \frac{1}{v_2}\Gamma_{\phiI\phiI\phiII}+\frac{v_1^2 }{3 v_2^2}\Gamma_{\phiI\phiI\phiI\phiI}+\frac{v_1 }{3 v_2^2}\Gamma_{\phiI\phiI\phiI}-\frac{2 }{v_2^2}\Gamma_{\phiI\phiI}+\frac{2 }{v_2^2}\Gamma_{\phiIm\phiIp},\\
\Gamma_{\phiIp\phiII\phiII\phiIIm}=&{} -\frac{i v_1^2 }{v_2^3}\Gamma_{\chiI\phiI\phiI}+\frac{i v_1 }{v_2^3}\Gamma_{\chiI\phiI}-\frac{i }{v_2^2}\Gamma_{\chiI\phiII} \nonumber\\
& -\frac{v_1 }{v_2^2}\Gamma_{\phiI\phiI\phiII}+\frac{1}{v_2^2}\Gamma_{\phiI\phiII}-\frac{v_1^3 }{3 v_2^3}\Gamma_{\phiI\phiI\phiI\phiI}+\frac{v_1 }{v_2^3}\Gamma_{\phiI\phiI}+\frac{i }{v_2^3}\Gamma_{\chiI}-\frac{1}{v_2^3}\Gamma_{\phiI},\\
\Gamma_{\phiIm\phiII\phiII\phiIIp}=&{} \frac{i v_1^2 }{v_2^3}\Gamma_{\chiI\phiI\phiI}-\frac{i v_1 }{v_2^3}\Gamma_{\chiI\phiI}+\frac{i }{v_2^2}\Gamma_{\chiI\phiII}-\frac{v_1 }{v_2^2}\Gamma_{\phiI\phiI\phiII}+\frac{1}{v_2^2}\Gamma_{\phiI\phiII}-\frac{v_1^3 }{3 v_2^3}\Gamma_{\phiI\phiI\phiI\phiI} \nonumber\\
& +\frac{v_1 }{v_2^3}\Gamma_{\phiI\phiI}-\frac{i }{v_2^3}\Gamma_{\chiI}-\frac{1}{v_2^3}\Gamma_{\phiI},\\
\Gamma_{\phiII\phiII\phiIIm\phiIIp}=&{} \frac{2 v_1^2 }{v_2^3}\Gamma_{\phiI\phiI\phiII}-\frac{2 v_1 }{v_2^3}\Gamma_{\phiI\phiII}+\frac{v_1^4 }{3 v_2^4}\Gamma_{\phiI\phiI\phiI\phiI} \nonumber\\
& +\frac{2 v_1^3 }{3 v_2^4}\Gamma_{\phiI\phiI\phiI}-\frac{3 v_1^2 }{v_2^4}\Gamma_{\phiI\phiI}+\frac{1}{v_2^2}\Gamma_{\phiII\phiII}+\frac{3 v_1 }{v_2^4}\Gamma_{\phiI}-\frac{1}{v_2^3}\Gamma_{\phiII},\\
\Gamma_{\chiI\chiI\chiI\phiII}=&{} \frac{3 }{v_2}\Gamma_{\chiI\phiI\phiI},\\
\Gamma_{\chiI\chiI\chiII\phiII}=&{} \frac{2 }{v_2^2}\Gamma_{\chiI\phiI}-\frac{2 v_1 }{v_2^2}\Gamma_{\chiI\phiI\phiI},\\
\Gamma_{\chiI\chiII\chiII\phiII}=&{} \frac{v_1^2 }{v_2^3}\Gamma_{\chiI\phiI\phiI}-\frac{v_1 }{v_2^3}\Gamma_{\chiI\phiI}+\frac{1}{v_2^2}\Gamma_{\chiI\phiII}-\frac{1}{v_2^3}\Gamma_{\chiI},\\
\Gamma_{\chiI\phiIm\phiIp\phiII}=&{} \frac{1}{v_2}\Gamma_{\chiI\phiI\phiI},\\
\Gamma_{\chiI\phiIp\phiII\phiIIm}=&{} -\frac{v_1 }{v_2^2}\Gamma_{\chiI\phiI\phiI}+\frac{1}{v_2^2}\Gamma_{\chiI\phiI}-\frac{i }{v_2^2}\Gamma_{\chiI\chiI}+\frac{i }{v_2^2}\Gamma_{\phiIm\phiIp},\\
\Gamma_{\chiI\phiIm\phiII\phiIIp}=&{} -\frac{v_1 }{v_2^2}\Gamma_{\chiI\phiI\phiI}+\frac{1}{v_2^2}\Gamma_{\chiI\phiI}+\frac{i }{v_2^2}\Gamma_{\chiI\chiI}-\frac{i }{v_2^2}\Gamma_{\phiIm\phiIp},\\
\Gamma_{\chiI\phiII\phiIIm\phiIIp}=&{} \frac{v_1^2 }{v_2^3}\Gamma_{\chiI\phiI\phiI}-\frac{v_1 }{v_2^3}\Gamma_{\chiI\phiI}+\frac{1}{v_2^2}\Gamma_{\chiI\phiII}-\frac{1}{v_2^3}\Gamma_{\chiI},\\
\Gamma_{\chiII\chiII\chiII\phiII}=&{} 0,\\
\Gamma_{\chiII\phiIm\phiIp\phiII}=&{} 0,\\
\Gamma_{\chiII\phiIp\phiII\phiIIm}=&{} 0,\\
\Gamma_{\chiII\phiIm\phiII\phiIIp}=&{} 0,\\
\Gamma_{\chiII\phiII\phiIIm\phiIIp}=&{} 0,\\
\Gamma_{\chiI\chiI\chiI\chiI}=&{} \Gamma_{\phiI\phiI\phiI\phiI},\\
\Gamma_{\chiI\chiI\chiI\chiII}=&{} \frac{1}{v_2}\Gamma_{\phiI\phiI\phiI}-\frac{v_1 }{v_2}\Gamma_{\phiI\phiI\phiI\phiI},\\
\Gamma_{\chiI\chiI\chiII\chiII}=&{} \frac{1}{v_2}\Gamma_{\phiI\phiI\phiII}+\frac{v_1^2 }{v_2^2}\Gamma_{\phiI\phiI\phiI\phiI}-\frac{v_1 }{v_2^2}\Gamma_{\phiI\phiI\phiI},\\
\Gamma_{\chiI\chiI\phiIm\phiIp}=&{} \frac{1}{3} \Gamma_{\phiI\phiI\phiI\phiI},\\
\Gamma_{\chiI\chiI\phiIp\phiIIm}=&{} -\frac{i }{v_2}\Gamma_{\chiI\phiI\phiI}-\frac{v_1 }{3 v_2}\Gamma_{\phiI\phiI\phiI\phiI}+\frac{1}{3 v_2}\Gamma_{\phiI\phiI\phiI},\\
\Gamma_{\chiI\chiI\phiIm\phiIIp}=&{} \frac{i }{v_2}\Gamma_{\chiI\phiI\phiI}-\frac{v_1 }{3 v_2}\Gamma_{\phiI\phiI\phiI\phiI}+\frac{1}{3 v_2}\Gamma_{\phiI\phiI\phiI},\\
\Gamma_{\chiI\chiI\phiIIm\phiIIp}=&{} \frac{1}{v_2}\Gamma_{\phiI\phiI\phiII}+\frac{v_1^2 }{3 v_2^2}\Gamma_{\phiI\phiI\phiI\phiI}+\frac{v_1 }{3 v_2^2}\Gamma_{\phiI\phiI\phiI}-\frac{2 }{v_2^2}\Gamma_{\phiI\phiI}+\frac{2 }{v_2^2}\Gamma_{\phiIm\phiIp},\\
\Gamma_{\chiI\chiII\chiII\chiII}=&{} -\frac{3 v_1 }{v_2^2}\Gamma_{\phiI\phiI\phiII}+\frac{3 }{v_2^2}\Gamma_{\phiI\phiII}-\frac{v_1^3 }{v_2^3}\Gamma_{\phiI\phiI\phiI\phiI}+\frac{3 v_1 }{v_2^3}\Gamma_{\phiI\phiI}-\frac{3 }{v_2^3}\Gamma_{\phiI},\\
\Gamma_{\chiI\chiII\phiIm\phiIp}=&{} \frac{1}{3 v_2}\Gamma_{\phiI\phiI\phiI}-\frac{v_1 }{3 v_2}\Gamma_{\phiI\phiI\phiI\phiI},\\
\Gamma_{\chiI\chiII\phiIp\phiIIm}=&{} \frac{i v_1 }{v_2^2}\Gamma_{\chiI\phiI\phiI}-\frac{i }{v_2^2}\Gamma_{\chiI\phiI}+\frac{v_1^2 }{3 v_2^2}\Gamma_{\phiI\phiI\phiI\phiI}-\frac{2 v_1 }{3 v_2^2}\Gamma_{\phiI\phiI\phiI} \nonumber\\
& +\frac{1}{v_2^2}\Gamma_{\phiI\phiI}-\frac{1}{v_2^2}\Gamma_{\phiIm\phiIp},\\
\Gamma_{\chiI\chiII\phiIm\phiIIp}=&{} -\frac{i v_1 }{v_2^2}\Gamma_{\chiI\phiI\phiI}+\frac{i }{v_2^2}\Gamma_{\chiI\phiI}+\frac{v_1^2 }{3 v_2^2}\Gamma_{\phiI\phiI\phiI\phiI}-\frac{2 v_1 }{3 v_2^2}\Gamma_{\phiI\phiI\phiI} \nonumber\\
& +\frac{1}{v_2^2}\Gamma_{\phiI\phiI}-\frac{1}{v_2^2}\Gamma_{\phiIm\phiIp},\\
\Gamma_{\chiI\chiII\phiIIm\phiIIp}=&{} -\frac{v_1 }{v_2^2}\Gamma_{\phiI\phiI\phiII}+\frac{1}{v_2^2}\Gamma_{\phiI\phiII}-\frac{v_1^3 }{3 v_2^3}\Gamma_{\phiI\phiI\phiI\phiI}+\frac{v_1 }{v_2^3}\Gamma_{\phiI\phiI}-\frac{1}{v_2^3}\Gamma_{\phiI},\\
\Gamma_{\chiII\chiII\chiII\chiII}=&{} \frac{6 v_1^2 }{v_2^3}\Gamma_{\phiI\phiI\phiII}-\frac{6 v_1 }{v_2^3}\Gamma_{\phiI\phiII}+\frac{v_1^4 }{v_2^4}\Gamma_{\phiI\phiI\phiI\phiI}+\frac{2 v_1^3 }{v_2^4}\Gamma_{\phiI\phiI\phiI}-\frac{9 v_1^2 }{v_2^4}\Gamma_{\phiI\phiI} \nonumber\\
& +\frac{3 }{v_2^2}\Gamma_{\phiII\phiII}+\frac{9 v_1 }{v_2^4}\Gamma_{\phiI}-\frac{3 }{v_2^3}\Gamma_{\phiII},\\
\Gamma_{\chiII\chiII\phiIm\phiIp}=&{} \frac{1}{v_2}\Gamma_{\phiI\phiI\phiII}+\frac{v_1^2 }{3 v_2^2}\Gamma_{\phiI\phiI\phiI\phiI}+\frac{v_1 }{3 v_2^2}\Gamma_{\phiI\phiI\phiI}-\frac{2 }{v_2^2}\Gamma_{\phiI\phiI}+\frac{2 }{v_2^2}\Gamma_{\phiIm\phiIp},\\
\Gamma_{\chiII\chiII\phiIp\phiIIm}=&{} -\frac{i v_1^2 }{v_2^3}\Gamma_{\chiI\phiI\phiI}+\frac{i v_1 }{v_2^3}\Gamma_{\chiI\phiI}-\frac{i }{v_2^2}\Gamma_{\chiI\phiII}-\frac{v_1 }{v_2^2}\Gamma_{\phiI\phiI\phiII}+\frac{1}{v_2^2}\Gamma_{\phiI\phiII} \nonumber\\
& -\frac{v_1^3 }{3 v_2^3}\Gamma_{\phiI\phiI\phiI\phiI}+\frac{v_1 }{v_2^3}\Gamma_{\phiI\phiI}+\frac{i }{v_2^3}\Gamma_{\chiI}-\frac{1}{v_2^3}\Gamma_{\phiI},\\
\Gamma_{\chiII\chiII\phiIm\phiIIp}=&{} \frac{i v_1^2 }{v_2^3}\Gamma_{\chiI\phiI\phiI}-\frac{i v_1 }{v_2^3}\Gamma_{\chiI\phiI}+\frac{i }{v_2^2}\Gamma_{\chiI\phiII}-\frac{v_1 }{v_2^2}\Gamma_{\phiI\phiI\phiII}+\frac{1}{v_2^2}\Gamma_{\phiI\phiII} \nonumber\\
& -\frac{v_1^3 }{3 v_2^3}\Gamma_{\phiI\phiI\phiI\phiI}+\frac{v_1 }{v_2^3}\Gamma_{\phiI\phiI}-\frac{i }{v_2^3}\Gamma_{\chiI}-\frac{1}{v_2^3}\Gamma_{\phiI},\\
\Gamma_{\chiII\chiII\phiIIm\phiIIp}=&{} \frac{2 v_1^2 }{v_2^3}\Gamma_{\phiI\phiI\phiII}-\frac{2 v_1 }{v_2^3}\Gamma_{\phiI\phiII}+\frac{v_1^4 }{3 v_2^4}\Gamma_{\phiI\phiI\phiI\phiI}+\frac{2 v_1^3 }{3 v_2^4}\Gamma_{\phiI\phiI\phiI}-\frac{3 v_1^2 }{v_2^4}\Gamma_{\phiI\phiI} \nonumber\\
& +\frac{1}{v_2^2}\Gamma_{\phiII\phiII}+\frac{3 v_1 }{v_2^4}\Gamma_{\phiI}-\frac{1}{v_2^3}\Gamma_{\phiII},\\
\Gamma_{\phiIm\phiIm\phiIp\phiIp}=&{} \frac{2}{3} \Gamma_{\phiI\phiI\phiI\phiI},\\
\Gamma_{\phiIm\phiIp\phiIp\phiIIm}=&{} -\frac{2 i }{v_2}\Gamma_{\chiI\phiI\phiI}-\frac{2 v_1 }{3 v_2}\Gamma_{\phiI\phiI\phiI\phiI}+\frac{2 }{3 v_2}\Gamma_{\phiI\phiI\phiI},\\
\Gamma_{\phiIp\phiIp\phiIIm\phiIIm}=&{} \frac{4 i v_1 }{v_2^2}\Gamma_{\chiI\phiI\phiI}-\frac{4 i }{v_2^2}\Gamma_{\chiI\phiI}-\frac{2 }{v_2^2}\Gamma_{\chiI\chiI}+\frac{2 v_1^2 }{3 v_2^2}\Gamma_{\phiI\phiI\phiI\phiI}-\frac{4 v_1 }{3 v_2^2}\Gamma_{\phiI\phiI\phiI} \nonumber\\
& +\frac{2 }{v_2^2}\Gamma_{\phiI\phiI},\\
\Gamma_{\phiIm\phiIm\phiIp\phiIIp}=&{} \frac{2 i }{v_2}\Gamma_{\chiI\phiI\phiI}-\frac{2 v_1 }{3 v_2}\Gamma_{\phiI\phiI\phiI\phiI}+\frac{2 }{3 v_2}\Gamma_{\phiI\phiI\phiI},\\
\Gamma_{\phiIm\phiIp\phiIIm\phiIIp}=&{} \frac{1}{v_2^2}\Gamma_{\chiI\chiI}+\frac{1}{v_2}\Gamma_{\phiI\phiI\phiII}+\frac{2 v_1^2 }{3 v_2^2}\Gamma_{\phiI\phiI\phiI\phiI}-\frac{v_1 }{3 v_2^2}\Gamma_{\phiI\phiI\phiI}-\frac{1}{v_2^2}\Gamma_{\phiI\phiI},\\
\Gamma_{\phiIp\phiIIm\phiIIm\phiIIp}=&{} -\frac{2 i v_1^2 }{v_2^3}\Gamma_{\chiI\phiI\phiI}+\frac{2 i v_1 }{v_2^3}\Gamma_{\chiI\phiI}-\frac{2 i }{v_2^2}\Gamma_{\chiI\phiII}-\frac{2 v_1 }{v_2^2}\Gamma_{\phiI\phiI\phiII}+\frac{2 }{v_2^2}\Gamma_{\phiI\phiII} \nonumber\\
& -\frac{2 v_1^3 }{3 v_2^3}\Gamma_{\phiI\phiI\phiI\phiI}+\frac{2 v_1 }{v_2^3}\Gamma_{\phiI\phiI}+\frac{2 i }{v_2^3}\Gamma_{\chiI}-\frac{2 }{v_2^3}\Gamma_{\phiI},\\
\Gamma_{\phiIm\phiIm\phiIIp\phiIIp}=&{} -\frac{4 i v_1 }{v_2^2}\Gamma_{\chiI\phiI\phiI}+\frac{4 i }{v_2^2}\Gamma_{\chiI\phiI}-\frac{2 }{v_2^2}\Gamma_{\chiI\chiI}+\frac{2 v_1^2 }{3 v_2^2}\Gamma_{\phiI\phiI\phiI\phiI}-\frac{4 v_1 }{3 v_2^2}\Gamma_{\phiI\phiI\phiI} \nonumber\\
& +\frac{2 }{v_2^2}\Gamma_{\phiI\phiI},\\
\Gamma_{\phiIm\phiIIm\phiIIp\phiIIp}=&{} \frac{2 i v_1^2 }{v_2^3}\Gamma_{\chiI\phiI\phiI}-\frac{2 i v_1 }{v_2^3}\Gamma_{\chiI\phiI}+\frac{2 i }{v_2^2}\Gamma_{\chiI\phiII}-\frac{2 v_1 }{v_2^2}\Gamma_{\phiI\phiI\phiII}+\frac{2 }{v_2^2}\Gamma_{\phiI\phiII}-\frac{2 v_1^3 }{3 v_2^3}\Gamma_{\phiI\phiI\phiI\phiI} \nonumber\\
& +\frac{2 v_1 }{v_2^3}\Gamma_{\phiI\phiI}-\frac{2 i }{v_2^3}\Gamma_{\chiI}-\frac{2 }{v_2^3}\Gamma_{\phiI},\\
\Gamma_{\phiIIm\phiIIm\phiIIp\phiIIp}=&{} \frac{4 v_1^2 }{v_2^3}\Gamma_{\phiI\phiI\phiII}-\frac{4 v_1 }{v_2^3}\Gamma_{\phiI\phiII}+\frac{2 v_1^4 }{3 v_2^4}\Gamma_{\phiI\phiI\phiI\phiI}+\frac{4 v_1^3 }{3 v_2^4}\Gamma_{\phiI\phiI\phiI}-\frac{6 v_1^2 }{v_2^4}\Gamma_{\phiI\phiI} \nonumber\\
& +\frac{2 }{v_2^2}\Gamma_{\phiII\phiII}+\frac{6 v_1 }{v_2^4}\Gamma_{\phiI}-\frac{2 }{v_2^3}\Gamma_{\phiII}.
\end{align}


\section{Matching of THDM scalar four-point couplings: Explicit field normalization contributions}
\label{sec:THDMWFR}

Assuming that the scalar four-point couplings of the THDM are zero at the tree level, the two-loop contribution to their matching condition due to the different normalization of the Higgs fields in the THDM and the HET is given by
\begin{subequations}
\begin{align}
\Delta_\text{WFR}^{(2)}\lambda_1 ={}& 2 \delta^{(1)} Z_{11} \Delta^{(1)}\lambda_1 + 2\Re(\delta^{(1)} Z_{12}\Delta^{(1)}\lambda_6^*), \\
\Delta_\text{WFR}^{(2)}\lambda_2 ={}& 2 \delta^{(1)} Z_{22} \Delta^{(1)}\lambda_2 + 2  \Re(\delta^{(1)} Z_{12}\delta^{(1)}\lambda_7^*), \\
\Delta_\text{WFR}^{(2)}\lambda_3 ={}& (\delta^{(1)} Z_{11} + \delta^{(1)} Z_{22}) \Delta^{(1)}\lambda_3 + \Re\left(\delta^{(1)} Z_{12}(\Delta^{(1)}\lambda_6^* + \Delta^{(1)}\lambda_7^*)\right), \\
\Delta_\text{WFR}^{(2)}\lambda_4 ={}& (\delta^{(1)} Z_{11} + \delta^{(1)} Z_{22}) \Delta^{(1)}\lambda_4 + \Re\left(\delta^{(1)} Z_{12}(\Delta^{(1)}\lambda_6^* + \Delta^{(1)}\lambda_7^*)\right), \\
\Delta_\text{WFR}^{(2)}\lambda_5 ={}& (\delta^{(1)} Z_{11} + \delta^{(1)} Z_{22}) \Delta^{(1)}\lambda_5 + \delta^{(1)} Z_{21}(\Delta^{(1)}\lambda_6^* + \Delta^{(1)}\lambda_7^*), \\
\Delta_\text{WFR}^{(2)}\lambda_6 ={}& \frac{1}{2}(3\delta^{(1)} Z_{11} + \delta^{(1)} Z_{22}) \Delta^{(1)}\lambda_6 + \frac{1}{2}\delta^{(1)} Z_{12}(\Delta^{(1)}\lambda_1 + \Delta^{(1)}\lambda_3 + \Delta^{(1)}\lambda_4) \nonumber\\
& + \frac{1}{2}\delta^{(1)} Z_{21}\Delta^{(1)}\lambda_5^*\\
\Delta_\text{WFR}^{(2)}\lambda_7 ={}& \frac{1}{2}(\delta^{(1)} Z_{11} + 3\delta^{(1)} Z_{22}) \Delta^{(1)}\lambda_7 + \frac{1}{2}\delta^{(1)} Z_{12}(\Delta^{(1)}\lambda_2 + \Delta^{(1)}\lambda_3 + \Delta^{(1)}\lambda_4) \nonumber\\
& + \frac{1}{2}\delta^{(1)} Z_{21}\Delta^{(1)}\lambda_5^*,
\end{align}
\end{subequations}
where $\delta^{(1)}\lambda_i$ are the one-loop threshold corrections and
\begin{align}
\delta^{(1)} Z_{ij} = -\frac{\partial}{\partial p^2}\left[\Sigma^{\HET,(1)}_{\phi_i\phi_j}(p^2) - \Sigma^{\THDM,(1)(}_{\phi_i\phi_j}(p^2)\right]_{p^2 = 0, \text{fin}}.
\end{align}
Explicit expressions for the $\delta^{(1)} Z_{ij}$'s for the case of the MSSM as HET can be found in~\cite{Bahl:2018ykj}.



\newpage

\bibliographystyle{JHEP.bst}
\bibliography{bibliography}{}

\providecommand{\href}[2]{#2}\begingroup\raggedright\begin{thebibliography}{10}

\bibitem{Staub:2009bi}
F.~Staub, \emph{{From Superpotential to Model Files for FeynArts and
  CalcHep/CompHep}},
  \href{http://dx.doi.org/10.1016/j.cpc.2010.01.011}{\emph{Comput.Phys.Commun.}
  {\bf 181} (2010) 1077--1086}, [\href{http://arxiv.org/abs/0909.2863}{{\tt
  0909.2863}}].

\bibitem{Staub:2010jh}
F.~Staub, \emph{{Automatic Calculation of supersymmetric Renormalization Group
  Equations and Self Energies}},
  \href{http://dx.doi.org/10.1016/j.cpc.2010.11.030}{\emph{Comput.Phys.Commun.}
  {\bf 182} (2011) 808--833}, [\href{http://arxiv.org/abs/1002.0840}{{\tt
  1002.0840}}].

\bibitem{Staub:2012pb}
F.~Staub, \emph{{SARAH 3.2: Dirac Gauginos, UFO output, and more}},
  \href{http://dx.doi.org/10.1016/j.cpc.2013.02.019}{\emph{Comput. Phys.
  Commun.} {\bf 184} (2013) 1792--1809},
  [\href{http://arxiv.org/abs/1207.0906}{{\tt 1207.0906}}].

\bibitem{Staub:2013tta}
F.~Staub, \emph{{SARAH 4 : A tool for (not only SUSY) model builders}},
  \href{http://dx.doi.org/10.1016/j.cpc.2014.02.018}{\emph{Comput. Phys.
  Commun.} {\bf 185} (2014) 1773--1790},
  [\href{http://arxiv.org/abs/1309.7223}{{\tt 1309.7223}}].

\bibitem{Lyonnet:2013dna}
F.~Lyonnet, I.~Schienbein, F.~Staub and A.~Wingerter, \emph{{PyR@TE:
  Renormalization Group Equations for General Gauge Theories}},
  \href{http://dx.doi.org/10.1016/j.cpc.2013.12.002}{\emph{Comput. Phys.
  Commun.} {\bf 185} (2014) 1130--1152},
  [\href{http://arxiv.org/abs/1309.7030}{{\tt 1309.7030}}].

\bibitem{Lyonnet:2016xiz}
F.~Lyonnet and I.~Schienbein, \emph{{PyR@TE 2: A Python tool for computing RGEs
  at two-loop}},
  \href{http://dx.doi.org/10.1016/j.cpc.2016.12.003}{\emph{Comput. Phys.
  Commun.} {\bf 213} (2017) 181--196},
  [\href{http://arxiv.org/abs/1608.07274}{{\tt 1608.07274}}].

\bibitem{1808891}
L.~Sartore and I.~Schienbein, \emph{{PyR@TE 3}},
  \href{http://arxiv.org/abs/2007.12700}{{\tt 2007.12700}}.

\bibitem{Machacek:1983tz}
M.~E. Machacek and M.~T. Vaughn, \emph{{Two Loop Renormalization Group
  Equations in a General Quantum Field Theory. 1. Wave Function
  Renormalization}},
  \href{http://dx.doi.org/10.1016/0550-3213(83)90610-7}{\emph{Nucl. Phys. B}
  {\bf 222} (1983) 83--103}.

\bibitem{Machacek:1983fi}
M.~E. Machacek and M.~T. Vaughn, \emph{{Two Loop Renormalization Group
  Equations in a General Quantum Field Theory. 2. Yukawa Couplings}},
  \href{http://dx.doi.org/10.1016/0550-3213(84)90533-9}{\emph{Nucl. Phys. B}
  {\bf 236} (1984) 221--232}.

\bibitem{Machacek:1984zw}
M.~E. Machacek and M.~T. Vaughn, \emph{{Two Loop Renormalization Group
  Equations in a General Quantum Field Theory. 3. Scalar Quartic Couplings}},
  \href{http://dx.doi.org/10.1016/0550-3213(85)90040-9}{\emph{Nucl. Phys. B}
  {\bf 249} (1985) 70--92}.

\bibitem{Luo:2002ti}
M.-x. Luo, H.-w. Wang and Y.~Xiao, \emph{{Two loop renormalization group
  equations in general gauge field theories}},
  \href{http://dx.doi.org/10.1103/PhysRevD.67.065019}{\emph{Phys. Rev. D} {\bf
  67} (2003) 065019}, [\href{http://arxiv.org/abs/hep-ph/0211440}{{\tt
  hep-ph/0211440}}].

\bibitem{Sperling:2013eva}
M.~Sperling, D.~St{\"o}ckinger and A.~Voigt, \emph{{Renormalization of vacuum
  expectation values in spontaneously broken gauge theories}},
  \href{http://dx.doi.org/10.1007/JHEP07(2013)132}{\emph{JHEP} {\bf 07} (2013)
  132}, [\href{http://arxiv.org/abs/1305.1548}{{\tt 1305.1548}}].

\bibitem{Sperling:2013xqa}
M.~Sperling, D.~St{\"o}ckinger and A.~Voigt, \emph{{Renormalization of vacuum
  expectation values in spontaneously broken gauge theories: Two-loop
  results}}, \href{http://dx.doi.org/10.1007/JHEP01(2014)068}{\emph{JHEP} {\bf
  01} (2014) 068}, [\href{http://arxiv.org/abs/1310.7629}{{\tt 1310.7629}}].

\bibitem{Bednyakov:2018cmx}
A.~Bednyakov, \emph{{On three-loop RGE for the Higgs sector of 2HDM}},
  \href{http://dx.doi.org/10.1007/JHEP11(2018)154}{\emph{JHEP} {\bf 11} (2018)
  154}, [\href{http://arxiv.org/abs/1809.04527}{{\tt 1809.04527}}].

\bibitem{Schienbein:2018fsw}
I.~Schienbein, F.~Staub, T.~Steudtner and K.~Svirina, \emph{{Revisiting RGEs
  for general gauge theories}},
  \href{http://dx.doi.org/10.1016/j.nuclphysb.2018.12.001}{\emph{Nucl. Phys. B}
  {\bf 939} (2019) 1--48}, [\href{http://arxiv.org/abs/1809.06797}{{\tt
  1809.06797}}].

\bibitem{Poole:2019kcm}
C.~Poole and A.~E. Thomsen, \emph{{Constraints on 3- and 4-loop
  $\beta$-functions in a general four-dimensional Quantum Field Theory}},
  \href{http://dx.doi.org/10.1007/JHEP09(2019)055}{\emph{JHEP} {\bf 09} (2019)
  055}, [\href{http://arxiv.org/abs/1906.04625}{{\tt 1906.04625}}].

\bibitem{Henning:2014wua}
B.~Henning, X.~Lu and H.~Murayama, \emph{{How to use the Standard Model
  effective field theory}},
  \href{http://dx.doi.org/10.1007/JHEP01(2016)023}{\emph{JHEP} {\bf 01} (2016)
  023}, [\href{http://arxiv.org/abs/1412.1837}{{\tt 1412.1837}}].

\bibitem{Drozd:2015rsp}
A.~Drozd, J.~Ellis, J.~Quevillon and T.~You, \emph{{The Universal One-Loop
  Effective Action}},
  \href{http://dx.doi.org/10.1007/JHEP03(2016)180}{\emph{JHEP} {\bf 03} (2016)
  180}, [\href{http://arxiv.org/abs/1512.03003}{{\tt 1512.03003}}].

\bibitem{delAguila:2016zcb}
F.~del Aguila, Z.~Kunszt and J.~Santiago, \emph{{One-loop effective lagrangians
  after matching}},
  \href{http://dx.doi.org/10.1140/epjc/s10052-016-4081-1}{\emph{Eur. Phys. J.
  C} {\bf 76} (2016) 244}, [\href{http://arxiv.org/abs/1602.00126}{{\tt
  1602.00126}}].

\bibitem{Boggia:2016asg}
M.~Boggia, R.~Gomez-Ambrosio and G.~Passarino, \emph{{Low energy behaviour of
  standard model extensions}},
  \href{http://dx.doi.org/10.1007/JHEP05(2016)162}{\emph{JHEP} {\bf 05} (2016)
  162}, [\href{http://arxiv.org/abs/1603.03660}{{\tt 1603.03660}}].

\bibitem{Henning:2016lyp}
B.~Henning, X.~Lu and H.~Murayama, \emph{{One-loop Matching and Running with
  Covariant Derivative Expansion}},
  \href{http://dx.doi.org/10.1007/JHEP01(2018)123}{\emph{JHEP} {\bf 01} (2018)
  123}, [\href{http://arxiv.org/abs/1604.01019}{{\tt 1604.01019}}].

\bibitem{Ellis:2016enq}
S.~A.~R. Ellis, J.~Quevillon, T.~You and Z.~Zhang, \emph{{Mixed heavy--light
  matching in the Universal One-Loop Effective Action}},
  \href{http://dx.doi.org/10.1016/j.physletb.2016.09.016}{\emph{Phys. Lett. B}
  {\bf 762} (2016) 166--176}, [\href{http://arxiv.org/abs/1604.02445}{{\tt
  1604.02445}}].

\bibitem{Fuentes-Martin:2016uol}
J.~Fuentes-Martin, J.~Portoles and P.~Ruiz-Femenia, \emph{{Integrating out
  heavy particles with functional methods: a simplified framework}},
  \href{http://dx.doi.org/10.1007/JHEP09(2016)156}{\emph{JHEP} {\bf 09} (2016)
  156}, [\href{http://arxiv.org/abs/1607.02142}{{\tt 1607.02142}}].

\bibitem{Zhang:2016pja}
Z.~Zhang, \emph{{Covariant diagrams for one-loop matching}},
  \href{http://dx.doi.org/10.1007/JHEP05(2017)152}{\emph{JHEP} {\bf 05} (2017)
  152}, [\href{http://arxiv.org/abs/1610.00710}{{\tt 1610.00710}}].

\bibitem{Ellis:2017jns}
S.~A.~R. Ellis, J.~Quevillon, T.~You and Z.~Zhang, \emph{{Extending the
  Universal One-Loop Effective Action: Heavy-Light Coefficients}},
  \href{http://dx.doi.org/10.1007/JHEP08(2017)054}{\emph{JHEP} {\bf 08} (2017)
  054}, [\href{http://arxiv.org/abs/1706.07765}{{\tt 1706.07765}}].

\bibitem{Summ:2018oko}
B.~Summ and A.~Voigt, \emph{{Extending the Universal One-Loop Effective Action
  by Regularization Scheme Translating Operators}},
  \href{http://dx.doi.org/10.1007/JHEP08(2018)026}{\emph{JHEP} {\bf 08} (2018)
  026}, [\href{http://arxiv.org/abs/1806.05171}{{\tt 1806.05171}}].

\bibitem{Bakshi:2018ics}
S.~Das~Bakshi, J.~Chakrabortty and S.~K. Patra, \emph{{CoDEx: Wilson
  coefficient calculator connecting SMEFT to UV theory}},
  \href{http://dx.doi.org/10.1140/epjc/s10052-018-6444-2}{\emph{Eur. Phys. J.
  C} {\bf 79} (2019) 21}, [\href{http://arxiv.org/abs/1808.04403}{{\tt
  1808.04403}}].

\bibitem{Kramer:2019fwz}
M.~Krämer, B.~Summ and A.~Voigt, \emph{{Completing the scalar and fermionic
  Universal One-Loop Effective Action}},
  \href{http://dx.doi.org/10.1007/JHEP01(2020)079}{\emph{JHEP} {\bf 01} (2020)
  079}, [\href{http://arxiv.org/abs/1908.04798}{{\tt 1908.04798}}].

\bibitem{Cohen:2019btp}
T.~Cohen, M.~Freytsis and X.~Lu, \emph{{Functional Methods for Heavy Quark
  Effective Theory}},
  \href{http://dx.doi.org/10.1007/JHEP06(2020)164}{\emph{JHEP} {\bf 06} (2020)
  164}, [\href{http://arxiv.org/abs/1912.08814}{{\tt 1912.08814}}].

\bibitem{Ellis:2020ivx}
S.~A. Ellis, J.~Quevillon, P.~N.~H. Vuong, T.~You and Z.~Zhang, \emph{{The
  Fermionic Universal One-Loop Effective Action}},
  \href{http://arxiv.org/abs/2006.16260}{{\tt 2006.16260}}.

\bibitem{Braathen:2018htl}
J.~Braathen, M.~D. Goodsell and P.~Slavich, \emph{{Matching renormalisable
  couplings: simple schemes and a plot}},
  \href{http://dx.doi.org/10.1140/epjc/s10052-019-7093-9}{\emph{Eur. Phys. J.
  C} {\bf 79} (2019) 669}, [\href{http://arxiv.org/abs/1810.09388}{{\tt
  1810.09388}}].

\bibitem{Gabelmann:2018axh}
M.~Gabelmann, M.~Mühlleitner and F.~Staub, \emph{{Automatised matching between
  two scalar sectors at the one-loop level}},
  \href{http://dx.doi.org/10.1140/epjc/s10052-019-6570-5}{\emph{Eur. Phys. J.
  C} {\bf 79} (2019) 163}, [\href{http://arxiv.org/abs/1810.12326}{{\tt
  1810.12326}}].

\bibitem{Kwasnitza:2020wli}
T.~Kwasnitza, D.~St\"ockinger and A.~Voigt, \emph{{Improved MSSM Higgs mass
  calculation using the 3-loop FlexibleEFTHiggs approach including
  $x_{t}$-resummation}},
  \href{http://dx.doi.org/10.1007/JHEP07(2020)197}{\emph{JHEP} {\bf 07} (2020)
  197}, [\href{http://arxiv.org/abs/2003.04639}{{\tt 2003.04639}}].

\bibitem{Giudice:2011cg}
G.~F. Giudice and A.~Strumia, \emph{{Probing high-scale and split supersymmetry
  with Higgs mass measurements}},
  \href{http://dx.doi.org/10.1016/j.nuclphysb.2012.01.001}{\emph{Nucl. Phys.}
  {\bf B858} (2012) 63--83}, [\href{http://arxiv.org/abs/1108.6077}{{\tt
  1108.6077}}].

\bibitem{Draper:2013oza}
P.~Draper, G.~Lee and C.~E.~M. Wagner, \emph{{Precise estimates of the Higgs
  mass in heavy supersymmetry}},
  \href{http://dx.doi.org/10.1103/PhysRevD.89.055023}{\emph{Phys. Rev.} {\bf
  D89} (2014) 055023}, [\href{http://arxiv.org/abs/1312.5743}{{\tt
  1312.5743}}].

\bibitem{Bagnaschi:2014rsa}
E.~Bagnaschi, G.~F. Giudice, P.~Slavich and A.~Strumia, \emph{{Higgs mass and
  unnatural supersymmetry}},
  \href{http://dx.doi.org/10.1007/JHEP09(2014)092}{\emph{JHEP} {\bf 09} (2014)
  092}, [\href{http://arxiv.org/abs/1407.4081}{{\tt 1407.4081}}].

\bibitem{Bagnaschi:2017xid}
E.~Bagnaschi, J.~Pardo~Vega and P.~Slavich, \emph{{Improved determination of
  the Higgs mass in the MSSM with heavy superpartners}},
  \href{http://dx.doi.org/10.1140/epjc/s10052-017-4885-7}{\emph{Eur. Phys. J.}
  {\bf C77} (2017) 334}, [\href{http://arxiv.org/abs/1703.08166}{{\tt
  1703.08166}}].

\bibitem{Vega:2015fna}
J.~P. Vega and G.~Villadoro, \emph{{SusyHD: Higgs mass determination in
  supersymmetry}}, \href{http://dx.doi.org/10.1007/JHEP07(2015)159}{\emph{JHEP}
  {\bf 07} (2015) 159}, [\href{http://arxiv.org/abs/1504.05200}{{\tt
  1504.05200}}].

\bibitem{Bagnaschi:2019esc}
E.~Bagnaschi, G.~Degrassi, S.~Paßehr and P.~Slavich, \emph{{Full two-loop QCD
  corrections to the Higgs mass in the MSSM with heavy superpartners}},
  \href{http://dx.doi.org/10.1140/epjc/s10052-019-7417-9}{\emph{Eur. Phys. J.
  C} {\bf 79} (2019) 910}, [\href{http://arxiv.org/abs/1908.01670}{{\tt
  1908.01670}}].

\bibitem{Bahl:2020tuq}
H.~Bahl, I.~Sobolev and G.~Weiglein, \emph{{The light MSSM Higgs boson mass for
  large $\tan\beta$ and complex input parameters}},
  \href{http://arxiv.org/abs/2009.07572}{{\tt 2009.07572}}.

\bibitem{Harlander:2018yhj}
R.~V. Harlander, J.~Klappert, A.~D. Ochoa~Franco and A.~Voigt, \emph{{The light
  CP-even MSSM Higgs mass resummed to fourth logarithmic order}},
  \href{http://dx.doi.org/10.1140/epjc/s10052-018-6351-6}{\emph{Eur. Phys. J.}
  {\bf C78} (2018) 874}, [\href{http://arxiv.org/abs/1807.03509}{{\tt
  1807.03509}}].

\bibitem{Allanach:2018fif}
B.~C. Allanach and A.~Voigt, \emph{{Uncertainties in the Lightest $CP$ Even
  Higgs Boson Mass Prediction in the Minimal Supersymmetric Standard Model:
  Fixed Order Versus Effective Field Theory Prediction}},
  \href{http://dx.doi.org/10.1140/epjc/s10052-018-6046-z}{\emph{Eur. Phys. J.}
  {\bf C78} (2018) 573}, [\href{http://arxiv.org/abs/1804.09410}{{\tt
  1804.09410}}].

\bibitem{Bahl:2019hmm}
H.~Bahl, S.~Heinemeyer, W.~Hollik and G.~Weiglein, \emph{{Theoretical
  uncertainties in the MSSM Higgs boson mass calculation}},
  \href{http://dx.doi.org/10.1140/epjc/s10052-020-8079-3}{\emph{Eur. Phys. J.
  C} {\bf 80} (2020) 497}, [\href{http://arxiv.org/abs/1912.04199}{{\tt
  1912.04199}}].

\bibitem{Haber:1993an}
H.~E. Haber and R.~Hempfling, \emph{{The renormalization group improved Higgs
  sector of the minimal supersymmetric model}},
  \href{http://dx.doi.org/10.1103/PhysRevD.48.4280}{\emph{Phys. Rev.} {\bf D48}
  (1993) 4280--4309}, [\href{http://arxiv.org/abs/hep-ph/9307201}{{\tt
  hep-ph/9307201}}].

\bibitem{Lee:2015uza}
G.~Lee and C.~E.~M. Wagner, \emph{{Higgs bosons in heavy supersymmetry with an
  intermediate m$_A$}},
  \href{http://dx.doi.org/10.1103/PhysRevD.92.075032}{\emph{Phys. Rev.} {\bf
  D92} (2015) 075032}, [\href{http://arxiv.org/abs/1508.00576}{{\tt
  1508.00576}}].

\bibitem{Bagnaschi:2015pwa}
E.~Bagnaschi, F.~Brümmer, W.~Buchmüller, A.~Voigt and G.~Weiglein,
  \emph{{Vacuum stability and supersymmetry at high scales with two Higgs
  doublets}}, \href{http://dx.doi.org/10.1007/JHEP03(2016)158}{\emph{JHEP} {\bf
  03} (2016) 158}, [\href{http://arxiv.org/abs/1512.07761}{{\tt 1512.07761}}].

\bibitem{Athron:2017fvs}
P.~Athron, M.~Bach, D.~Harries, T.~Kwasnitza, J.~Park, D.~Stöckinger et~al.,
  \emph{{FlexibleSUSY 2.0: Extensions to investigate the phenomenology of SUSY
  and non-SUSY models}},
  \href{http://dx.doi.org/10.1016/j.cpc.2018.04.016}{\emph{Comput. Phys.
  Commun.} {\bf 230} (2018) 145--217},
  [\href{http://arxiv.org/abs/1710.03760}{{\tt 1710.03760}}].

\bibitem{Benakli:2018vqz}
K.~Benakli, M.~D. Goodsell and S.~L. Williamson, \emph{{Higgs alignment from
  extended supersymmetry}},
  \href{http://dx.doi.org/10.1140/epjc/s10052-018-6125-1}{\emph{Eur. Phys. J.
  C} {\bf 78} (2018) 658}, [\href{http://arxiv.org/abs/1801.08849}{{\tt
  1801.08849}}].

\bibitem{Bahl:2018jom}
H.~Bahl and W.~Hollik, \emph{{Precise prediction of the MSSM Higgs boson masses
  for low $M_{A}$}},
  \href{http://dx.doi.org/10.1007/JHEP07(2018)182}{\emph{JHEP} {\bf 07} (2018)
  182}, [\href{http://arxiv.org/abs/1805.00867}{{\tt 1805.00867}}].

\bibitem{Murphy:2019qpm}
N.~Murphy and H.~Rzehak, \emph{{Higgs-Boson Masses and Mixings in the MSSM with
  CP Violation and Heavy SUSY Particles}},
  \href{http://arxiv.org/abs/1909.00726}{{\tt 1909.00726}}.

\bibitem{Heinemeyer:1998yj}
S.~Heinemeyer, W.~Hollik and G.~Weiglein, \emph{{FeynHiggs: A Program for the
  calculation of the masses of the neutral CP even Higgs bosons in the MSSM}},
  \href{http://dx.doi.org/10.1016/S0010-4655(99)00364-1}{\emph{Comput. Phys.
  Commun.} {\bf 124} (2000) 76--89},
  [\href{http://arxiv.org/abs/hep-ph/9812320}{{\tt hep-ph/9812320}}].

\bibitem{Heinemeyer:1998np}
S.~Heinemeyer, W.~Hollik and G.~Weiglein, \emph{{The masses of the neutral
  CP-even Higgs bosons in the MSSM: Accurate analysis at the two loop level}},
  \href{http://dx.doi.org/10.1007/s100529900006,
  10.1007/s100520050537}{\emph{Eur. Phys. J.} {\bf C9} (1999) 343--366},
  [\href{http://arxiv.org/abs/hep-ph/9812472}{{\tt hep-ph/9812472}}].

\bibitem{Hahn:2009zz}
T.~Hahn, S.~Heinemeyer, W.~Hollik, H.~Rzehak and G.~Weiglein, \emph{{FeynHiggs:
  A program for the calculation of MSSM Higgs-boson observables - Version
  2.6.5}}, \href{http://dx.doi.org/10.1016/j.cpc.2009.02.014}{\emph{Comput.
  Phys. Commun.} {\bf 180} (2009) 1426--1427}.

\bibitem{Degrassi:2002fi}
G.~Degrassi, S.~Heinemeyer, W.~Hollik, P.~Slavich and G.~Weiglein,
  \emph{{Towards high precision predictions for the MSSM Higgs sector}},
  \href{http://dx.doi.org/10.1140/epjc/s2003-01152-2}{\emph{Eur. Phys. J.} {\bf
  C28} (2003) 133--143}, [\href{http://arxiv.org/abs/hep-ph/0212020}{{\tt
  hep-ph/0212020}}].

\bibitem{Frank:2006yh}
M.~Frank, T.~Hahn, S.~Heinemeyer, W.~Hollik, H.~Rzehak and G.~Weiglein,
  \emph{{The Higgs boson masses and mixings of the complex MSSM in the
  Feynman-diagrammatic approach}},
  \href{http://dx.doi.org/10.1088/1126-6708/2007/02/047}{\emph{JHEP} {\bf 02}
  (2007) 047}, [\href{http://arxiv.org/abs/hep-ph/0611326}{{\tt
  hep-ph/0611326}}].

\bibitem{Hahn:2013ria}
T.~Hahn, S.~Heinemeyer, W.~Hollik, H.~Rzehak and G.~Weiglein,
  \emph{{High-precision predictions for the light CP-even Higgs boson mass of
  the Minimal Supersymmetric Standard Model}},
  \href{http://dx.doi.org/10.1103/PhysRevLett.112.141801}{\emph{Phys. Rev.
  Lett.} {\bf 112} (2014) 141801}, [\href{http://arxiv.org/abs/1312.4937}{{\tt
  1312.4937}}].

\bibitem{Bahl:2016brp}
H.~Bahl and W.~Hollik, \emph{{Precise prediction for the light MSSM Higgs boson
  mass combining effective field theory and fixed-order calculations}},
  \href{http://dx.doi.org/10.1140/epjc/s10052-016-4354-8}{\emph{Eur. Phys. J.}
  {\bf C76} (2016) 499}, [\href{http://arxiv.org/abs/1608.01880}{{\tt
  1608.01880}}].

\bibitem{Bahl:2017aev}
H.~Bahl, S.~Heinemeyer, W.~Hollik and G.~Weiglein, \emph{{Reconciling EFT and
  hybrid calculations of the light MSSM Higgs-boson mass}},
  \href{http://dx.doi.org/10.1140/epjc/s10052-018-5544-3}{\emph{Eur. Phys. J.}
  {\bf C78} (2018) 57}, [\href{http://arxiv.org/abs/1706.00346}{{\tt
  1706.00346}}].

\bibitem{Bahl:2018qog}
H.~Bahl, T.~Hahn, S.~Heinemeyer, W.~Hollik, S.~Paßehr, H.~Rzehak et~al.,
  \emph{{Precision calculations in the MSSM Higgs-boson sector with FeynHiggs
  2.14}}, \href{http://dx.doi.org/10.1016/j.cpc.2019.107099}{\emph{Comput.
  Phys. Commun.} {\bf 249} (2020) 107099},
  [\href{http://arxiv.org/abs/1811.09073}{{\tt 1811.09073}}].

\bibitem{Appelquist:1974tg}
T.~Appelquist and J.~Carazzone, \emph{{Infrared Singularities and Massive
  Fields}}, \href{http://dx.doi.org/10.1103/PhysRevD.11.2856}{\emph{Phys. Rev.}
  {\bf D11} (1975) 2856}.

\bibitem{Muhlleitner:2008yw}
M.~Mühlleitner, H.~Rzehak and M.~Spira, \emph{{MSSM Higgs Boson Production via
  Gluon Fusion: The Large Gluino Mass Limit}},
  \href{http://dx.doi.org/10.1088/1126-6708/2009/04/023}{\emph{JHEP} {\bf 04}
  (2009) 023}, [\href{http://arxiv.org/abs/0812.3815}{{\tt 0812.3815}}].

\bibitem{Bahl:2019wzx}
H.~Bahl, I.~Sobolev and G.~Weiglein, \emph{{Precise prediction for the mass of
  the light MSSM Higgs boson for the case of a heavy gluino}},
  \href{http://dx.doi.org/10.1016/j.physletb.2020.135644}{\emph{Phys. Lett. B}
  {\bf 808} (2020) 135644}, [\href{http://arxiv.org/abs/1912.10002}{{\tt
  1912.10002}}].

\bibitem{Fuchs:2016swt}
E.~Fuchs and G.~Weiglein, \emph{{Breit-Wigner approximation for propagators of
  mixed unstable states}},
  \href{http://dx.doi.org/10.1007/JHEP09(2017)079}{\emph{JHEP} {\bf 09} (2017)
  079}, [\href{http://arxiv.org/abs/1610.06193}{{\tt 1610.06193}}].

\bibitem{Fuchs:2017wkq}
E.~Fuchs and G.~Weiglein, \emph{{Impact of CP-violating interference effects on
  MSSM Higgs searches}},
  \href{http://dx.doi.org/10.1140/epjc/s10052-018-5543-4}{\emph{Eur. Phys. J.
  C} {\bf 78} (2018) 87}, [\href{http://arxiv.org/abs/1705.05757}{{\tt
  1705.05757}}].

\bibitem{Bahl:2018ykj}
H.~Bahl, \emph{{Pole mass determination in presence of heavy particles}},
  \href{http://dx.doi.org/10.1007/JHEP02(2019)121}{\emph{JHEP} {\bf 02} (2019)
  121}, [\href{http://arxiv.org/abs/1812.06452}{{\tt 1812.06452}}].

\bibitem{Steinhauser:2002rq}
M.~Steinhauser, \emph{{Results and techniques of multiloop calculations}},
  \href{http://dx.doi.org/10.1016/S0370-1573(02)00017-0}{\emph{Phys. Rept.}
  {\bf 364} (2002) 247--357}, [\href{http://arxiv.org/abs/hep-ph/0201075}{{\tt
  hep-ph/0201075}}].

\bibitem{Denner:1994xt}
A.~Denner, G.~Weiglein and S.~Dittmaier, \emph{{Application of the background
  field method to the electroweak standard model}},
  \href{http://dx.doi.org/10.1016/0550-3213(95)00037-S}{\emph{Nucl. Phys. B}
  {\bf 440} (1995) 95--128}, [\href{http://arxiv.org/abs/hep-ph/9410338}{{\tt
  hep-ph/9410338}}].

\bibitem{Grassi:1999nb}
P.~A. Grassi, \emph{{Renormalization of nonsemisimple gauge models with the
  background field method}},
  \href{http://dx.doi.org/10.1016/S0550-3213(99)00457-5}{\emph{Nucl. Phys. B}
  {\bf 560} (1999) 499--550}, [\href{http://arxiv.org/abs/hep-th/9908188}{{\tt
  hep-th/9908188}}].

\bibitem{Collins:2016aya}
J.~C. Collins and J.~Vermaseren, \emph{{Axodraw Version 2}},
  \href{http://arxiv.org/abs/1606.01177}{{\tt 1606.01177}}.

\bibitem{Collins:1984xc}
J.~C. Collins, \emph{{Renormalization}}, vol.~26 of \emph{Cambridge Monographs
  on Mathematical Physics}.
\newblock Cambridge University Press, Cambridge, 1986,
  \href{http://dx.doi.org/10.1017/CBO9780511622656}{10.1017/CBO9780511622656}.

\bibitem{Luthe:2017ttg}
T.~Luthe, A.~Maier, P.~Marquard and Y.~Schroder, \emph{{The five-loop Beta
  function for a general gauge group and anomalous dimensions beyond Feynman
  gauge}}, \href{http://dx.doi.org/10.1007/JHEP10(2017)166}{\emph{JHEP} {\bf
  10} (2017) 166}, [\href{http://arxiv.org/abs/1709.07718}{{\tt 1709.07718}}].

\bibitem{Bahl:2020mjy}
H.~Bahl, N.~Murphy and H.~Rzehak, \emph{{Hybrid calculation of the MSSM Higgs
  boson masses using the complex THDM as EFT}},
  \href{http://arxiv.org/abs/2010.04711}{{\tt 2010.04711}}.

\bibitem{Kublbeck:1990xc}
J.~Küblbeck, M.~Böhm and A.~Denner, \emph{{Feyn Arts: Computer Algebraic
  Generation of Feynman Graphs and Amplitudes}},
  \href{http://dx.doi.org/10.1016/0010-4655(90)90001-H}{\emph{Comput. Phys.
  Commun.} {\bf 60} (1990) 165--180}.

\bibitem{Eck:1992ms}
H.~Eck and J.~Kublbeck, \emph{{Computeralgebraic generation of Feynman graphs
  and amplitudes}},  in \emph{{2nd International Workshop on Software
  Engineering, Artificial Intelligence and Expert Systems for High-energy and
  Nuclear Physics}}, pp.~677--682, 1992.

\bibitem{Hahn:2000kx}
T.~Hahn, \emph{{Generating Feynman diagrams and amplitudes with FeynArts 3}},
  \href{http://dx.doi.org/10.1016/S0010-4655(01)00290-9}{\emph{Comput. Phys.
  Commun.} {\bf 140} (2001) 418--431},
  [\href{http://arxiv.org/abs/hep-ph/0012260}{{\tt hep-ph/0012260}}].

\bibitem{Weiglein:1992hd}
G.~Weiglein, R.~Mertig, R.~Scharf and M.~B{\"o}hm{\emph{,\, New computing
  techniques in physics research II, ed. D.Perret-Gallix (World Scientific,
  Singapore, 1992)} 617}.

\bibitem{Weiglein:1993hd}
G.~Weiglein, R.~Scharf and M.~Bohm, \emph{{Reduction of general two loop
  selfenergies to standard scalar integrals}},
  \href{http://dx.doi.org/10.1016/0550-3213(94)90325-5}{\emph{Nucl. Phys. B}
  {\bf 416} (1994) 606--644}, [\href{http://arxiv.org/abs/hep-ph/9310358}{{\tt
  hep-ph/9310358}}].

\bibitem{Hahn:1998yk}
T.~Hahn and M.~Perez-Victoria, \emph{{Automatized one loop calculations in
  four-dimensions and D-dimensions}},
  \href{http://dx.doi.org/10.1016/S0010-4655(98)00173-8}{\emph{Comput. Phys.
  Commun.} {\bf 118} (1999) 153--165},
  [\href{http://arxiv.org/abs/hep-ph/9807565}{{\tt hep-ph/9807565}}].

\bibitem{Hahn:2015gaa}
T.~Hahn and S.~Paßehr, \emph{{Implementation of the
  $\mathcal{O}{\left(\alpha_t^2\right)}$ MSSM Higgs-mass corrections in
  $\tt{FeynHiggs}$}},
  \href{http://dx.doi.org/10.1016/j.cpc.2017.01.026}{\emph{Comput. Phys.
  Commun.} {\bf 214} (2017) 91--97},
  [\href{http://arxiv.org/abs/1508.00562}{{\tt 1508.00562}}].

\bibitem{Carena:2015uoe}
M.~Carena, J.~Ellis, J.~S. Lee, A.~Pilaftsis and C.~E.~M. Wagner, \emph{{CP
  Violation in Heavy MSSM Higgs Scenarios}},
  \href{http://dx.doi.org/10.1007/JHEP02(2016)123}{\emph{JHEP} {\bf 02} (2016)
  123}, [\href{http://arxiv.org/abs/1512.00437}{{\tt 1512.00437}}].

\bibitem{Herren:2017uxn}
F.~Herren, L.~Mihaila and M.~Steinhauser, \emph{{Gauge and Yukawa coupling beta
  functions of two-Higgs-doublet models to three-loop order}},
  \href{http://dx.doi.org/10.1103/PhysRevD.97.015016}{\emph{Phys. Rev. D} {\bf
  97} (2018) 015016}, [\href{http://arxiv.org/abs/1712.06614}{{\tt
  1712.06614}}].

\bibitem{Bahl:2018zmf}
E.~Bagnaschi et~al., \emph{{MSSM Higgs Boson Searches at the LHC: Benchmark
  Scenarios for Run 2 and Beyond}},
  \href{http://dx.doi.org/10.1140/epjc/s10052-019-7114-8}{\emph{Eur. Phys. J.}
  {\bf C79} (2019) 617}, [\href{http://arxiv.org/abs/1808.07542}{{\tt
  1808.07542}}].

\bibitem{Bahl:2019ago}
H.~Bahl, S.~Liebler and T.~Stefaniak, \emph{{MSSM Higgs benchmark scenarios for
  Run 2 and beyond: the low $\tan \beta $ region}},
  \href{http://dx.doi.org/10.1140/epjc/s10052-019-6770-z}{\emph{Eur. Phys. J.
  C} {\bf 79} (2019) 279}, [\href{http://arxiv.org/abs/1901.05933}{{\tt
  1901.05933}}].

\end{thebibliography}\endgroup

\end{document}